\documentclass[aps,nofootinbib,showpacs,preprintnumbers,amsmath,amssymb]{revtex4}
\def\bes{\begin{subequations}}
\def\ees{\end{subequations}}
\def\be{\begin{equation}}
\def\ee{\end{equation}}
\def\bea{\begin{eqnarray}}
\def\eea{\end{eqnarray}}
\def\ba{\begin{eqnarray}}
\def\ea{\end{eqnarray}}
\def\bear{\begin{array}}
\def\eear{\end{array}}
\def\p1sl{\displaystyle{\not}p_1}
\def\p2sl{\displaystyle{\not}p_2}

\newcommand{\K}{{\widetilde {\cal K}}}
\newcommand{\G}{{\widetilde {\Gamma}}}
\newcommand{\bG}{{\overline{\Gamma}}}
\newcommand{\bL}{{\overline{L}}}
\newcommand{\bBr}{{\overline {\rm Br}}}

\usepackage{graphics}
\usepackage{graphicx}
\usepackage{dcolumn}
\usepackage{bm}
\usepackage{epsfig}
\usepackage{graphicx}
\usepackage{multirow}
\usepackage{dcolumn}
\usepackage{graphicx,epsfig}%

\begin{document}
\preprint{USM-TH-334}

\title{Probing the Majorana neutrinos and their CP violation in decays of 
charged scalar mesons $\pi, K, D, D_s, B, B_c$\footnote{1503.01358v3: 37 pages, 14 figs; minor typographical errors corrected; published in Symmetry 7 (2015) 726-773} }
\author{Gorazd Cveti\v{c}$^1$}
\email{gorazd.cvetic@usm.cl}
\author{Claudio Dib$^{1,2}$}
\email{claudio.dib@usm.cl}
\author{C.~S.~Kim$^3$}
\email{cskim@yonsei.ac.kr}
\author{Jilberto Zamora-Sa\'a$^1$}
\email{jilberto.zamora@usm.cl}
\affiliation{$^1$
Department of Physics, Universidad T\'ecnica Federico Santa Mar\'ia (UTFSM), Valpara\'iso, Chile\\
$^2$
Centro Cient\'{\i}fico Tecnol\'ogico de Valpara\'{\i}so, UTFSM, Valpara\'iso, Chile\\
$^3$Department of Physics and IPAP, Yonsei University, Seoul 120-749, Korea}

\date{\today}

\begin{abstract}
Some of the outstanding questions of
particle physics today concern the neutrino sector, in particular whether there are more neutrinos than those already known and whether they are Dirac or Majorana particles.
%
%
There are different ways to explore these issues. In this article we describe
neutrino-mediated decays of charged pseudoscalar mesons such as
$\pi^{\pm}$, $K^{\pm}$ and $B^{\pm}$, in scenarios where extra neutrinos
are heavy and can be on their mass shell. We discuss semileptonic and leptonic decays of such kinds.
We investigate possible ways of using
these decays in order to distinguish between the Dirac and Majorana character of
neutrinos. Further, we argue that there are significant
possibilities of detecting CP violation in such decays
when there are at least two almost
degenerate Majorana neutrinos involved.
This latter type of scenario fits well into the known
neutrino minimal standard model
($\nu$MSM)
which could simultaneously
explain the Dark Matter and Baryon Asymmetry of the Universe.
\end{abstract}

\maketitle


\section{Introduction}
\label{sec:intr}




To date it is unclear whether the neutrinos we know are Dirac of Majorana fermions.
Unlike Dirac fermions, Majorana fermions cannot be distinguished from their own antiparticles. As a consequence, processes involving Dirac neutrinos conserve charges such as Lepton Number,
while processes involving Majorana neutrinos will not conserve them.
There exist several processes which may clarify the
Majorana or Dirac nature of neutrinos. Among such processes
the most prominent are neutrinoless double beta decays
($0 \nu \beta \beta$) in nuclei \cite{0nubb,0nubb2,0nubb3,0nubb31,0nubb4,0nubb5,0nubb6,0nubb7,0nubb8,0nubb9}. Other such processes
are specific scattering processes~
\cite{scatt1,scatt12,scatt13,scatt14,scatt2,scatt3,scatt4,DevscattLR,DevscattLR2,Devscattssaw,Devscattssaw1},
%
and rare meson decays
\cite{LittSh,LittSh2,DGKS,Ali,IvKo,GoJe,Atre,HKS,QLD,QLD1,QLD2,Abada,Wang,Boya,CDKK,CDK,CKZ,CKZ2,DCK}. 
%

A related issue in neutrinos physics is the absolute mass values of the known neutrinos. While
the experimental evidence of neutrino oscillations within the known three flavor states \cite{oscatm,oscsol,oscsol2,oscsol3,oscsol4,oscnuc} clearly shows that these particles cannot all be massless, the oscillations are only sensitive to mass differences, not to their absolute values. In contrast, $0 \nu \beta \beta$ decays are sensitive to the absolute mass and may help in their determination, if neutrinos turn out to be Majorana particles. So far the best bounds on the absolute masses of the light neutrinos  come from Cosmology $m_\nu \gtrsim 0.23$ eV \cite{PlanckColl}.

A pending question is then why light neutrinos are so light, specifically so much lighter than all other  Standard Model (SM) fermions. Interesting enough, the existence of such very light neutrinos
can be explained via the seesaw mechanism \cite{seesaw,seesaw2,seesaw3,seesaw4,seesaw5} where more neutrinos are required
and where all of them are, in general, Majorana particles. In the simplest form of this mechanism,
the masses of the light neutrinos are
$\sim {\cal M}_D^2/{\cal M}_R$ ($\lesssim$1 eV),
where ${\cal M}_D$ is an electroweak scale or lower.
At the same time, additional neutrinos, usually much heavier (with masses ${\cal M}_R \gg 1$ TeV)
and sterile under electroweak interactions  except through small mixing with the SM flavors, are required.
This mixing is suppressed as $\sim {\cal M}_D/{\cal M}_R$ ($\ll$1).
Besides the simplest scenario, there are other seesaw scenarios in which the heavy neutrinos may have lower masses, namely near or below 1 TeV
%
\cite{Wyler1, Witten1, Mohapatra1, Malinsky1,Devseesaw,Devseesaw2,Devseesaw3}
%
and even near the 1 GeV scale or below \cite{scatt2,nuMSM,nuMSM2,HeAAS,HeAAS2,KS,AMP},
and at the same time their mixing with the SM flavors may not be so extremely suppressed as in the original scenarios.

In the first part of this work we
discuss lepton number violating (LNV)
semileptonic decays of charged pseudoscalar
mesons such as  $K^{\pm}$ and $B^{\pm}$, mediated by
a heavy Majorana neutrino on its mass shell, \textit{cf.}~Ref.~\cite{CDKK}.
The pions, which are the lightest mesons, can have only leptonic decays;
we discuss hypothetical leptonic decays that could be mediated by on-shell
heavy neutrinos.  Such decays could be either lepton number
conserving (LNC) or lepton number violating (LNV) when the mediating neutrinos are
Majorana particles, \textit{cf.}~Ref.~\cite{CDK}, while only LNC decays occur when the neutrino is of Dirac type. We present ways of determining the
nature of neutrinos using the differential decay rates of these processes.

Yet another interesting issue in neutrino physics is the possible existence of CP
violation in the lepton sector, a phenomenon that could be measured,
for example, in neutrino oscillations \cite{oscCP}.
Alternatively, as we present in the second part of this work, leptonic CP violation
may show in leptonic LNC and LNV decays of charged pions, \textit{cf.}~Ref.~\cite{CKZ},
or in semileptonic LNV
decays of  $K^{\pm}$ and $B^{\pm}$,
\textit{cf.}~Refs.~\cite{CKZ2,DCK}.
It turns out that such CP violation
becomes appreciable and possibly detectable in these decays if the scenario contains
at least two on-shell heavy neutrinos that are almost degenerate.
Interestingly, this scenario fits well into the neutrino
minimal standard model ($\nu$MSM) \cite{nuMSM,nuMSM2,Shapo,Shapo2,Shapo3,Shapo4,Shapo5,Shapo6} which
contains two almost degenerate Majorana neutrinos of mass
$\sim$ 1 GeV and another lighter neutrino of mass $\sim 10^1$ keV,
besides the three light neutrinos of mass  $\lesssim$1 eV.
This model can explain simultaneously the existence of neutrino
oscillations, dark matter and baryon asymmetry of the Universe.
Furthermore, in more general frameworks of low-scale seesaw, baryon asymmetry (but not dark matter) is
explained while keeping even larger values of the heavy-light mixing \cite{1404.7114} than in the $\nu$MSM,
and in such frameworks the case of almost degenerate Majorana neutrinos is preferred \cite{1502.00477}
since it allows larger mixings.
CP violation effects in the neutrino sector in scenarios with nearly degenerate heavy neutrinos have also been investigated earlier \cite{Pilaftsis,Pilaftsis2}
using a more detailed formalism, although it amounts to the same effect described here which is the interference of amplitudes with two slightly different dispersive and absorptive parts in the neutrino self energy.

%

In Section~\ref{sec:BrM} we
discuss the LNV semileptonic decays of mesons
$M^{\pm} \to \ell_1^{\pm} \ell_2^{\pm} M^{\prime \mp}$,
mediated by an on-shell Majorana neutrino (which henceforth we call $N$)
where $M$ is a heavy pseudoscalar \linebreak($M=K,D, D_s, B, B_c$), $M^{'}$ is
a lighter pseudoscalar, and $\ell_j$ ($j=1,2$) are charged leptons,
and we present there the corresponding branching ratios.
In Section~\ref{sec:BrPi} we present the expressions and values of the
branching ratios for the LNC and LNV leptonic decays of charged
pions mediated by on-shell $N$ sterile neutrino, {
$\pi^{\pm} \to e^{\pm} N \to e^{\pm} e^{\pm} \mu^{\mp} \nu$,}
as well as the differential
branching ratio $d {\rm Br}/d E_{\mu}$ for these decays. We discuss the
possibilities of detecting such branching ratios and to discern from them
the Majorana or Dirac nature of neutrinos.
In Section~\ref{sec:BrCPV} we then extend the analysis of the mentioned
leptonic and semileptonic decays to a scenario where we have at least
two heavy on-shell sterile neutrinos involved ($N_1$, $N_2$), and
we present an analysis of CP-violating asymmetries ${\cal A}_{\rm CP}
\equiv [\Gamma(M^-) -\Gamma(M^+)]/[\Gamma(M^-) +\Gamma(M^+)]$
for such processes.
In Appendices 1 and 2 we present
explicit formulas for our LNV semileptonic decays,
and in Appendices 4 and 5 explicit formulas needed for the analysis of our
LNC and LNV leptonic decays of the charged pion.
Appendix 3 contains formulas needed for evaluation of the
decay width of the heavy neutrino $N$,
and in Appendix 6 we derive an identity relevant for
CP violation asymmetry.
In Section~\ref{sec:summ} we discuss and summarize our results.

\section{Lepton Number Violating Semileptonic Decays of Scalar Mesons}
\label{sec:BrM}

\begin{figure}[htb] 
\begin{minipage}[b]{.49\linewidth}
\includegraphics[width=80mm]{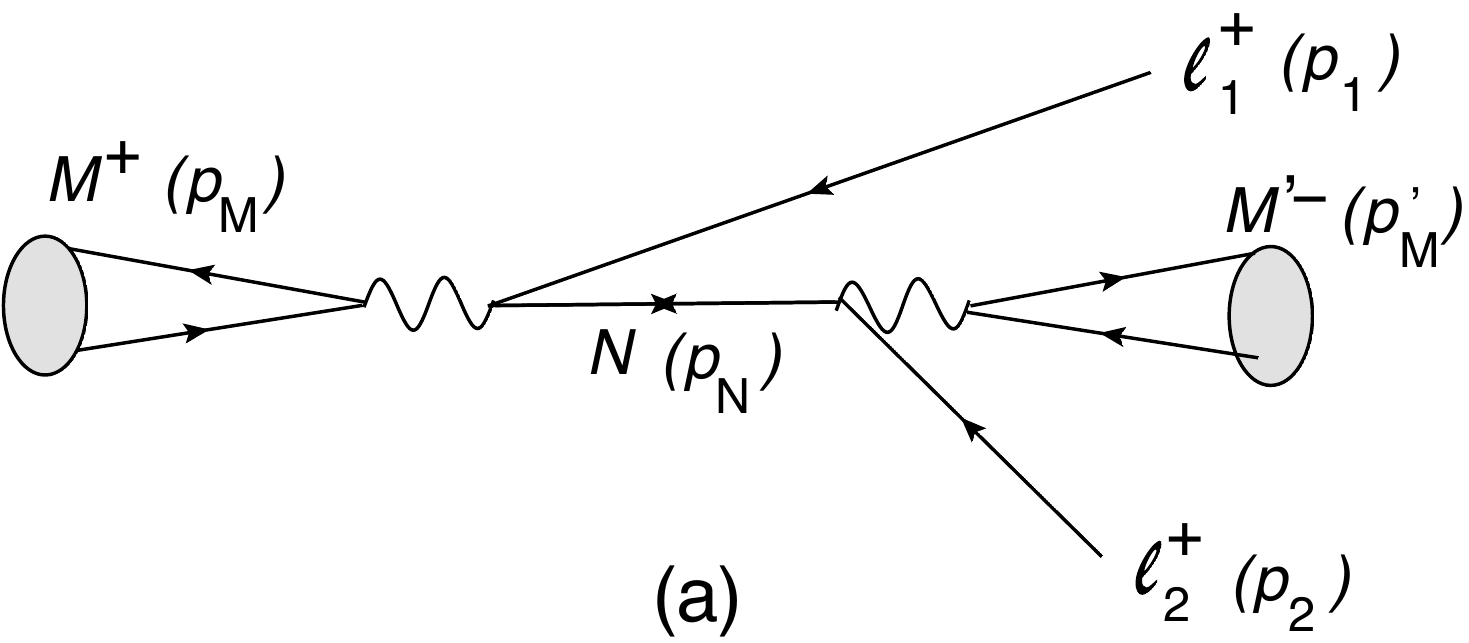}
\end{minipage}
\begin{minipage}[b]{.49\linewidth}
\includegraphics[width=80mm]{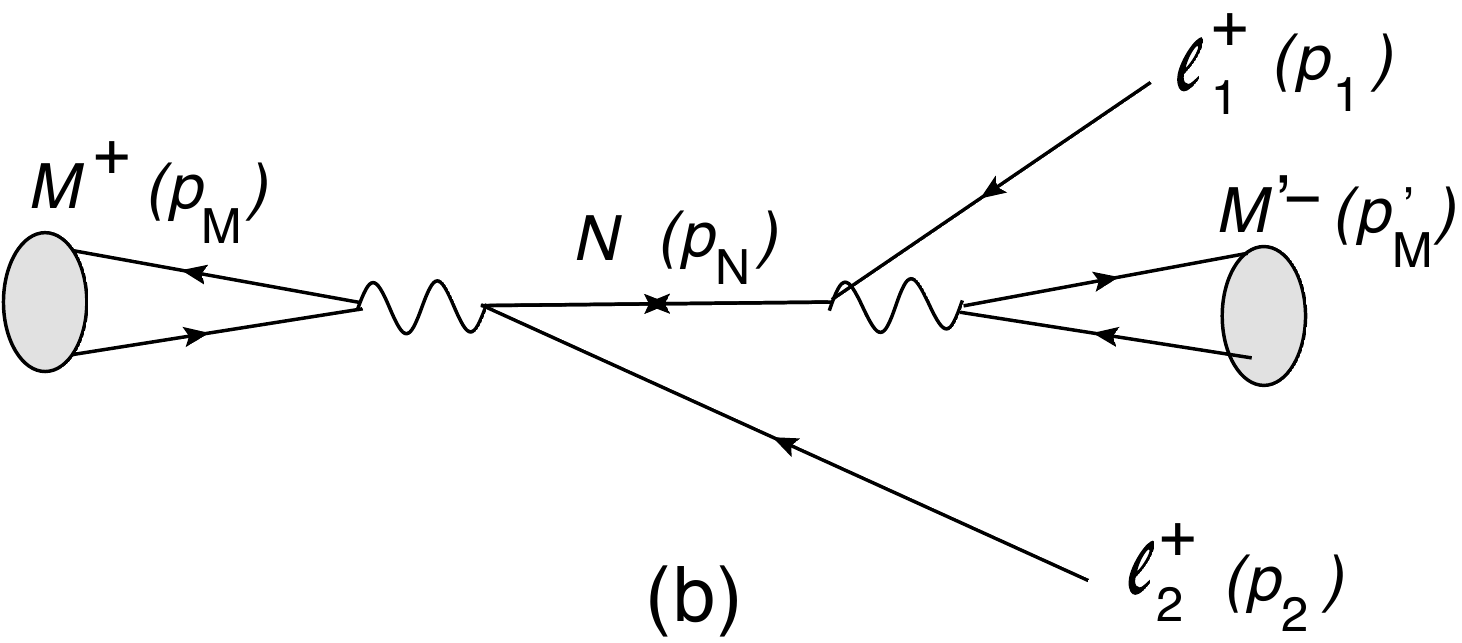}
\end{minipage}
\caption{The lepton number violating (LNV) semileptonic decay
$M^+(p_M) \to \ell^+_1(p_1) \ell^+_2(p_2) M^{' -}(p_{M'})$ mediated
by a Majorana neutrino $N$:
(\textbf{a}) the direct ($D$) channel; (\textbf{b}) the crossed ($C$) channel.}
\label{FigMMp}
\end{figure}

If there is a Majorana sterile neutrino $N$ with mass $M_N \sim 1$ GeV,
its existence could be discerned by detecting semileptonic
LNV decays of heavy mesons mediated by on-shell $N$. Here we will
consider such LNV decays $M^{\pm} \to \ell_1^{\pm} N \to
\ell_1^{\pm} \ell_2^{\pm} M^{' \mp}$, where $M$ and $M^{'}$ are
pseudoscalar mesons ($M=K, D, D_s, B, B_c$; $M^{'}= \pi, K, D, D_s$)
while $\ell_1$ and $\ell_2$ are charged leptons ($ e, \mu$ or $ \tau$), \textit{cf.}~Figure~\ref{FigMMp}.
Large part of this Section uses the results of Refs.~\cite{CDKK,CDK}, and
for certain general formulas, those of Refs.~\cite{CKZ2,DCK}.

We consider a scenario where we have at least one heavy sterile neutrino $N$,
which has (small) mixing $B_{\ell N}$ with the three known
neutrino flavors $\nu_{\ell}$ ($\ell=e, \mu, \tau$)
\be
\nu_{\ell} = \sum_{k=1}^3 B_{\ell \nu_k} \nu_k + B_{\ell N} N + \ldots \
\label{mixN}
\ee
Here, $\nu_k$ ($k=1,2,3$) are the three light mass neutrino eingenstates.
The considered decays may be appreciable only if $N$
can go  on its mass shell
(in the $s$-type channel, Figure~\ref{FigMMp}),  creating a very large resonant enhancement of order $m_N/\Gamma_N$, condition which is fulfilled if
\bea
(M_{M'} + M_{\ell_2}) &<& M_{N} < (M_{M}-M_{\ell_1}) \ , \ {\rm or/and}
\nonumber\\
(M_{M'} + M_{\ell_1}) &<& M_{N} < (M_{M}-M_{\ell_2}) \
\label{MNjint}
\eea

Consequently, only tree level resonant amplitudes need to be considered.
The mixing matrix $B$ appearing in Equation~(\ref{mixN}) should be unitary,
implying that the PMNS $3 \times 3$ block $B_{\ell \nu_k}$ ($\ell=e, \mu, \tau$
and $k=1,2,3$) is in general not unitary.
%
If one adds extra and heavy neutrinos---as in most seesaw \linebreak models---the unitarity of $B$ thus provides upper limit constraints on
the heavy-to-light mixing \mbox{elements~\cite{Akhmedov,Abada,Qian,Basso}}. This is in part one of the reasons for the high suppression suffered by all lepton flavour violating processes involving heavy neutrinos.
%


\subsection{Branching Ratio for $M^{\pm} \to \ell_1^{\pm} \ell_2^{\pm} M^{' \mp}$}
\label{sec:BrMsub}

The decay width of the considered decays can be written
as final particles' phase space integral of the
square of the reduced decay amplitude ${\cal T}(M^{\pm})$
(summed over helicities of charged leptons)
 \be
\Gamma(M^{\pm} \to \ell_1^{\pm} \ell_2^{\pm} M^{'\mp})
=  (2 - \delta_{\ell_1 \ell_2} ) \frac{1}{2!} \frac{1}{2 M_M} \frac{1}{(2 \pi)^5}
\int d_3 \; | {\cal T}(M^{\pm}) | ^2 \
\label{GM1}
\ee

Factor $1/2!$ above is the symmetry factor when the two produced
leptons are equal; $d_3$ is the integration differential of the
final three-particle phase space
\be
d_3 \equiv \frac{d^3 {\vec p}_1}{2 E_{\ell_1}({\vec p}_1)}
 \frac{d^3 {\vec p}_2}{2 E_{\ell_2}({\vec p}_2)}
 \frac{d^3 {\vec p}_{M'}}{2 E_{M'}({\vec p}_{M'})}
\delta^{(4)} \left( p_{M} - p_1 - p_2 - p_{M'} \right) \
\label{d3}
\ee

The resulting decay width can be written as
\bea
\Gamma(M^{\pm} \to \ell_1^{\pm} \ell_2^{\pm} M^{'\mp})
&=&  (2 - \delta_{\ell_1 \ell_2} )
|k|^2
\left[
\G(DD^{*}) + \G(CC^{*}) + \G_{\pm}(DC^{*}) + \G_{\pm}(CD^{*}) \right] \
\label{GM2N}
\ea
where $k$ is the mixing factor
\be
k =  B_{\ell_1 N} B_{\ell_2 N} \
\label{kN}
\ee
and $\G_{\pm}(XY^{*})$ are the normalized (\emph{i.e}., without the
explicit mixing) contributions
from the $X$ channel and the complex-conjugate of the $Y$ channel
($X,Y=D, C$, where $D$ and $C$ stand for the direct and crossed channels)
\be
\G_{\pm}(XY^{*}) \equiv K^2 \; \frac{1}{2!} \frac{1}{2 M_M} \frac{1}{(2 \pi)^5}
\int d_3 \; P(X) P(Y)^{*} M_{N}^2 T_{\pm}(X) T_{\pm}(Y)^{*} \
\label{bGXYN}
\ee

The expressions for $T_{\pm}(X) T_{\pm}(Y)^{*}$ ($X,Y=D,C$)
are given in Appendix 1.
$T_{\pm}(X)$ is the relevant part of the amplitude in the $X$ channel
and forms part of the total decay amplitude ${\cal T}(M^{\pm})$,
cf.~Appendix~1.
 {
In Equation~(\ref{GM2N}), notice that subscripts $\pm$ for the
contributions $\G(DD^{*})$ and $\G(CC^{*})$ are unnecessary because
$|T_{+}(D)|^2 = |T_{-}(D)|^2$ and  $|T_{+}(C)|^2 = |T_{-}(C)|^2$.}
%
%
$P(X)$ ($X=D,C$) are the propagator functions of the intermediate neutrino
$N$ in the two channels
\bes
\label{PN}
\ba
P(D) &=& \frac{1}{\left[ (p_{M}-p_1)^2 - M_{N}^2 + i \Gamma_{N} M_{N} \right]} \
\label{PDN}
\\
P(C) &=& \frac{1}{\left[ (p_{M}-p_2)^2 - M_{N}^2 + i \Gamma_{N} M_{N} \right]} \
\label{PCN}
\ea
\ees
The overall constant $K^2$ in Equation~(\ref{bGXYN}) is
\be
K^2 = G_F^4 f_{M}^2 f_{M'}^2  |V_{Q_u Q_d} V_{q_u q_d}|^2 \
\label{Ksqr}
\ee
Here, $f_{M}$ and $f_{M'}$ are the decay constants of $M^{\pm}$ and $M^{' \mp}$,
and $V_{Q_u Q_d}$ and $V_{q_u q_d}$ are the corresponding CKM matrix elements.
We denote the valence quark content of $M^+$ as $Q_u {\bar Q}_d$;
of $M^{' +}$ as $q_u {\bar q}_d$.

When the intermediate neutrino $N$ has such a mass that it is on mass shell,
Equation~(\ref{MNjint}), the squares of the propagators (\ref{PN}) are
reduced to Dirac delta functions because $\Gamma_N \ll M_N$
\ba
|P(X)|^2 &=&
\left | \frac{1}{(p_{M}-p_k)^2-M^{2}_{N}+i \Gamma_{N} M_{N}} \right | ^2
\nonumber\\
&= &
\frac{\pi}{M_{N} \Gamma_{N}} \delta((p_{M}-p_k)^2-M^{2}_{N})
\qquad ( \Gamma_{N} \ll M_{N}) \
\label{P1P1}
\ea
where $p_k=p_1, p_2$ for $X=D,C$. In this on-shell case, the
$DD^*$ and $CC^*$ contributions in Equation~(\ref{GM2N})
are large, and the interference contributions $DC^*$ and $CD^*$ are
negligible in comparison (\textit{cf.}~Ref.~\cite{CKZ2} for details on this point),
leading to
\bes
\label{GM2NOS}
\bea
\Gamma(M^{\pm} \to \ell_1^{\pm} \ell_2^{\pm} M^{'\mp})
&=&  (2 - \delta_{\ell_1 \ell_2} )
|k|^2
\left[ \G(DD^{*}) + \G(CC^{*}) \right]
\label{GM2NOSa}
\\
& \equiv &
\Gamma(M^{\pm} \to \ell_1^{\pm} \ell_2^{\pm} M^{'\mp}; DD^*) +
\Gamma(M^{\pm} \to \ell_1^{\pm} \ell_2^{\pm} M^{'\mp}; CC^*) \
 \label{GM2NOSb}
\eea
\ees
\noindent
when $\ell_1=\ell_2$, we even have $\G(DD^{*}) = \G(CC^{*})$.
The normalized decay width $\G(DD^{*})$ can be calculated
explicitly, and it turns out to be
\ba
\G(DD^{*}) & = &
 \frac{K^2 M_M^5}{128 \pi^2} \frac{M_{N}}{\Gamma_{N} }
\lambda^{1/2}(1, y_N,y_{\ell_1})
\lambda^{1/2} \left( 1, \frac{y'}{y_N},\frac{y_{\ell_2}}{y_N} \right)
Q(y_N; y_{\ell_1}, y_{\ell_2},y') \
\label{GDDN}
\ea
and $\G(CC^{*})$ is obtained by the
simple exchange  $y_{\ell_1} \leftrightarrow y_{\ell_2}$
\ba
\G(CC^{*}) & = & \G(DD^{*})(y_{\ell_1} \leftrightarrow y_{\ell_2}) \
\label{GCCN}
\ea
The notations used in Equations~(\ref{GDDN}) and (\ref{GCCN}) are
\bes
\label{notGDDN}
\ba
\lambda(y_1,y_2,y_3) & = & y_1^2 + y_2^2 + y_3^2 - 2 y_1 y_2 - 2 y_2 y_3 - 2 y_3 y_1 \
\label{lambdaN}
\\
y_N &=& \frac{M_{N}^2}{M_M^2} \ , \quad
y_{\ell_s} =  \frac{M_{\ell_s}^2}{M_M^2} \ , \quad
y' =\frac{M_{M'}^2}{M_M^2} \ ,  \quad (\ell_s=\ell_1, \ell_2) \
\label{yNs}
\ea
\ees
and  the function $Q(y_N; y_{\ell_1}, y_{\ell_2},y')$ is given in
Appendix 2. In the limit of massless charged leptons
($y_{\ell_1}=y_{\ell_2}=0$), the expression (\ref{GDDN}) reduces to
\ba
\G(DD^{*}){\big |}_{M_{\ell_1}=M_{\ell_2}=0} & = &
 \frac{K^2 M_M^5}{256 \pi^2} \frac{M_{N}}{\Gamma_{N} }
y_N^2 (1 - y_N)^2 \left( 1 - \frac{y'}{y_N} \right)^2
\label{GDDNM0}
\ea

We note that the expression (\ref{GDDN}), although
having the explicit mixing dependence factored out [\textit{cf.}~Equation~(\ref{GM2NOS})],
contains the dependence on the mixing coefficients $B_{\ell N}$ in the
denominator due to the $N$-decay width there $\Gamma_N \propto |B_{\ell N}|^2$
($\ell=e, \mu, \tau$, see below). This factor $1/\Gamma_N$ in
$\G(DD^*)$ of Equation~(\ref{GDDN})
represents the $N$-on-shell effect Equation~(\ref{P1P1}).
As a result, the considered width
$\Gamma(M^{\pm} \to \ell_1^{\pm} \ell_2^{\pm} M^{'\mp})$ is
by many orders of magnitude larger when $N$ is on-shell
than it would be if $N$ were off-shell.
For more quantitative analyses, it is thus important
to have an expression for $\Gamma_N$ as a function of mass $M_N$.
Using the results of Ref.~\cite{CKZ2}, we can write this decay width as
\begin{equation}
\Gamma_{N} = \K \bG_N(M_{N}) \
\label{GNwidth}
\end{equation}
where the corresponding canonical (\emph{i.e}., without any mixing dependence)
decay width is
\begin{equation}
 \bG_N(M_{N}) \equiv \frac{G_F^2 M_{N}^5}{96 \pi^3} \
\label{barGN}
\end{equation}
and the factor $\K$ contains the dependence on
the heavy-light mixing factors
\begin{equation}
\K(M_{N}) \equiv \K = {\cal N}_{e N} \; |B_{e N}|^2 + {\cal N}_{\mu N} \; |B_{\mu N}|^2 + {\cal N}_{\tau N} \; |B_{\tau N}|^2  \
\label{calK}
\end{equation}

\begin{figure}[htb] 
\begin{minipage}[b]{.49\linewidth}
\includegraphics[width=85mm]{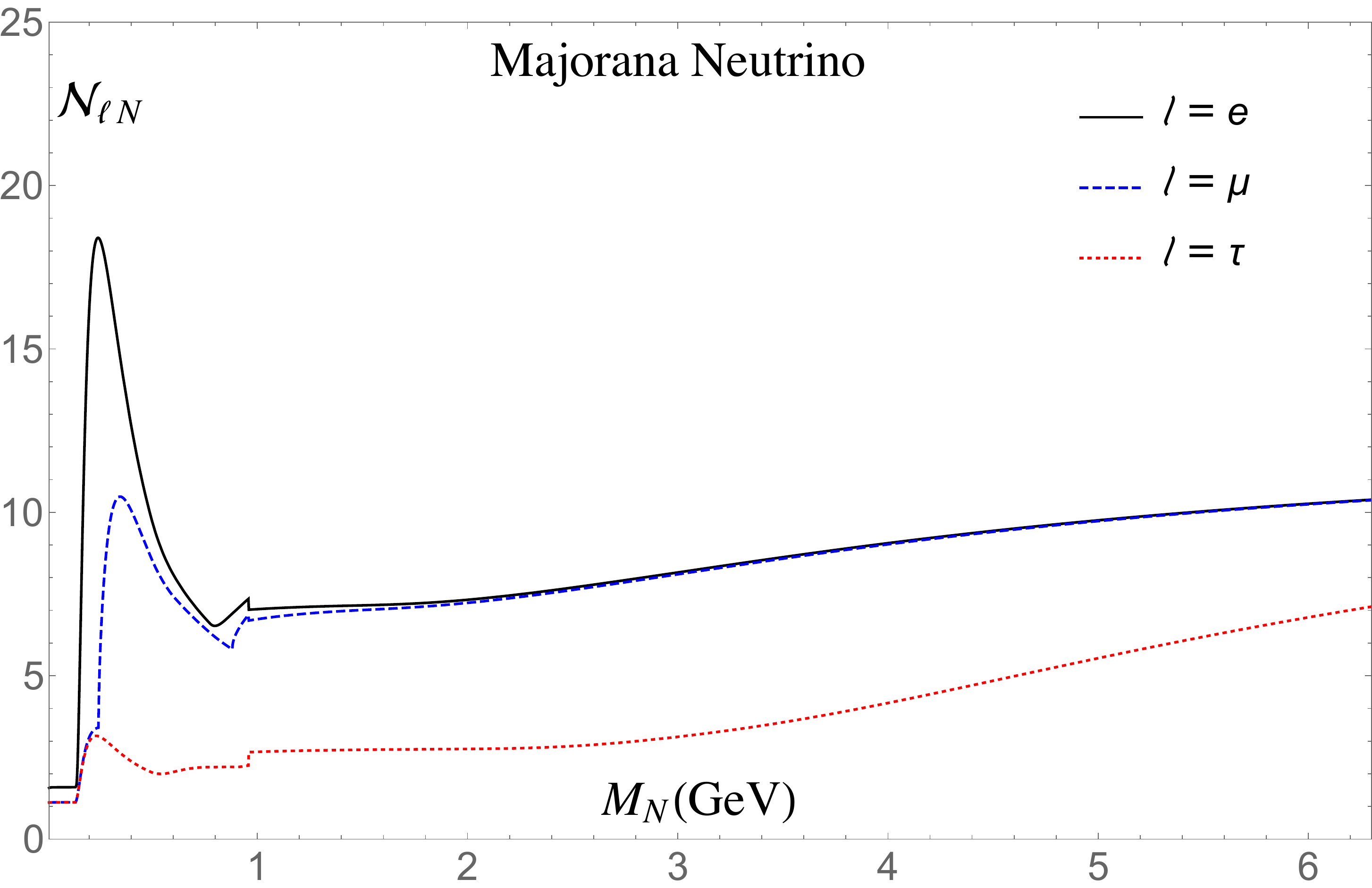}
\end{minipage}
\begin{minipage}[b]{.49\linewidth}
\includegraphics[width=85mm]{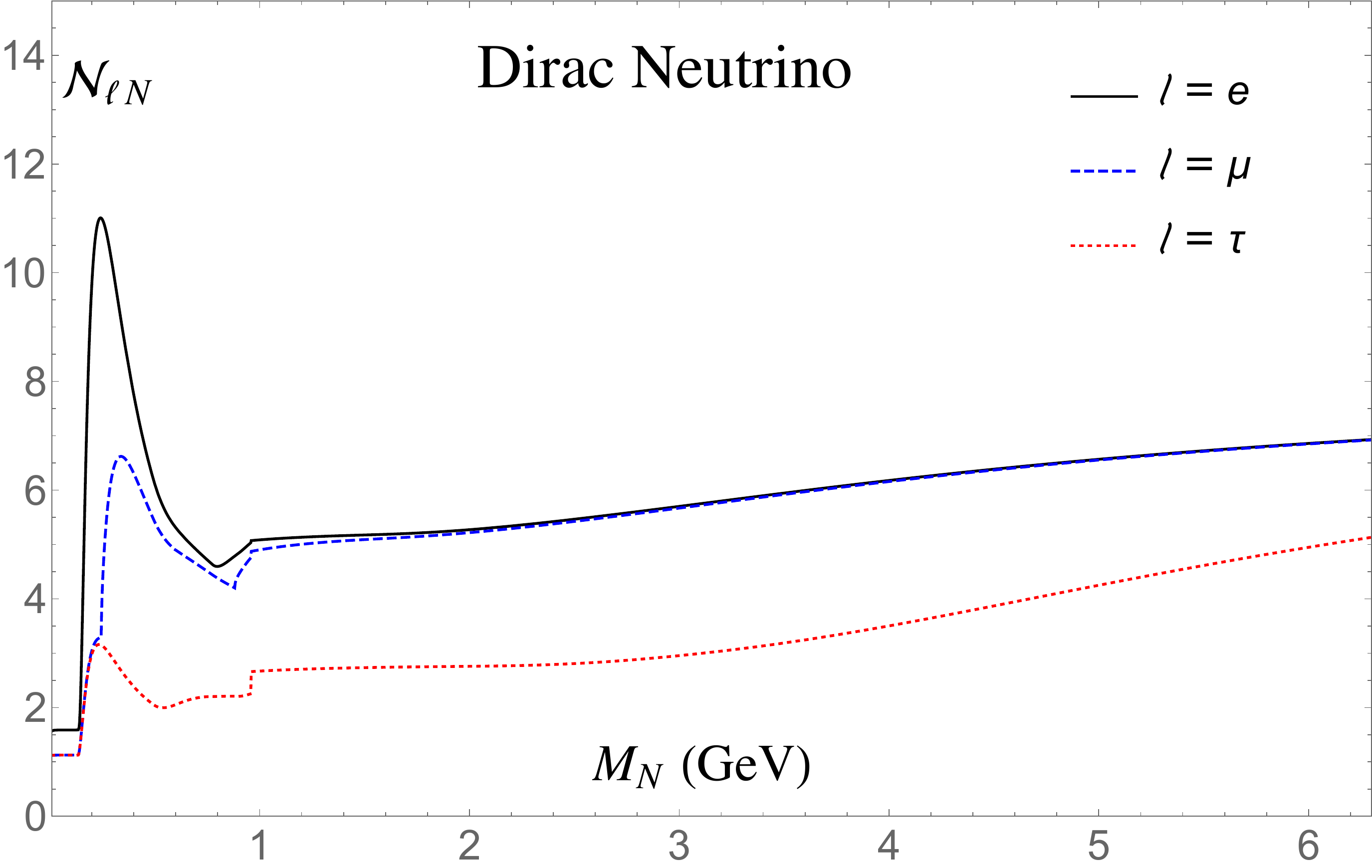}
\end{minipage} \vspace{12pt}
\caption{The effective mixing coefficients
${\cal N}_{\ell N}$ ($\ell = e, \mu, \tau$)
appearing in Equations~(\ref{GNwidth})--(\ref{calK}),
as a function of the mass $M_N$ of
the neutrino $N$. The left-hand figure is for the case of Majorana
neutrino; the right-hand figure for the case of Dirac neutrino.
For details we refer to Appendix 3.}
\label{FigcNellN}
\end{figure}

In this expression,
${\cal N}_{\ell N}(M_N) \equiv {\cal N}_{\ell N}$ ($\ell = e, \mu, \tau$)
are the effective mixing coefficients; these are numbers $\sim 10^0$--$10^1$
which depend on the mass $M_N$.
In Appendix 3 we write down the relevant formulas
for the evaluation of these
coefficients. The results of these evaluations are presented in
Figure~\ref{FigcNellN}, for the case of Majorana and Dirac neutrino $N$,
in the entire neutrino mass interval
$0.1 \ {\rm GeV} < M_N < 6.3 \ {\rm GeV}$ which will be
of interest in this work. For further clarifying remarks
we refer to Appendix 3.
Equations~(\ref{GM2NOS}) and (\ref{GDDN}) imply that
$\Gamma(M^{\pm} \to \ell_1^{\pm} \ell_2^{\pm} M^{'\mp})$ is
proportional to $1/\K$ ($\propto 1/|B_{\ell N}|^2$).
Hence, we can define a
canonical branching ratio ${\overline {\rm Br}}$,
being the part of the branching ratio
${\rm Br}(DD^*) \equiv
\Gamma(M^{\pm} \to \ell_1^{\pm} \ell_2{\pm} M^{' \mp}; DD^*)/\Gamma(M^{\pm} \to {\rm all})$
with no explicit or implicit heavy-light mixing factors
\bes
\label{BrMDD}
\bea
{\rm Br}(M^{\pm} \to \ell_1^{\pm} \ell_2^{\pm} M^{' \mp}; DD^*) & \equiv
& \frac{\Gamma(M^{\pm} \to \ell_1^{\pm} \ell_2^{\pm} M^{' \mp}; DD^*)}
{\Gamma(M^{\pm} \to {\rm all})} =
\frac{(2 - \delta_{\ell_1 \ell_2}) |k|^2}{\Gamma(M^{\pm} \to {\rm all})}
\G(DD^*)
\label{BrMDDa}
\\
& = & (2 - \delta_{\ell_1 \ell_2}) \frac{|k|^2}{\K} 2 {\overline {\rm Br}}(y_N; y_{\ell_1}, y_{\ell_2}; y') \
\label{BrMDDb}
\eea
\ees

Use of the expressions (\ref{GDDN}) and Equations (\ref{GNwidth}) and (\ref{barGN}) then
gives for the canonical branching ratio the following expression:
\bea
{\overline {\rm Br}}(DD^*) &\equiv&
{\overline {\rm Br}}(y_N; y_{\ell_1}, y_{\ell_2}, y')
\nonumber\\
&=&
\frac{3 \pi}{8} \frac{K^2 M_M}{G_F^2 \Gamma(M^{\pm} \to {\rm all})}
\frac{1}{y_N^2}  \lambda^{1/2}(1, y_N, y_{\ell_1})
\lambda^{1/2}\left( 1, \frac{y'}{y_N}, \frac{y_{\ell_2}}{y_N} \right)
Q(y_N; y_{\ell_1}, y_{\ell_2}, y') \
\label{bBrM}
\eea
where the notations (\ref{notGDDN}) and (\ref{Ksqr}) are used.
In the limit of massless charged leptons ($M_{\ell_1}=M_{\ell_2}=0$)
this expression becomes simpler
\be
{\overline {\rm Br}}(y_N; 0, 0, y') =
\frac{3 \pi}{16} \frac{K^2 M_M}{G_F^2 \Gamma(M^{\pm} \to {\rm all})}
(1 - y_N)^2 \left( 1 - \frac{y'}{y_N} \right)^2
\label{bBrMM0}
\ee


\subsection{Effect of the Long Neutrino Lifetime on the Observability of $M^{\pm} \to \ell_1^{\pm} \ell_2^{\pm} M^{' \mp}$}
\label{sec:effBrMsub}

In the mentioned branching ratios, an often important effect of suppression
due to the decay (\emph{i.e}., nonsurvival) probability was not included.
Namely, if the detector for the considered decays has a certain length $L$,
the produced (on-shell) massive neutrino $N$ could survive
during its flight through the detector, and would decay later outside it.
Such decays are thus not detected and should be eliminated
from the width and the branching ratio of the considered process
$M^{\pm} \to \ell_1^{\pm} \ell_2^{\pm} M^{'\mp}$,
by introducing a suppression factor
(nonsurvival probability) $P_N =1- \exp[-t/(\tau_N \gamma_N)]$,
where $t \approx L/\beta_N$ is the time of flight of $N$ through the detector
($\beta_N$ is the velocity of $N$ in the lab frame),
and $\gamma_N=(1 - \beta_N^2)^{-1/2}$
is the Lorentz time dilation factor.
Hence, the suppression factor, which should multiply the branching ratio, is
\be
P_N = 1 - \exp \left[ - \frac{L}{\tau_N \gamma_N \beta_N} \right]
\approx 1- \exp \left[ - \frac{L \Gamma_N}{\gamma_N} \right] \
\label{PN1}
\ee

In the last relation, we used $\beta_N \approx 1$ and
$\tau_N = 1/\Gamma_N$ [$ \equiv 1/\Gamma(N \to {\rm all})$],
in the units used here ($c=1=\hbar$).
This decay-within-the-detector probability $P_N$
has been discussed and presented
for the processes with intermediate on-shell particle (such as $N$)
in Refs.~\cite{CDK,scatt3,CKZ,CKZ2,CERN-SPS,CERN-SPS2,commKim}.
In this respect, here we follow mostly the notations
of Ref.~\cite{CKZ2}. Usually, the quantity $P_N$ is small
and is then written as
\be
\label{PN2}
P_N \approx  L/(\tau_N \gamma_N \beta_N) \quad  \left( \approx
 L/(\tau_N \gamma_N) \right)  \qquad {\rm if} \; P_N\ll 1
\ee
which agrees with Equation~(\ref{PN1}) in the limit of small $P_N$.
The suppression factor (\ref{PN1}) can be rewritten~as
\bes
\label{PN3}
\bea
P_N &=& 1- \exp \left(- \frac{L}{L_N} \right) =
1 - \exp \left( - \frac{L}{{\bL}_N} \K \right)
\label{PN3a}
\\
& \approx & \frac{L}{{\bL}_N} \K  \quad
{\rm if} \; P_N \ll 1
\label{PN3b}
\eea
\ees
Here, $L_N$ is the decay length, and ${\bL}_N$ is the
canonical decay length (independent of
the mixing parameters $B_{\ell^{'} N}$)
\bes
\label{LNbLN}
\bea
L_N^{-1} &=& {\bL}_N^{-1} \K \
\label{LN}
\\
{\bL}_N^{-1} & = & \frac{\bG_N(M_N)}{\gamma_N} =
\frac{1}{\gamma_N} \frac{G_F^2 M_N^5}{96 \pi^3} \
\label{bLN}
\eea
\ees
where $\K$ and $\bG_N(M_N)$ are from Equations~(\ref{GNwidth})--(\ref{calK}),
\textit{cf.}~also Figure~\ref{FigcNellN}. Equation~(\ref{PN3b}) suggests that it is
convenient to define a canonical (\emph{i.e}., independent of mixing)
probability ${\overline P}_N$ for the decay of $N$ within the detector
as
\be
{\overline P}_N =  \frac{1 {\rm m}}{{\bL}_N}  \; \Rightarrow
P_N \approx {\overline P}_N \left( \frac{L}{1 {\rm m}} \right) \K
\label{bPN}
\ee

\begin{figure}[htb]
\centering\includegraphics[width=120mm]{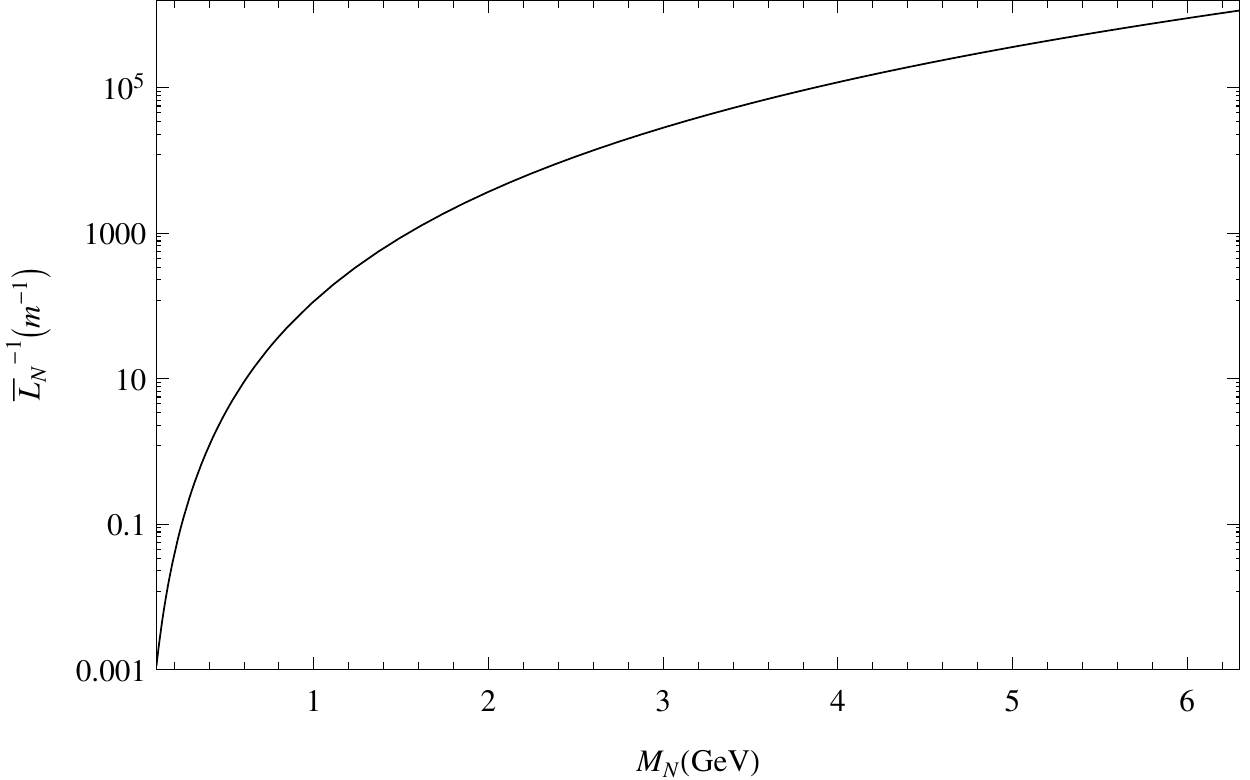}
\caption{The inverse canonical decay length
$\bL_N^{-1} \equiv \bG_N(M_{N})/\gamma_{N}$, Equation~(\ref{bLN}),
in units of inverse meters ($m^{-1}$),
as a function of the neutrino mass $M_N$, with the Lorentz lab time dilation
factor chosen to be $\gamma_{N}$ [$\equiv (1 - \beta_N^2)^{-1/2}$] $=2$.
The $y$ axis can also be interpreted as the canonical probability
${\overline P}_N \equiv (1 {\rm m})/\bL_N$ (dimensionless), Equation~(\ref{bPN}).}
\label{bLNfig}
\end{figure}

We present the inverse canonical
decay length, $\bL_N^{-1}$, for $\gamma_N =2$, in Figure~\ref{bLNfig}
as a function of $M_N$.
We note that $\bL_N^{-1}$ increases very fast (as $M_N^5$) when $M_N$ increases.
Therefore, the supression due to the factor $P_N$
may not necessarily be strong (\emph{i.e}., $P_N \not \ll 1$)
for semileptonic LNV decays of heavier mesons $M^{\pm}$, such as $B^{\pm}$.
If we use eyeball estimates for the coefficients
${\cal N}_{\ell N}$ of the left-hand Figure~\ref{FigcNellN},
approximate expressions for the factor $\K$ of Equation~(\ref{calK})
for Majorana neutrinos can be written
\bes
\label{calKappr}
\bea
\K &\approx& 15 |B_{e N_j}|^2 + 8 |B_{\mu N_j}|^2 + 2  |B_{\tau N_j}|^2
\quad (K \; {\rm decays}) \
\label{calKapprK}
\\
\K &\approx& 7  (|B_{e N_j}|^2 + |B_{\mu N_j}|^2) + 2  |B_{\tau N_j}|^2 \quad (D, D_s \; {\rm decays}) \
\label{calKapprD}
\\
\K &\approx& 8  (|B_{e N_j}|^2 + |B_{\mu N_j}|^2) + 3  |B_{\tau N_j}|^2 \quad (B, B_c \; {\rm decays}) \
\label{calKapprB}
\eea
\ees

In order to estimate better the values of $\K$ Equations~(\ref{calKappr})
and thus the suppression factor $P_N$ Equation~(\ref{PN3}), we need
to know the present upper bounds for the squares $|B_{\ell N}|^2$
as a function of $M_N^2$. These upper bounds we take from compilation of
values of Ref.~\cite{Atre}, based in turn on upper bound values obtained
in Refs.~\cite{Benes,Belanger,Belanger2,Kusenko,beamdump,beamdump2,beamdump3,beamdump4,beamdump5,l3etal,l3etal2,delphi,noch,noch2}.
We present them in Table \ref{T1},
 for specific chosen values of $M_N$ in the
mass range of interest.
\begin{table}
\small
\centering
\caption{Presently known upper bounds for the squares $|B_{\ell N}|^2$ of the
heavy-light mixing matrix elements, for various specific values of $M_N$.}
\label{T1}
\begin{tabular}{| c | c | c | c |}
\hline
\bf{$M_N [GeV]$} & $|B_{eN}|^2$ & $|B_{\mu N}|^2$ & $|B_{\tau N}|^2$ \\
\hline
0.1 & $(1.5\pm 0.5)\times10^{-8} $ & $(6.0\pm 0.5)\times10^{-6}$ & $(8.0\pm 0.5)\times10^{-4}$ \\
\hline
0.3 & $(2.5\pm 0.5)\times10^{-9}$ & $(3.0\pm 0.5)\times10^{-9}$ & $(1.5\pm 0.5)\times10^{-1}$ \\
\hline
0.5 & $(2.0\pm 0.5)\times10^{-8}$ & $(6.5\pm 0.5)\times10^{-7}$ & $(2.5\pm 0.5)\times10^{-2}$ \\
\hline
0.7 & $(3.5\pm 0.5)\times10^{-8}$ & $(2.5\pm 0.5)\times10^{-7}$ & $(9.0\pm 0.5)\times10^{-3}$ \\
\hline
1.0 & $(4.5\pm 0.5)\times10^{-8}$ & $(1.5\pm 0.5)\times10^{-7}$ & $(3.0\pm 0.5)\times10^{-3}$ \\
\hline
2.0 & $(1.0\pm 0.5)\times10^{-7}$ & $(2.5\pm 0.5)\times10^{-5}$ & $(3.0\pm 0.5)\times10^{-4}$ \\
\hline
3.0 & $(1.5\pm 0.5)\times10^{-7}$ & $(2.5\pm 0.5)\times10^{-5}$ & $(4.5\pm 0.5)\times10^{-5}$ \\
\hline
4.0 & $(2.5\pm 0.5)\times10^{-7}$ & $(1.5\pm 0.5)\times10^{-5}$ & $(1.5\pm 0.5)\times10^{-5}$ \\
\hline
5.0 & $(3.0\pm 0.5)\times10^{-7}$ & $(1.5\pm 0.5)\times10^{-5}$ & $(1.5\pm 0.5)\times10^{-5}$ \\
\hline
6.0 & $(3.5\pm 0.5)\times10^{-7}$ & $(1.5\pm 0.5)\times10^{-5}$ & $(1.5\pm 0.5)\times10^{-5}$ \\
\hline
\end{tabular}
\end{table}
\begin{table} \centering
\caption{Rough estimates of upper bounds for $|B_{\ell N}|^2$
($\ell = e, \mu, \tau$), for  $M_N$ in three different ranges around the values
$0.25$, $1$, $3$ GeV; and
the inverse of the canonical decay length, ${\bL}_N^{-1}$ (in units of $m^{-1}$
and for $\gamma_N=2$).}
\label{T2}
\begin{tabular}{|l|lll|l|}
\hline
$M_N$ [GeV] & $|B_{e N}|^2$ &  $|B_{\mu N}|^2$ & $|B_{\tau N}|^2$ &
${\bL}_N^{-1} [m^{-1}]$
\\
\hline
$\approx 0.25$ & $10^{-8}$ & $10^{-7}$ & $10^{-4}$ & 0.11
\\
$\approx 1.0$ &  $10^{-7}$ & $10^{-7}$ & $10^{-2}$ & $1.1\times 10^2$
\\
$\approx 3.0$ &  $10^{-6}$ & $10^{-4}$ & $10^{-4}$ & $3 \times 10^4$
\\
\hline
\end{tabular}
\end{table}
We remark that the upper bounds have in some cases strong dependence on the
precise values of $M_N$, see Ref.~\cite{Atre} for further details.
In order to use only rough estimates for the values of $\K$, we
present in Table \ref{T2} order of magnitude values for upper
bounds of $|B_{\ell N}|^2$.
These rough upper bounds
are given for three typical ranges of our
interest: $M_N$ around $0.25$; $1$; $3$ GeV. They are
relevant for the decays of
$K$; ($D, D_s$); ($B, B_c$), respectively.
The corresponding values of the inverse of the canonical
decay length, ${\bL}_N^{-1}$, are included.
As seen in Tables \ref{T1} and \ref{T2}, the upper bounds for
$|B_{\tau N}|^2$ are at present significantly less stringent and are
expected to become more stringent in the future.
When we combine Equations~(\ref{PN3b}) with (\ref{calKappr}) and Table \ref{T2},
we obtain for the decay-within-the-detector probability
$P_N \equiv {\overline P}_N \K$
the following estimates and upper bounds,
relevant for the $K$  decays ($M_N \approx 0.25$ GeV),
$D$ and $D_s$ decays ($M_N \approx 1$ GeV), and $B$ and $B_c$ decays
($M_N \approx 3$ GeV), all when $L=1$ m and $\gamma_N=2$:
\bes
\label{uppPN}
\bea
P_{N}(M_N\approx 0.25{\rm GeV}) & \approx &
1.7 |B_{e N_j}|^2 + 0.9 |B_{\mu N_j}|^2  \quad (+  0.2 |B_{\tau N_j}|^2)
\nonumber\\
&\lesssim& 10^{-8} + 10^{-7} \quad (+ 10^{-5}) \
\label{uppPNK}
\\
P_{N}(M_N\approx 1{\rm GeV}) & \approx &
0.8 \cdot 10^3 |B_{e N_j}|^2 + 0.8 \cdot 10^3 |B_{\mu N_j}|^2
\quad (+  2 \cdot 10^2 |B_{\tau N_j}|^2)
\nonumber\\
&\lesssim& 10^{-4} + 10^{-4} \quad (+ 10^{0}) \
\label{uppPND}
\\
P_{N}(M_N\approx 3{\rm GeV}) & \approx &
3 \cdot 10^5 |B_{e N_j}|^2 + 3 \cdot 10^5 |B_{\mu N_j}|^2
\quad (+  1 \cdot 10^5 |B_{\tau N_j}|^2)
\nonumber\\
&\lesssim& 10^{0} + 10^{0} \quad (+ 10^{0}) \
\label{uppPNB}
\eea
\ees

In order to have the analysis and the formulas simpler,
in the rest of this Section we will assume
that one mixing parameter, $|B_{\ell N}|$, dominates over the
other two mixing parameters:
\be
\label{assu}
|B_{\ell N}| \gg |B_{\ell' N}| \qquad (\ell^{'} \not= \ell)
\ee

For example, it may well be that $\ell=\mu$, \emph{i.e}., that
$|B_{\mu N}| \gg |B_{e N}|, |B_{\tau N}|$.
Then, of the branching ratios
${\rm Br}(M^{\pm} \to \ell^{\pm}_1 \ell^{\pm}_2 M^{' \mp})$
the largest
will be $M^{\pm} \to \ell^{\pm} \ell^{\pm} M^{' \mp}$ which,
 according to Equations~(\ref{assu}) and (\ref{BrMDD})
(note that $DD^*$ and $CC^*$ give the same contribution since
$\ell_1=\ell_2$ now), is:
\be
{\rm Br}(M^{\pm} \to \ell^{\pm} \ell^{\pm} M^{' \mp}) =
4 \frac{|B_{\ell N}|^4}{\K} {\overline {\rm Br}}
\label{Brassu}
\ee

Multiplying this expression by the probability $P_N$ of the decay in the detector,
Equation~(\ref{bPN}), we obtain the effective
branching ratio ${\rm Br}_{\rm eff}$
\bes
\label{Breff}
\bea
{\rm Br}_{\rm eff}(M^{\pm} \to \ell^{\pm} \ell^{\pm} M^{' \mp}) & = &
P_N  {\rm Br}(M^{\pm} \to \ell^{\pm} \ell^{\pm} M^{' \mp})
=  \left[ {\overline P}_N \left( \frac{L}{1 {\rm m}} \right) \K \right]
\times \left[ 4 \frac{|B_{\ell N}|^4}{\K} {\overline {\rm Br}} \right]
\label{Breffa}
\\
& = &  \left( \frac{L}{1 {\rm m}} \right)
4   |B_{\ell N}|^4  {\overline P}_N {\overline {\rm Br}} \equiv
 |B_{\ell N}|^4
\left( \frac{L}{1 {\rm m}} \right) {\overline {\rm Br}}_{\rm eff}
\label{Breffb}
\eea
\ees

We see that in the effective  branching ratio, ${\rm Br}_{\rm eff}$,
the complicated dependence on mixing parameters encoded in $\K$
[\textit{cf.}~Equation~(\ref{calK})] disappeared because factors $\K$ cancel here.
All the mixing effects in  ${\rm Br}_{\rm eff}$ are in the simple
factor $|B_{\ell N}|^4$. Unfortunately, this factor represents
a strong suppression, in comparison with ${\rm Br}$ of Equations~(\ref{BrMDD})
where ${\rm Br} \propto |k|^2/\K = |B_{\ell_1 N} B_{\ell_2 N}|^2/\K \sim
|B_{\ell N}|^2$.

In the last identity (\ref{Breffb}) we introduced the canonical
(\emph{i.e}., without any mixing dependence) effective branching ratio
${\overline {\rm Br}}_{\rm eff}$
\be
{\overline {\rm Br}}_{\rm eff} \equiv 4  {\overline P}_N {\overline {\rm Br}}
 = 4 \left( \frac{1 {\rm m}}{{\bL}_N} \right) {\overline {\rm Br}} \
\label{bBreff}
\ee
where we recall that ${\overline {\rm Br}}$ was defined
in Equations~(\ref{BrMDD}) and (\ref{bBrM}).
{
Here $\overline{\rm Br_{eff}}$ is half the value of $\overline{\rm Br_{eff}}$ in Ref.~\cite{CKZ2} because the latter expression referred to the exchange of two Majorana neutrinos instead of one.
}

\begin{figure}[htb]
\centering\includegraphics[width=90mm]{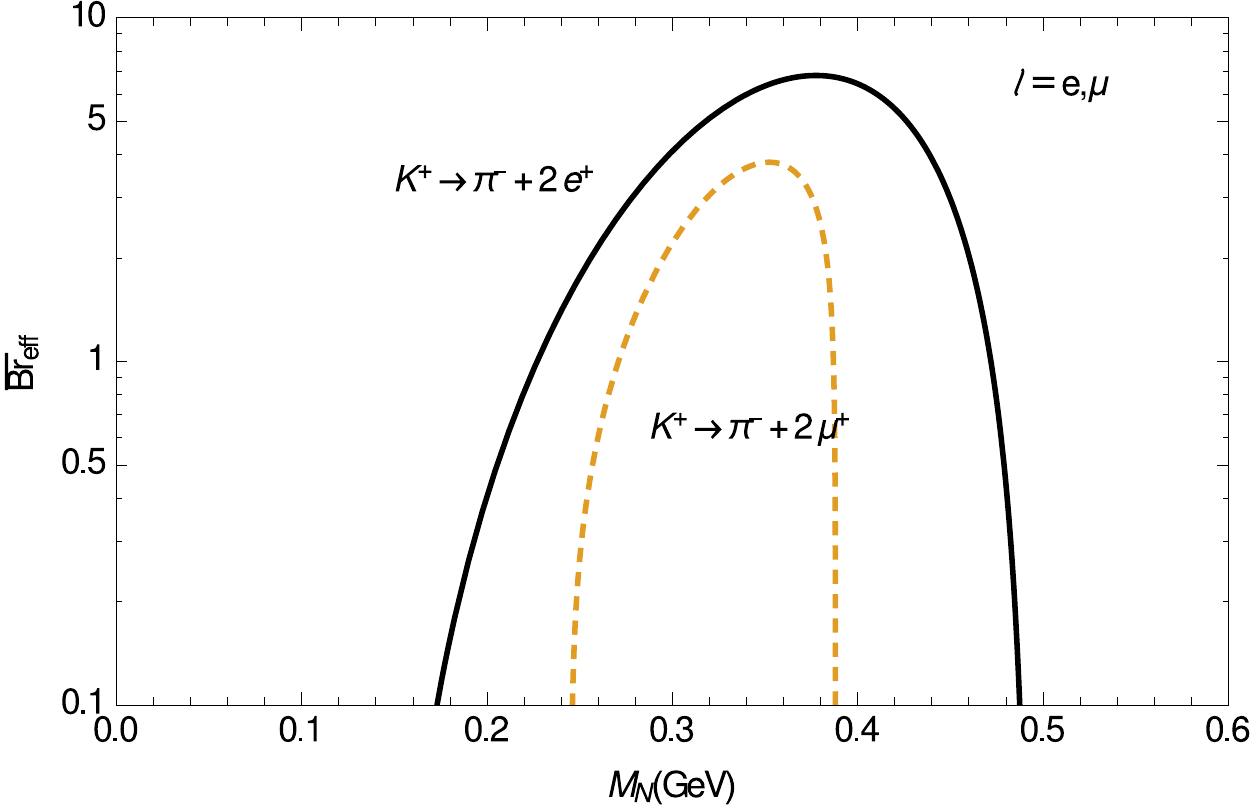}
\caption{The effective canonical branching ratio (\ref{bBreff})
for the  decays $K^{\pm} \to \ell^{\pm} \ell^{\pm} \pi^{\mp}$
($\ell=e, \mu$) as a function of the Majorana neutrino mass $M_N$.}
\label{brKfig}
\end{figure}

\begin{figure}[htb] 
\begin{minipage}[b]{.49\linewidth}
\includegraphics[width=85mm]{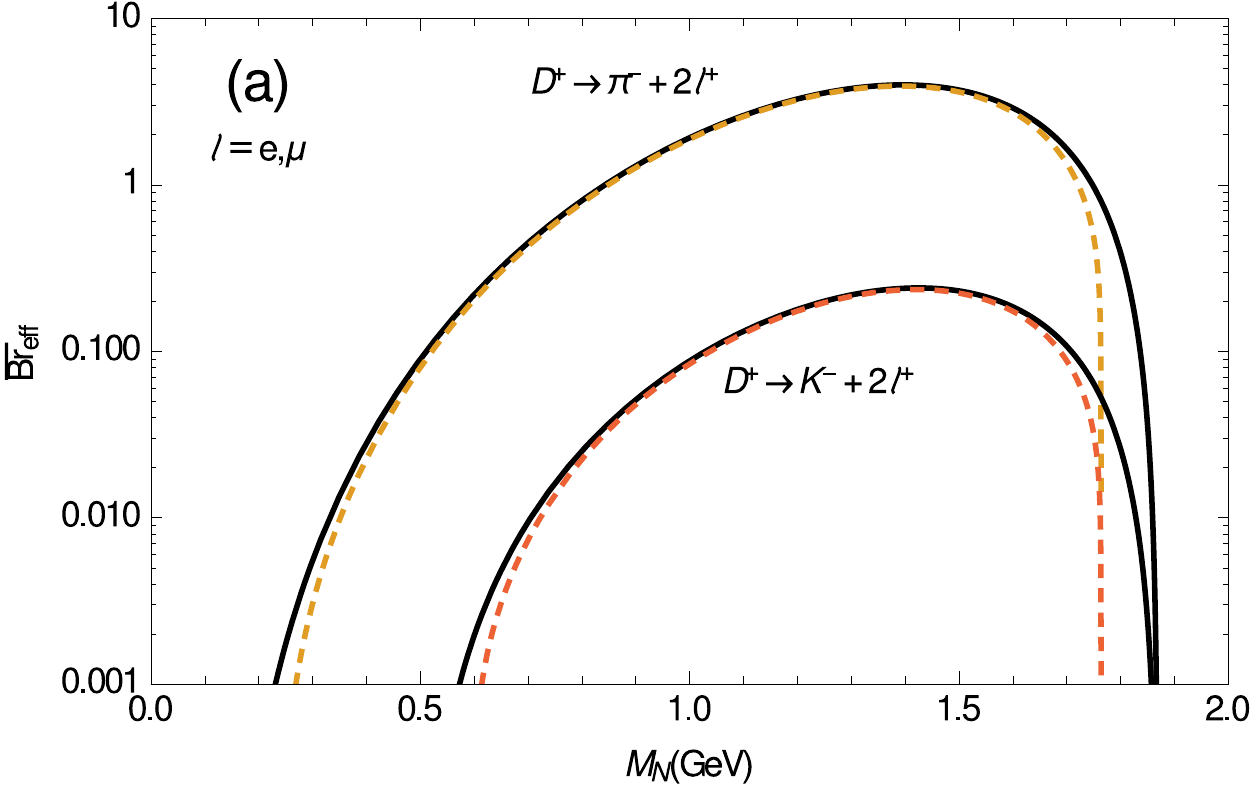}
\end{minipage}
\begin{minipage}[b]{.49\linewidth}
\includegraphics[width=85mm]{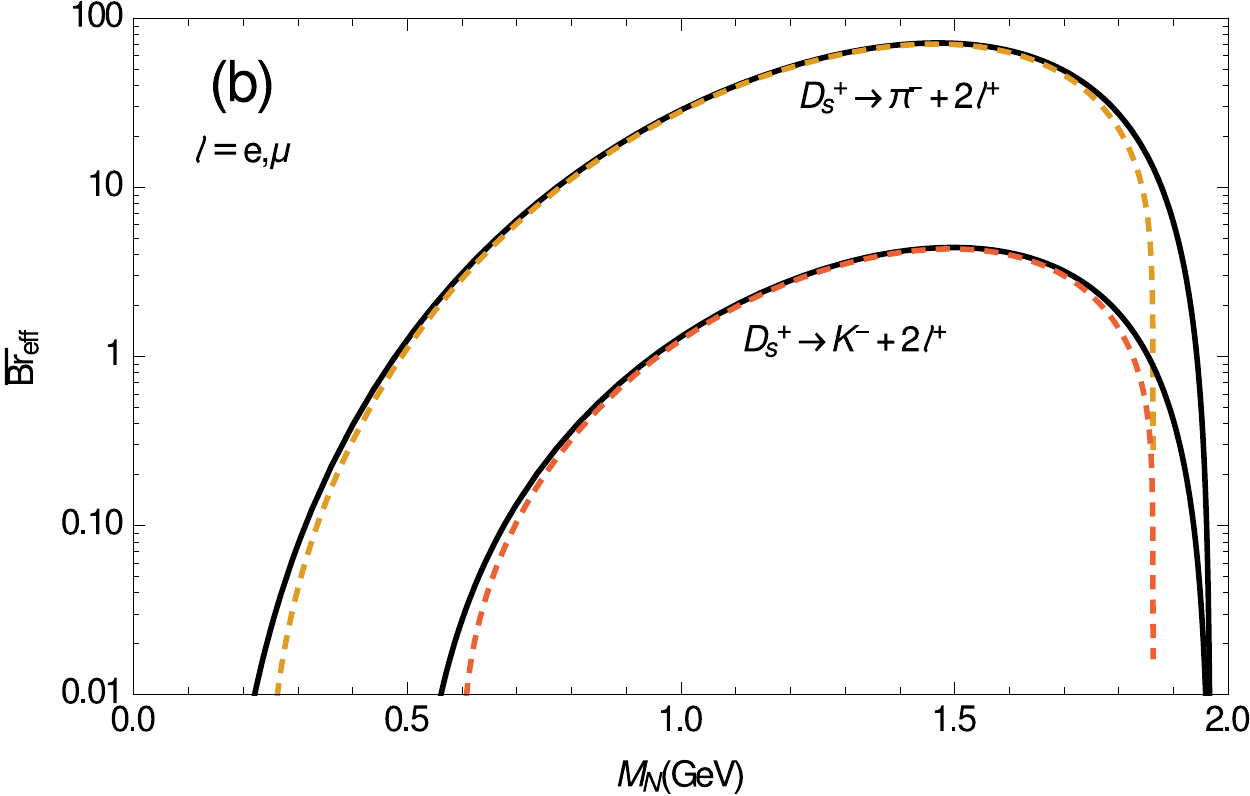}
\end{minipage}
\caption{The same as in Figure~\ref{brKfig}, but now for the decay of
(\textbf{a}) $D^{\pm}$ mesons; (\textbf{b}) $D_s^{\pm}$ mesons.
The solid lines are for $\ell=e$, and the dashed lines for $\ell=\mu$.}
\label{brDDsfig}
 \end{figure}

\begin{figure}[htb] 
\begin{minipage}[b]{.49\linewidth}%
\includegraphics[width=85mm]{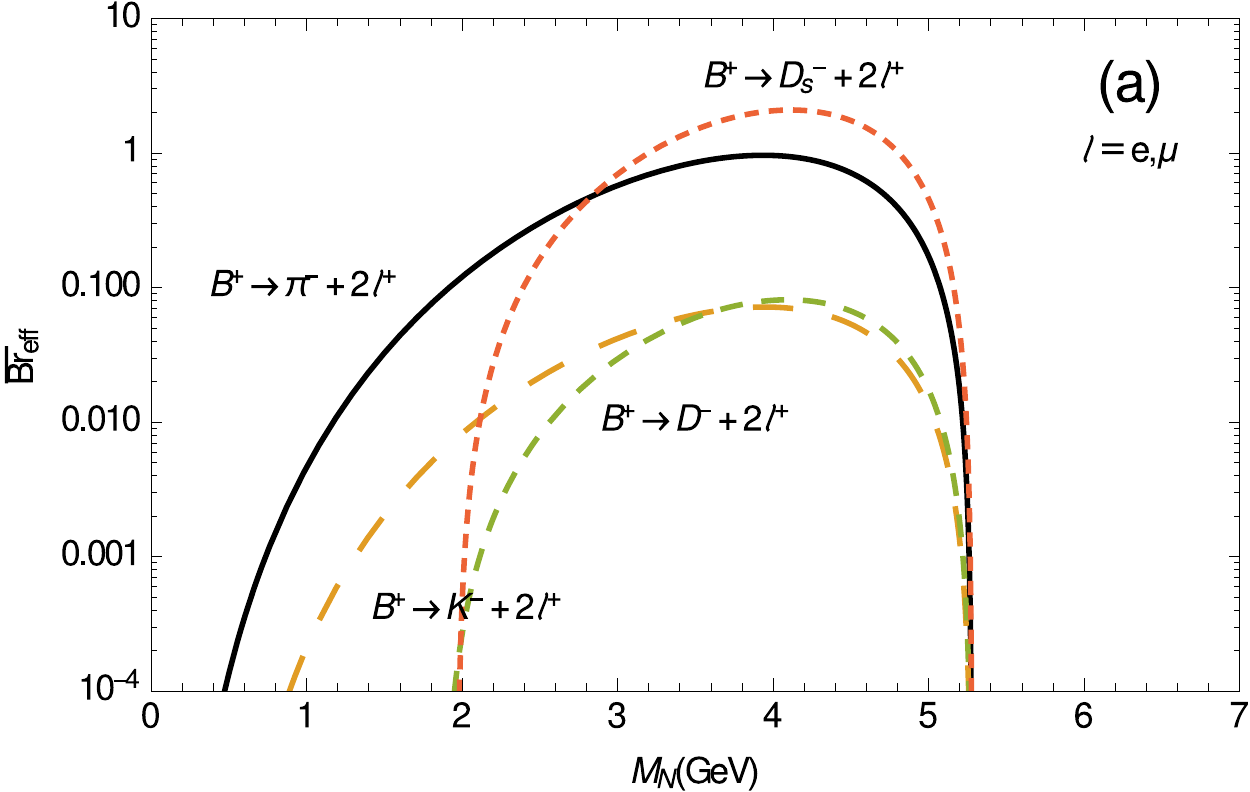}
\end{minipage}
\begin{minipage}[b]{.49\linewidth}
\includegraphics[width=85mm]{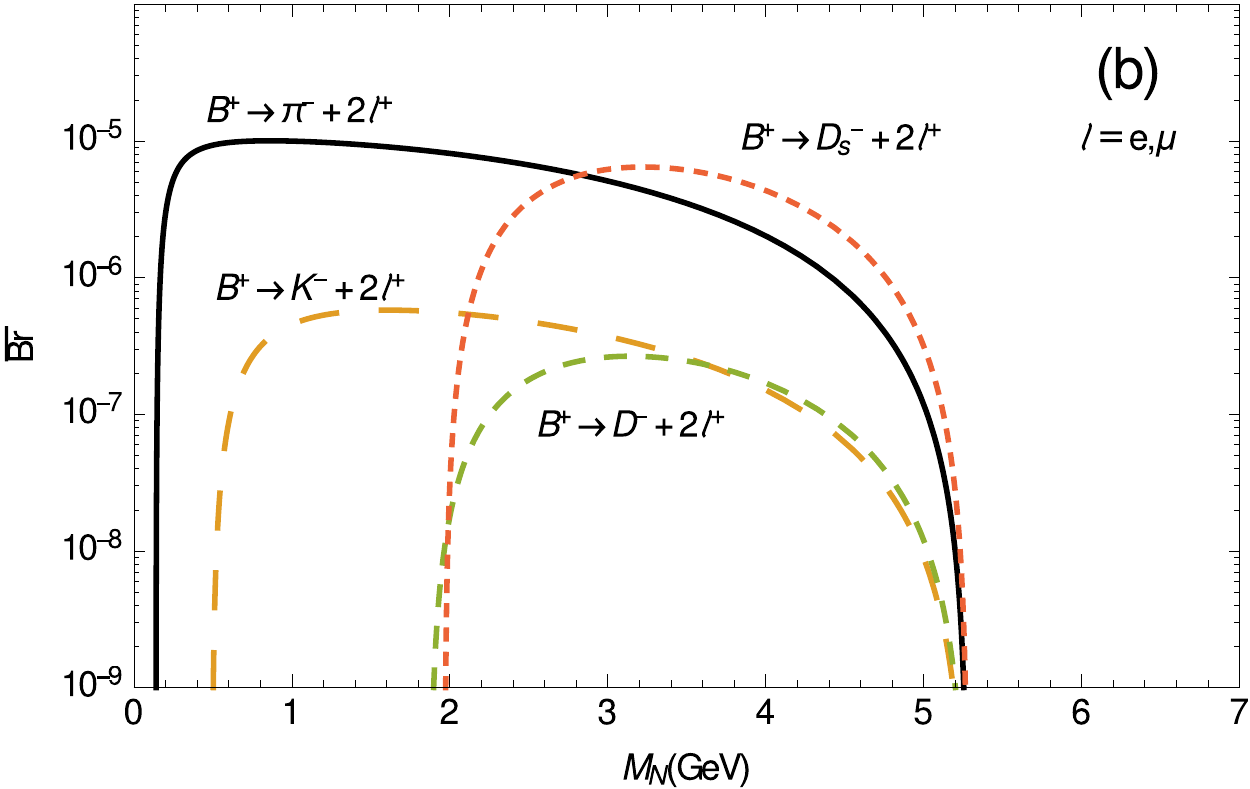}
\end{minipage}
\caption{(\textbf{a}) The effective canonical branching ratio (\ref{bBreff})
as a function of $M_N$
for the lepton number violating (LNV) decays $B^{\pm} \to \ell^{\pm} \ell^{\pm} M^{' \mp}$,
where $\ell=e, \mu$ (no visible difference between $\ell=e$ and $\ell = \mu$);
 (\textbf{b}) the theoretical canonical branching ratio ${\overline {\rm Br}}$,
Equations~(\ref{BrMDD}) and (\ref{bBrM}), for these decays.}
\label{brBfig}
 \end{figure}

\begin{figure}[htb] 
\begin{minipage}[b]{.49\linewidth}
\includegraphics[width=85mm]{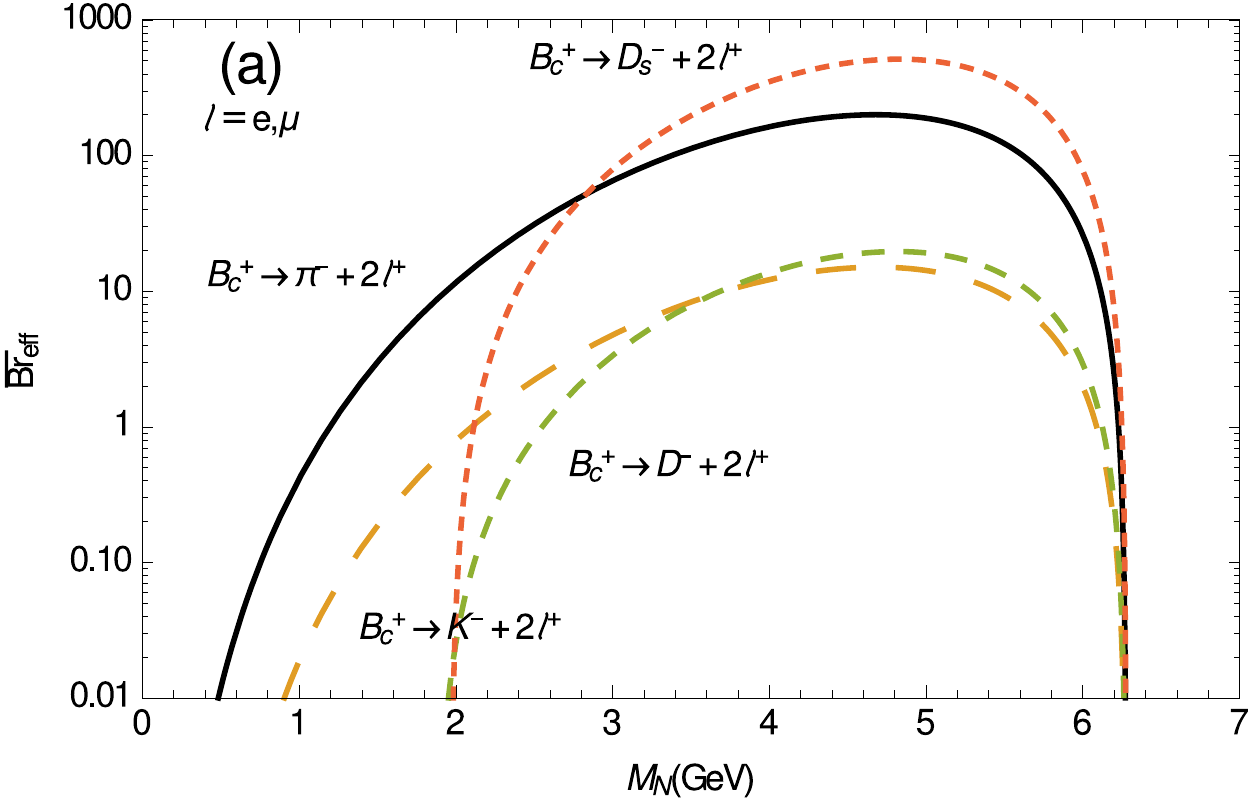}
\end{minipage}
\begin{minipage}[b]{.49\linewidth}
\includegraphics[width=85mm]{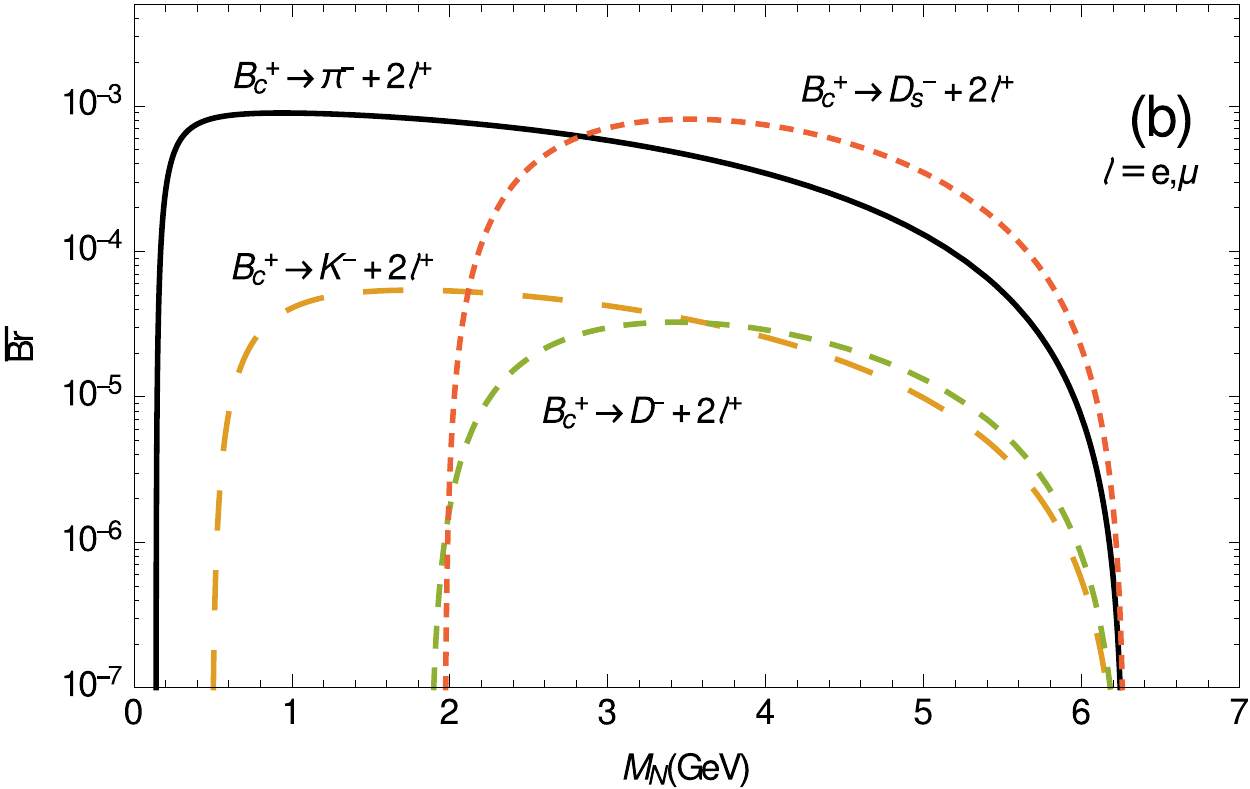}
\end{minipage}
\caption{The same as in Figure~\ref{brBfig}, but for the LNV decays of
the charmed mesons $B_c^{\pm}$.}
\label{brBcfig}
 \end{figure}

Only when $M^{\pm} =B^{\pm}$ or $B_c^{\pm}$, \emph{i.e}.,
when the mass of the on-shell $N$ can be high ($M_N \gtrsim 1$ GeV),
would it be possible to have $P_N \sim 1$ [Equation~(\ref{uppPNB})];
and in such a case Equations~(\ref{Breff}) do not apply,
but rather Equations~(\ref{Brassu}),
\emph{i.e}., ${\rm Br}_{\rm eff} = {\rm Br}$ in this case.
Figures~\ref{brKfig}--\ref{brBcfig} show the
effective canonical branching ratios (\ref{bBreff}) as a function of the
neutrino mass $M_N$, for various considered LNV decays
of the type $M^{\pm} \to \ell^{\pm} \ell^{\pm} M^{' \mp}$:
Figure~\ref{brKfig} for $M=K$; Figure~\ref{brDDsfig}a, b for $M=D, D_s$,
respectively;
Figures~\ref{brBfig}a and \ref{brBcfig}a for $M=B, B_c$, respectively.
We took $\ell=e, \mu$, and $L=1$ m and $\gamma_N=2$.
For the case when $P_N \sim 1$ and
hence the estimates Equations~(\ref{Brassu}) apply,
Figures~\ref{brBfig}b and \ref{brBcfig}b
present the branching ratios
${\overline {\rm Br}}(x)$ as a function of $M_N$,
for $M^{\pm}=B^{\pm}$ and $B_c^{\pm}$, respectively.
For the meson decay constants and CKM matrix elements,
needed for the evaluation of $K^2$ factor of Equation~(\ref{Ksqr}),
and for the masses and lifetimes of the mesons, we used the values
of Ref.~\cite{PDG2014}. The values of the decay constants $f_B$ and $f_{B_c}$
were taken from Ref.~\cite{CKWN}: $f_B = 0.196$ GeV,
$f_{B_c}=0.322$ GeV.
We note that the presented formulas for
${\overline {\rm Br}}_{\rm eff}$
and  ${\overline {\rm Br}}$ can be evaluated also
for the decays
$M^{\pm} \to \ell_1^{\pm} \ell_2^{\pm} M^{' \mp}$ when $\ell_1 \not= \ell_2$.
Furthermore, when the final leptons
are $\tau$ leptons (and $M^{\pm} = B^{\pm}$ or $B_c^{\pm}$), the values
of branching ratios
turn out to be similar to those in Figures~\ref{brBfig} and \ref{brBcfig},
but the range of $M_N$ in this case is shorter:
$M_{M^{'}} + M_{\tau} < M_N < M_M - M_{\tau}$.
Table \ref{T3} displays values of
${\overline {\rm Br}}_{\rm eff}$ for
representative values of $M_N$ in the decays
 $M^{\pm} \to \ell^{\pm} \ell^{\pm} M^{' \mp}$.
\begin{table} \centering
\caption{Values of the factor ${\overline {\rm Br}}_{\rm eff}$,
Equation~(\ref{bBreff}), with $L=1$ m and $\gamma_N=2$,
for some of the LNV decays
$M^{\pm} \to \ell^{\pm} \ell^{\pm} \pi^{\mp}$.
The value of $M_N$ is chosen such
that the maximal value of ${\overline {\rm Br}}_{\rm eff}$
is obtained (the value
of $M_N$ is given in parentheses, in GeV).
For $M^{\pm}=K^{\pm}$, two different values are
given, for $\ell=e$ and $\ell=\mu$.
For all other cases, $\ell=\mu$ is taken
(when $\ell=e$ the values are similar).}
\label{T3}
\begin{tabular}{|l|l|l|l|l|l|l|}
\hline
$M^{\pm}$: & $K^{\pm}$ ($\ell=e$)  & $K^{\pm}$ ($\ell=\mu$)  & $D^{\pm}$ & $D_s^{\pm}$ & $B^{\pm}$ & $B_c^{\pm}$
\\
\hline
${\overline {\rm Br}}_{\rm eff}$ & 6.8 (0.38) & 3.8 (0.35) &
 3.9 (1.39) & 70. (1.47) &  0.96 (3.9) & 199. (4.7)
\\
\hline
\end{tabular}
\end{table}
 \vspace{-6pt}

As an illustrative example, let us consider
the decays $D_s^{\pm} \to \mu^{\pm} \mu^{\pm} \pi^{\mp}$, which
is one of the preferred decay modes proposed at
CERN-SPS \cite{CERN-SPS,CERN-SPS2},
and, in addition, let us assume that
$|B_{\mu N}|^2$ is the dominant mixing.
In such a case, Equations~(\ref{Breff}) and Table \ref{T3}
imply for the experimentally measurable branching fraction
${\rm Br}_{\rm eff}$
\be
{\rm Br}_{\rm eff}(D^{\pm}_s \to \mu^{\pm} \mu^{\pm} \pi^{\mp})
\equiv P_N {\rm Br}(D^{\pm}_s \to \mu^{\pm} \mu^{\pm} \pi^{\mp}) \sim 10^2 |B_{\mu N}|^4 \
\label{BreffDs}
\ee
The present rough upper bound on the mixing
for $M_N \approx 1$ GeV is  $|B_{\mu N}|^2 \lesssim 10^{-7}$, \textit{cf.}~Table~\ref{T2}.
Equation~(\ref{BreffDs}) then implies that
${\rm Br}^{\rm (eff)} \lesssim 10^{-12}$ for such decays.
The proposed experiment at \mbox{CERN-SPS \cite{CERN-SPS,CERN-SPS2}} could
produce $D$ and $D_s$ mesons in numbers
by several orders of magnitude higher than $10^{12}$, which would open
the possibility to explore whether there is a production
of sterile Majorana neutrinos $N$ in the mass range $M_N \sim 1$ GeV.

If the decays  $B_c^{\pm} \to \mu^{\pm} \mu^{\pm} \pi^{\mp}$ are considered
(we do not  use $B^{\pm}$ decays as they are CKM-suppressed compared to $B_c^{\pm}$),
the results of Figure~\ref{brBcfig}a and Equation~(\ref{Breffb}) imply
%
an effective branching ratio
\be
{\rm Br}_{\rm eff}(B_c^{\pm} \to \mu^{\pm} \mu^{\pm} \pi^{\mp})
\sim 10^2 |B_{\mu N}|^4 \
\label{BreffBc}
\ee
which is similar to the case of $D_s$, Equation~(\ref{BreffDs}) in a detector of the same length ($L= 1$ m, in our example).
Since $D_s$ is significantly lighter than $B_c$, the relevant neutrinos that give a sizeable effect are also lighter and thus longer living, implying a smaller $P_N$ factor for the same detector length. On the other hand, $D_s$ decays have no CKM suppression compared to $B_c$ ($|V_{cs}| \approx 1$ while $|V_{cb}| \approx 0.04$) and the $B_c$ channels are more numerous, so that the true branching ratios of the latter are smaller. These two effects compensate in a given detector, as shown in Equations
(\ref{BreffDs}) and (\ref{BreffBc}). However for a longer detector the observable $D_s$ branching ratio increases considerably \cite{CERN-SPS,CERN-SPS2}.

%
%


\section{Charged Pion Decays Mediated by On-Shell Massive Neutrinos}
\label{sec:BrPi}

\begin{figure}[htb] 
\begin{minipage}[b]{.49\linewidth}
\includegraphics[width=80mm]{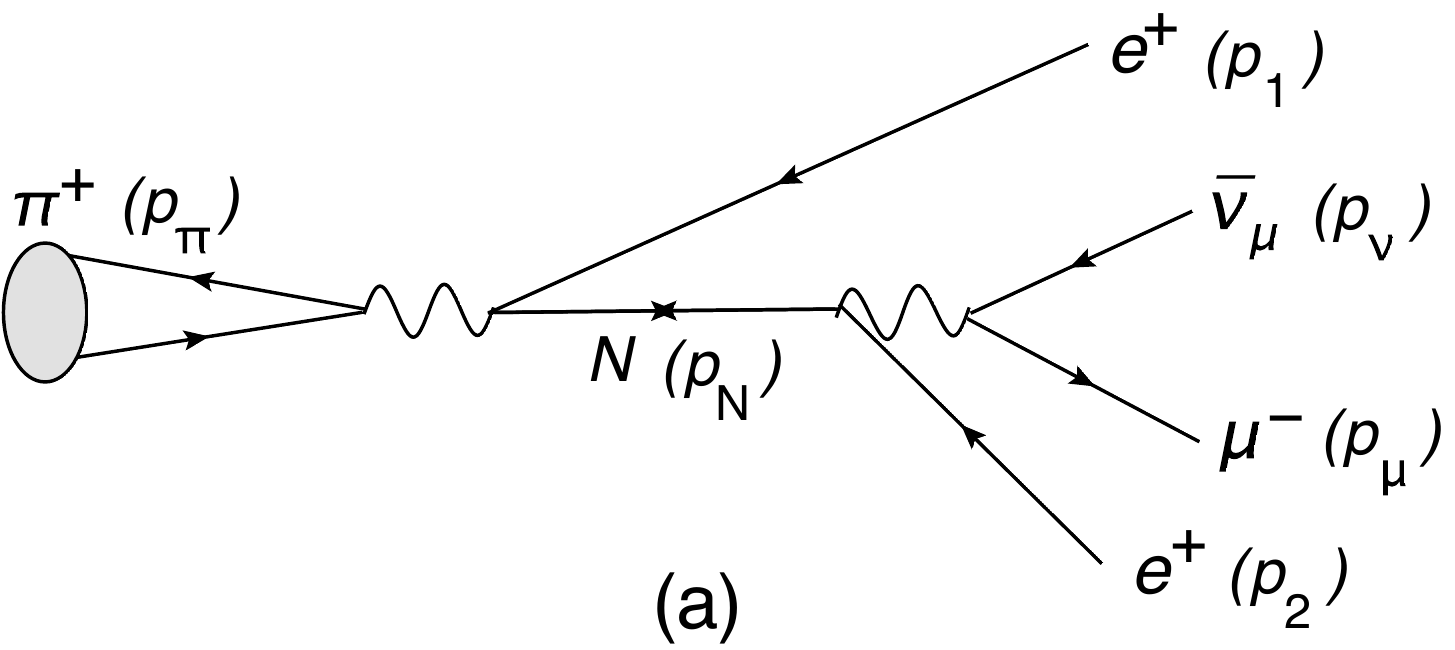}
\end{minipage}
\begin{minipage}[b]{.49\linewidth}
\includegraphics[width=80mm]{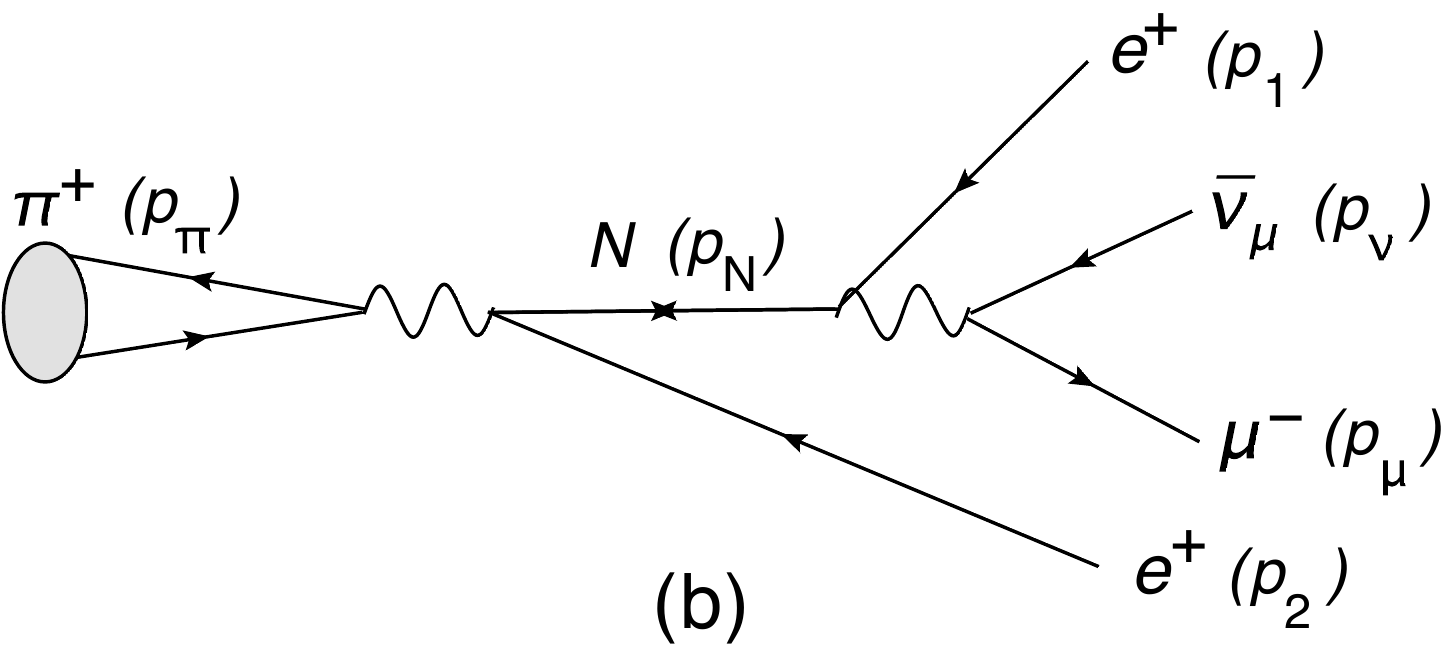}
\end{minipage}
\caption{The lepton number violating (LNV) process
$\pi^+ \to e^+ N \to e^+ e^+ \mu^- {\overline \nu}_{\mu}$ which can be
mediated by a neutrino $N$ only if $N$ is a Majorana particle:
(\textbf{a}) the direct (D) channel; (\textbf{b}) the crossed (C) channel.}
\label{FigLNVpi}
\end{figure}

In the previous Section we considered semileptonic decays of mesons heavier than the pion.
For pion decays, there are purely leptonic modes only, since the pion is the lightest meson.
In this Section we present and discuss the branching ratios
for the LNC decay
$\pi^{\pm} \to e^{\pm} N \to  e^{\pm} e^{\pm} \mu^{\mp} \nu_e$ and the LNV
decay $\pi^{\pm} \to e^{\pm} N \to  e^{\pm} e^{\pm} \mu^{\mp} \bar\nu_\mu$.
We also consider the differential branching ratios $d {\rm Br}/d E_{\mu}$, where $E_{\mu}$ is the energy of the produced $\mu^{\mp}$.
In contrast to the previous Section where the intermediate $N$ neutrino
has to be Majorana, here $N$ can be either Majorana or Dirac.
In the Dirac case, only the LNC mode is possible, while both LNC and LNV modes occur in the Majorana case. However, the experimental distinction between these two modes cannot be resolved by simply examining the final state, since the produced neutrino ($\nu_e$ or $\bar\nu_\mu$) is not detectable.
The major part of this Section refers to Ref.~\cite{CDK}, and for
certain details we use here results of Refs.~\cite{CKZ,CKZ2}.
The formalism is somewhat more complicated now because we have four
final particles (in the previous Section there were three). Nonetheless,
several features turn out to be similar as in the
previous Section.
\begin{figure}[htb] 
\begin{minipage}[b]{.49\linewidth}
\includegraphics[width=80mm]{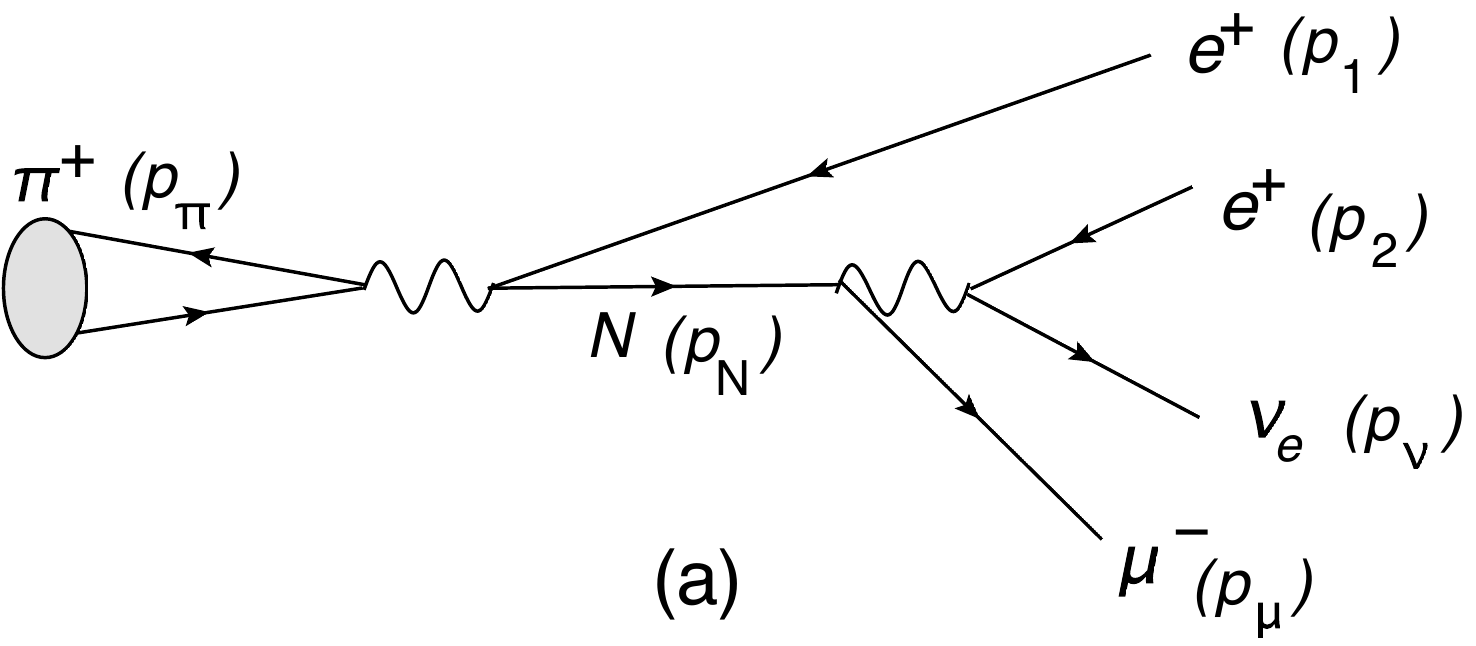}
\end{minipage}
\begin{minipage}[b]{.49\linewidth}
\includegraphics[width=80mm]{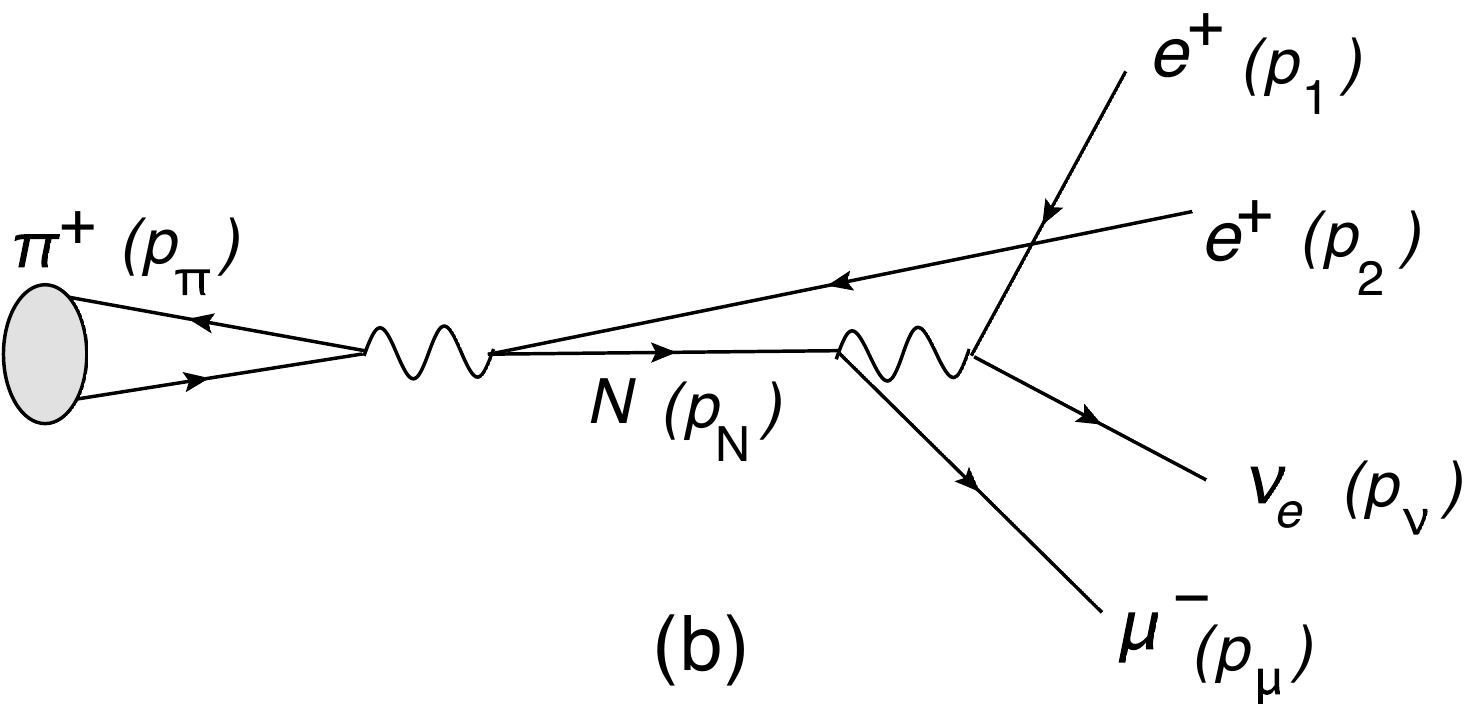}
\end{minipage}
\caption{The lepton number conserving (LNC)
process $\pi^+ \to e^+ N \to e^+ \mu^- e^+ \nu_e$ which can be
mediated by a Dirac or a Majorana on-shell neutrino $N$:
(\textbf{a}) the direct ($D$) channel; (\textbf{b}) the crossed ($C$) channel.}
\label{FigLNCpi}
 \end{figure}
The considered processes are presented in Figures~\ref{FigLNCpi} and \ref{FigLNVpi}.
As in the previous Section, we consider a scenario with
at least one heavy sterile neutrino $N$, which has suppressed
heavy-light mixing coefficients $B_{\ell N}$  with the first three
neutrino flavors $\nu_{\ell}$ ($\ell=e, \mu, \tau$),
\textit{cf}.~Equation~(\ref{mixN}).
The mentioned decay rates may be nonnegligible only if the intermediate
neutrino $N$ is on-shell, \emph{i.e}.,
\be
(M_{\mu} + M_e) < M_N   < (M_{\pi}-M_e) \
\label{MNint}
\ee
and the process is of the $s$-type, Figures~\ref{FigLNCpi} and \ref{FigLNVpi}.
Specifically, this means $106.2 \ {\rm MeV} < M_N < 139 \ {\rm MeV}$.


\subsection{Branching Ratios for $\pi^{\pm} \to e^{\pm} e^{\pm} \mu^{\mp} \nu$}
\label{sec:BrPisub}

The decay widths $\Gamma^{{\rm (X)}}(\pi^{\pm} \to e^{\pm} e^{\pm} \mu^{\mp} \nu)$
(X=LNC, LNV) can be written in terms of the corresponding reduced
decay amplitudes ${\cal T}_{\pi,\pm}^{{\rm (X)}}$
\be
\Gamma^{\rm (X)}(\pi^{\pm} \to e^{\pm} e^{\pm} \mu^{\mp} \nu) =
\frac{1}{2!} \frac{1}{2 M_{\pi}} \frac{1}{(2 \pi)^8}
\int d_4 \; | {\cal T}_{\pi,\pm}^{\rm (X)} | ^2 \
\label{GpiX1}
\ee
Here, $1/2!$ represents the symmetry factor from two final state electrons,
and $d_4$ is the integration element of the phase space of the four final
particles
\be
d_4 =\left( \prod_{j=1}^2  \frac{d^3 {\vec p}_j}{2 E_e({\vec p}_j)} \right)
 \frac{d^3 {\vec p}_{\mu}}{2 E_{\mu}({\vec p}_{\mu})}
 \frac{d^3 {\vec p}_{\nu}}{2 |{\vec p}_{\nu}|}
\delta^{(4)} \left( p_{\pi} - p_1 - p_2 - p_{\mu} - p_{\nu} \right) \
\label{d4}
\ee
Here, $p_1$ and $p_2$ are the momenta of $e$;
in the direct channel, the $e$ momentum at the first
(left-hand) vertex is $p_1$, and in
the crossed channel it is $p_2$,
\textit{cf.}~Figures~\ref{FigLNCpi} and \ref{FigLNVpi}.
When we use the expressions for the amplitudes
${\cal T}_{\pi,\pm}^{{\rm (X)}}$
of Appendix 4, for the specific considered case
$N_1=N$ (and no $N_2$), the decay width (\ref{GpiX1}) can be written as
\ba
\Gamma^{\rm (X)}(\pi^{\pm} \to e^{\pm} e^{\pm} \mu^{\mp} \nu) &=&
|k^{\rm (X)}|^2
{\big [}
\G_{\pi}^{\rm (X)}(DD^{*}) + \G_{\pi}^{\rm (X)}(CC^{*})
+ \G_{{\pi},\pm}^{\rm (X)}(DC^{*}) + \G_{{\pi},\pm}^{\rm (X)}(CD^{*} {\big ]} \
\label{GpiX2}
\ea
where X = LNC, LNV; $k_{\pm}^{\rm (X)}$ are the corresponding mixing factors
\be
\label{kpi}
k^{{\rm (LNV)}}  =  B_{e N}^2 \ , \qquad k^{{\rm (LNC)}} = B_{e N} B^{*}_{\mu N} \
\ee
and $\G_{\pi}^{{\rm (X)}}(YZ^{*})$ are the normalized (\emph{i.e}., without
explicit mixing dependence) decay widths
\be
\label{GpiX}
\G_{{\pi},\pm}^{{\rm (X)}}(YZ^{*}) =
K_{\pi}^2 \; \frac{1}{2!} \frac{1}{2 M_{\pi}} \frac{1}{(2 \pi)^8} \int d_4 \;
P^{{\rm (X)}}(Y) P^{{\rm (X)}}(Z)^{*} \; T_{{\pi},\pm}^{{\rm (X)}}(YZ^{*}) \
\ee

The expressions for $T_{{\pi},\pm}^{{\rm (X)}}(YZ^{*})$ 
are given in Appendix 4, for the direct ($YZ^{*}=DD^{*}$),
crossed ($YZ^{*}=CC^{*}$) and
direct-crossed interference contributions ($YZ^{*}=DC^{*}$, $CD^{*}$).
We have $T_{{\pi},+}^{{\rm (X)}}(DD^{*})=T_{{\pi},-}^{{\rm (X)}}(DD^{*})$ and
$T_{{\pi},+}^{{\rm (X)}}(CC^{*})=T_{{\pi},-}^{{\rm (X)}}(CC^{*})$, hence the terms
$\G_{\pi}^{{\rm (X)}}(DD^{*})$ and $\G_{\pi}^{{\rm (X)}}(CC^{*})$
in Equation~(\ref{GpiX2}) have
no subscripts $\pm$. Further, in Equation~(\ref{GpiX}),
$P^{{\rm (X)}}(Y)$  
represent the  $N$ propagator functions
of the direct and crossed channels ($Y=D, C$)
\bes
\label{Ps}
\ba
P^{\rm ({\rm LNC})}(D) &=&
\frac{1}{\left[ (p_{\pi}-p_1)^2 - M_{N}^2 + i \Gamma_{N} M_{N} \right]} , \qquad
P^{\rm ({\rm LNV})}(D) = M_{N} P^{\rm ({\rm LNC})}(D)
\label{PD}
\\
P^{\rm ({\rm LNC})}(C) &=&
\frac{1}{\left[ (p_{\pi}-p_2)^2 - M_{N}^2 + i \Gamma_{N} M_{N} \right]} , \qquad
P^{\rm ({\rm LNV})}(C) = M_{N} P^{\rm ({\rm LNC})}(C)
\label{PC}
\ea
\ees
where $\Gamma_N \equiv \Gamma(N \to {\rm all})$,
and $K_{\pi}^2$ is the following constant:
\be
K_{\pi}^2 =
G_F^4 f_{\pi}^2 |V_{ud}|^2 \approx 2.983 \times 10^{-22}   \ {\rm GeV}^{-6} \
\label{Kpisqr}
\ee

It turns out that in the case when the intermediate $N$ neutrino is on-shell,
\emph{i.e}., when its mass is in the interval of Equation~(\ref{MNint}), the
squares of the propagators (\ref{Ps}) reduce to simple delta functions
due to the inequality $\Gamma_N \ll M_N$
\ba
|P^{\rm (LNC)}(X)|^2 &=&
\left | \frac{1}{(p_{\pi}-p_k)^2-M^{2}_{N}+i \Gamma_{N} M_{N}} \right | ^2
\nonumber\\
&= &
\frac{\pi}{M_{N} \Gamma_{N}} \delta((p_{\pi}-p_k)^2-M^{2}_{N})
\quad (\Gamma_{N} \ll M_{N} ) \
\label{PP}
\ea
where $p_k = p_1, p_2$ for $X=D,C$.
In this on-shell case, the  $D D^{*}$ and $C C^{*}$
contributions in Equation~(\ref{GpiX2}) are large and equal, and the interference
contributions $D C^{*}$ and $C D^{*}$ are negligible in comparison;
we refer to \cite{CKZ} for details on this point. Hence
the decay width (\ref{GpiX2}) can be written in the on-shell case as
\ba
\Gamma^{{\rm (X)}}(\pi^{\pm} \to e^{\pm} e^{\pm} \mu^{\mp} \nu) &=&
2 |k^{\rm (X)}|^2 \G(\pi^{\pm} \to e^{\pm} e^{\pm} \mu^{\mp} \nu) \
\label{GpiX2OS}
\ea
Here, the normalized decay width
$\G_{\pi}^{{\rm (X)}}(DD^{*}) = \G_{\pi}^{{\rm (X)}}(CC^{*}) \equiv
\G(\pi^{\pm} \to e^{\pm} e^{\pm} \mu^{\mp} \nu)$ turns out to be the same
for X = LNC and X = LNV,
\ba
\lefteqn{
\G(\pi^{\pm} \to e^{\pm} e^{\pm} \mu^{\mp} \nu) \equiv  \G_{\pi}^{{\rm (X)}}(DD^{*})
\equiv \G_{\pi}(CC^*)}
\nonumber\\
&&
= \frac{K_{\pi}^2}{192 (2 \pi)^4} \frac{M_{N}^{11}}
{ M_{\pi}^3 \Gamma_{N} } \lambda^{1/2}(x_{\pi}, 1, x_{e})
\left[ x_{\pi} -1
+ x_{e}(x_{\pi} + 2 - x_{e}) \right] {\cal F}(x_{\mu}, x_{e})
\label{GXDDN}
\ea
where the following notations are used:
\bes
\label{notGXDD}
\ba
\lambda(y_1,y_2,y_3) & = & y_1^2 + y_2^2 + y_3^2 - 2 y_1 y_2 - 2 y_2 y_3 - 2 y_3 y_1 \
\label{lambda}
\\
x_{\pi} &=& \frac{M_{\pi}^2}{M_{N}^2} \ , \quad
x_{e} =  \frac{M_e^2}{M_{N}^2} \ , \quad
x_{\mu} =\frac{M_{\mu}^2}{M_{N}^2} \
\label{xs}
\ea
\ees
and the function ${\cal F}(x_{\mu}, x_{e})$ is given in
Appendix 5.
In the approximation $M_e=0$, the above result becomes simpler
\be
\lim_{M_e \to 0} \G(\pi^{\pm} \to e^{\pm} e^{\pm} \mu^{\mp} \nu) =
\frac{K_{\pi}^2}{192 (2 \pi)^4} \frac{M_{N}^{11}}
{\Gamma_{N} M_{\pi}^3} ( x_{\pi} - 1 )^2
f (x_{\mu}) \
\label{GXDDN0}
\ee
where the function $f(x_{\mu}) = {\cal F}(x_{\mu},0)$ is
\be
f(x_{\mu}) = 1 - 8 x_{\mu} + 8 x_{\mu}^3 - x_{\mu}^4 - 12 x_{\mu}^2 \ln x_{\mu} \
\label{fxmu}
\ee
It can be checked that the decay rate (\ref{GpiX2OS}), with $N$
on shell, coincides with the factorized expression
\be
\Gamma(\pi^+ \to e^+ e^+ \mu^- \nu) =
\Gamma(\pi^+ \to e^+ N) {\rm Br}(N \to e^+ \mu^- \nu) \
\label{fact}
\ee

In Appendix 5 we also provide the differential decay rates
$d \Gamma^{{\rm (X)}}(\pi^{\pm} \to e^{\pm} e^{\pm} \mu^{\mp} \nu)/d E_{\mu}$
with respect to
the final muon energy in the rest frame of $N$ neutrino. They turn out to
have quite different forms for X = LNC and X = LNV cases (we will return to this
later in this Section).

In order to obtain the branching fractions of the considered processes,
we need to divide the decay width by the total decay width of
the charged pion, $\Gamma(\pi^{\pm} \to {\rm all})$
\bes
\label{GPiall}
\bea
\Gamma(\pi^+ \to {\rm all}) &=& 2.529 \times 10^{-17} \ {\rm GeV}
\label{GPialla}
\\
&\approx &
\frac{1}{8\pi} G^{2}_{F}f^{2}_{\pi} M^{2}_{\mu} M_{\pi} |V_{ud}|^2
\left ( 1-\frac{M^{2}_{\mu}}{M^{2}_{\pi}} \right )^2 \left(1 +
\delta g_{\pi} \right) \
\label{GPiallb}
\eea
\ees
where the expression (\ref{GPiallb}) represents the
by far most dominant decay mode $\pi^{\pm} \to \mu^{\pm} \nu_{\mu}$,
and
\be
\delta g_{\pi} = \frac{M_e^2}{M_{\mu}^2}
\frac{ (1 - M_e^2/M_{\pi}^2)^2}{(1 - M_{\mu}^2/M_{\pi}^2)^2} \
\label{dgPi}
\ee
represents a (very small) relative correction coming from the
$\pi^{\pm} \to e^{\pm} \nu_{e}$ decay
($\delta g_{\pi} \approx 1.3 \times 10^{-4}$).

As we can see in Equations~(\ref{GpiX2OS}) and (\ref{GXDDN}), the
decay width $\Gamma(\pi^{\pm} \to e^{\pm} e^{\pm} \mu^{\mp} \nu)$
and its normalized counterpart
$\G(\pi^{\pm} \to e^{\pm} e^{\pm} \mu^{\mp} \nu)$
are inversely proportional to the (very small)
decay width \mbox{$\Gamma(N \to {\rm all}) \equiv \Gamma_N$}, which in turn
is proportional to the mixings $\K \sim |B_{\ell N}|^2$
($\ell = e, \mu, \tau$), \textit{cf.}~Equations~(\ref{GNwidth})--(\ref{calK})
and Figure~\ref{FigcNellN}.
As in Section~\ref{sec:BrMsub},
this effect represents the $N$-on-shell effect Equation~(\ref{PP}) and it makes
the considered width $\Gamma(\pi^{\pm} \to e^{\pm} e^{\pm} \mu^{\mp} \nu)$
by many orders of magnitude larger than it would be in the case of
off-shell $N$.
In the (narrow) mass interval (\ref{MNint}) for on-shell $N$
in the considered pion decays, the factor $\K$ in the width
$\Gamma_N$  Equation~(\ref{GNwidth}) has the following approximate form
(\textit{cf.}~Appendix 3 and Figure~\ref{FigcNellN}):
\be
\K(\pi^{\pm} \to e^{\pm} e^{\pm} \mu^{\mp} \nu)
\approx 1.6 |B_{e N}|^2 + 1.1 (|B_{\mu N}|^2 +|B_{\tau N}|^2)
\label{calKpi}
\ee
This expression, within the precision given here, is valid
equally for the Majorana and for the Dirac $N$
{
and agrees with that given in Ref.~\cite{CKZ} for the Majorana case.
However, the affirmation in Ref.~\cite{CKZ} that
$\K$ for Dirac $N$ is smaller by a factor of two is not correct.}

\begin{figure}[htb]
\centering\includegraphics[width=110mm]{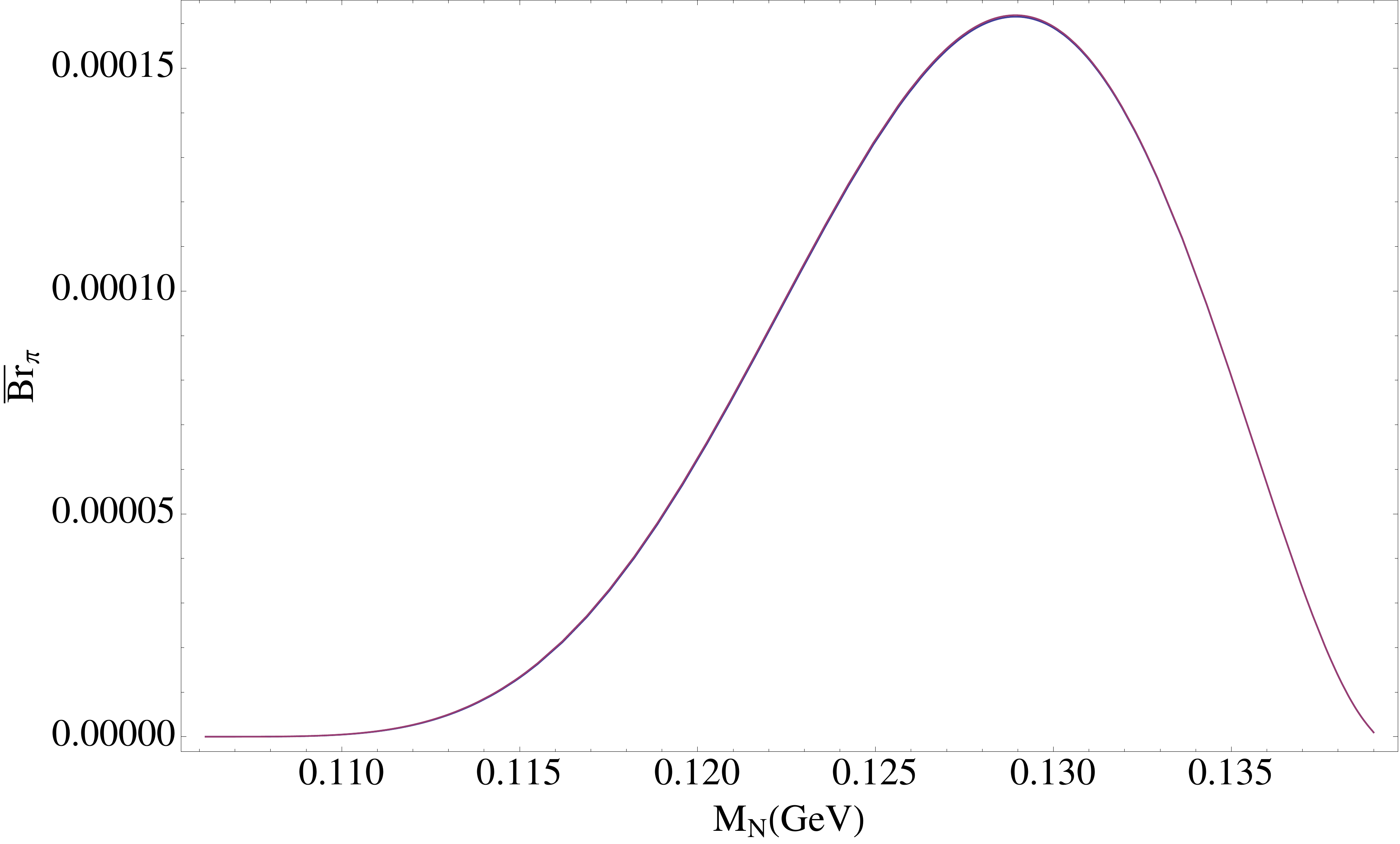}
\caption{The canonical branching ratio $\bBr_{\pi}$, Equation~(\ref{bBr}),
as a function of the mass
$M_N$. The full formula was used
(with $M_e=0.511 \times 10^{-3}$ GeV). The formula for $M_e=0$ case
gives a line which is in this Figure
indistinguishable from the depicted line.}
\label{bBrfig}
\end{figure}

\begin{figure}[htb] 
\begin{minipage}[b]{.49\linewidth}
\centering\includegraphics[width=88mm]{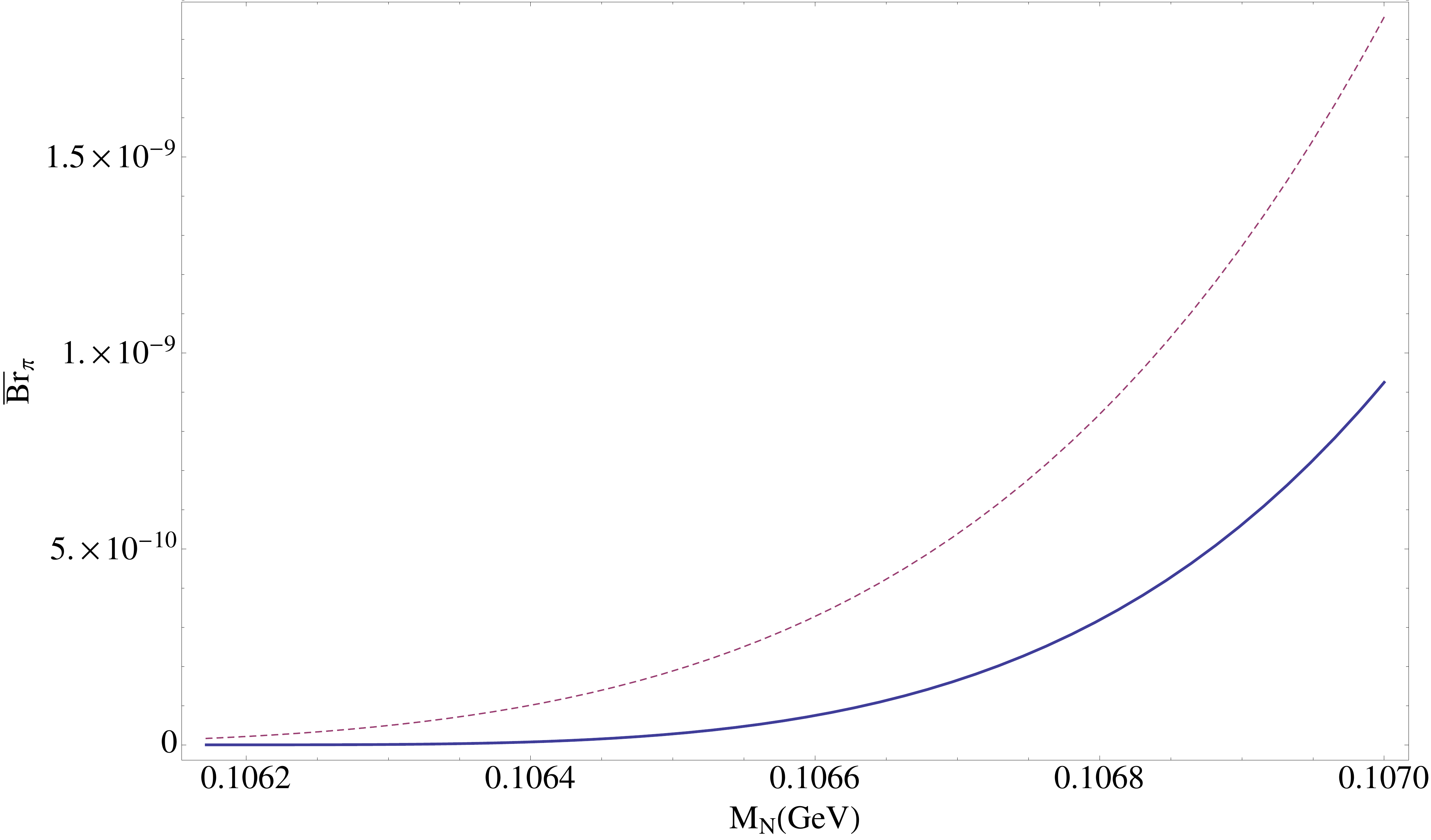}
\end{minipage}
\begin{minipage}[b]{.49\linewidth}
\centering\includegraphics[width=88mm]{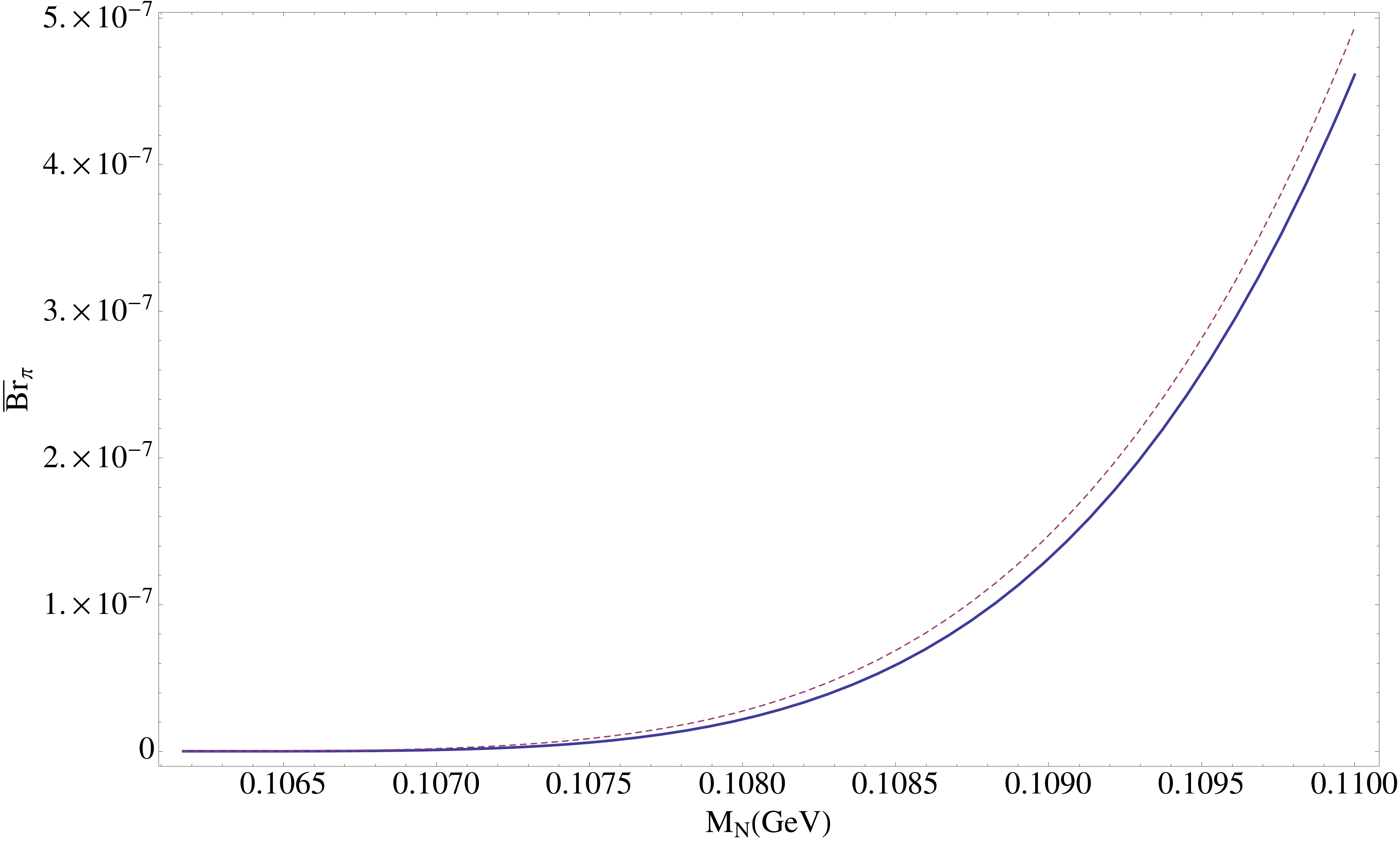}
\end{minipage}
 \caption{The canonical branching ratio $\bBr_{\pi}$ near the
lower end point $M_{\mu} +M_e$ (=$0.1062$~GeV): (\textbf{a}) in the interval below
$0.107$ GeV; (\textbf{b}) in the interval below $0.110$ GeV. The dashed line is
for $M_e=0$, the full line includes
the effects of $M_e = 0.511 \times 10^{-3}$ GeV.}
\label{bBrendfig}
 \end{figure}

In certain analogy with Section~\ref{sec:BrMsub}, we can define here the
canonical branching ratio ${\overline {\rm Br}}_{\pi}$
as the part of the branching ratio ${\rm Br}^{\rm (X)}_{\pi} \equiv
\Gamma(\pi^{\pm} \to e^{\pm} e^{\pm} \mu^{\mp} \nu)/\Gamma(\pi \to {\rm all})$
which contains no explicit or implicit heavy-light mixing dependence
(and is independent of X = LNC or LNV)
\bes
\label{bBr}
\ba
{\overline {\rm Br}}_{\pi} & \equiv & 2 \frac{{\K}}{\Gamma_{\pi}}
\G(\pi^{\pm} \to e^{\pm} e^{\pm} \mu^{\mp} \nu)
=  \frac{\K}{|k^{\rm (X)}|^2}
{\rm Br}^{\rm (X)}(\pi^{\pm} \to e^{\pm} e^{\pm} \mu^{\mp} \nu)
\label{bBra}
\\
& = &
 \frac{1}{16 \pi} \frac{K_{\pi}^2 M_{\pi}^3}{G_F^2 \Gamma(\pi^+ \to {\rm all})}
\frac{1}{x_{\pi}^3} \lambda^{1/2}(x_{\pi}, 1, x_{e})
\left[ x_{\pi} -1 + x_{e}(x_{\pi} + 2 - x_{e}) \right]
{\cal F}(x_{\mu}, x_{e})
\label{bBrb}
\\
& = &
\frac{1}{2} \frac{1}{x_{\mu} (x_{\pi}-x_{\mu})^2 (1 + \delta g_{\pi})}
\lambda^{1/2}(x_{\pi}, 1, x_{e})
\left[ x_{\pi} -1 + x_{e}(x_{\pi} + 2 - x_{e}) \right]
{\cal F}(x_{\mu}, x_{e}) \
\label{bBrc}
\ea
\ees
where the notations (\ref{notGXDD}) are used,
and in Equation~(\ref{bBrc}) we used the expression (\ref{const1})
in Appendix 5.
We note that $\K \sim |B_{\ell N}|^2$ and $|k^{\rm (X)}|^2 \sim |B_{\ell N}|^4$,
and ${\rm Br}^{\rm (X)}_{\pi} \propto |k^{\rm (X)}|^2/\K \sim |B_{\ell N}|^2$,
where $B_{\ell N}$ stands for a generic heavy-light mixing coefficient
($|B_{\ell N}| \sim |B_{e N}| \sim |B_{\mu N}|$). Naively, we should have
very strong suppression
${\rm Br}^{\rm (X)}_{\pi} \propto  |k^{\rm (X)}|^2 \propto |B_{\ell N}|^4$;
however, the
on-shellness (\ref{PP}) brings in factor $1/\Gamma_N \propto
1/\K \sim 1/|B_{\ell N}|^2$, reducing the suppression to
${\rm Br}^{\rm (X)}_{\pi} \propto |B_{\ell N}|^2$.
In Figure~\ref{bBrfig} we present the canonical branching fraction
$\bBr_{\pi}$ as a function of $M_N$ in the entire
interval of on-shellness (\ref{MNint}).
In Figure~\ref{bBrendfig} this canonical branching fraction is presented
at the lower edge of the on-shellness interval, where the
effects of $M_e \not= 0$ turn out to be appreciable.
We can see that the branching ratio is the largest when $M_N \approx 0.130$
GeV.
\vspace{-12pt}
The differential branching ratios
$d {\rm Br}^{\rm (X)}/d E_{\mu}$, where $E_{\mu}$ is the final muon
energy in the $N$-rest frame, are obtained directly from the
differential branching ratios $d \G_{\pi}^{\rm (X)}/d E_{\mu}$,
and the latter quantity for \linebreak X = LNV is given
explicitly in Appendix 5.
The canonical differential branching ratios, free of any mixing
dependence, can be defined in analogy with
the definition of the canonical total branching ratio
(\ref{bBra}), and, in contrast to canonical branching ratios
(\ref{bBr}) they do depend on whether X = LNC or \linebreak X = LNV
\ba
\frac{d {\overline {\rm Br}_{\pi}^{\rm (X)}}}{d E_{\mu}} &
\equiv & 2 \frac{{\K}}{\Gamma_{\pi}}
\frac{d \G(\pi^{\pm} \to e^{\pm} e^{\pm} \mu^{\mp} \nu)}{d E_{\mu}}
=  \frac{\K}{|k^{\rm (X)}|^2}
\frac{d {\rm Br}^{\rm (X)}(\pi^{\pm} \to e^{\pm} e^{\pm} \mu^{\mp} \nu)}{d E_{\mu}}
\label{dbBrX}
\ea

Explicit expressions for these quantities, when X = LNV and X = LNC, are
given in Appendix 5 in
Equations~(\ref{dbBrLVEmu}) and (\ref{dbBrLCEmu}), and in the
limit $M_e=0$ in Equation~(\ref{dbBrLVLCMe0}).

\begin{figure}[htb]
\begin{minipage}[b]{.49\linewidth}
\centering\includegraphics[width=\linewidth]{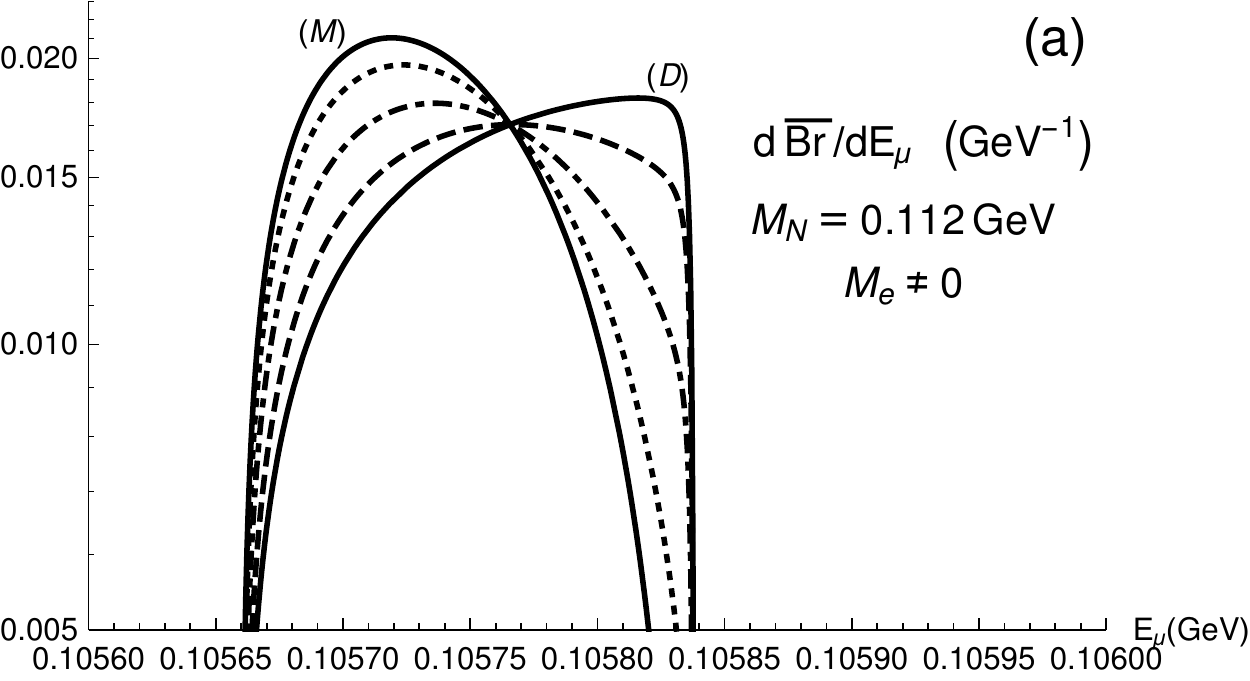}
\end{minipage}
\begin{minipage}[b]{.49\linewidth}
\centering\includegraphics[width=\linewidth]{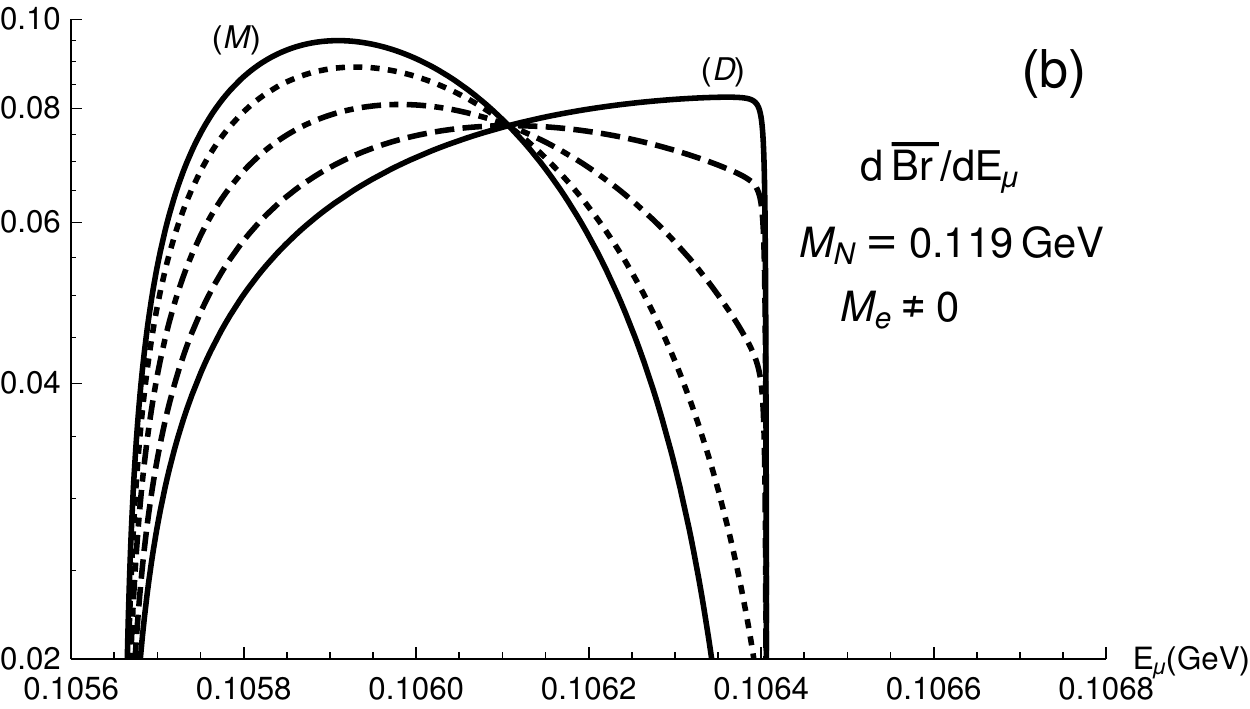}
\end{minipage} \vspace{6pt}
\begin{minipage}[b]{.49\linewidth}
\centering\includegraphics[width=\linewidth]{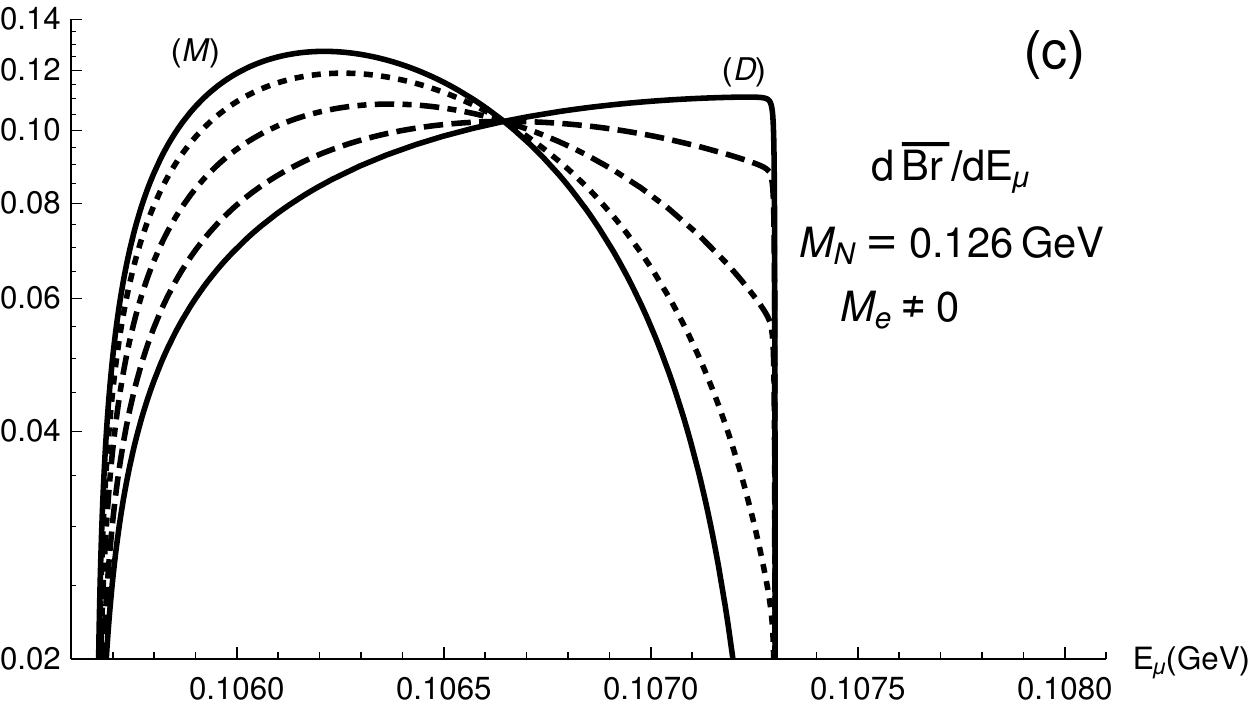}
\end{minipage}
\begin{minipage}[b]{.49\linewidth}
\centering\includegraphics[width=\linewidth]{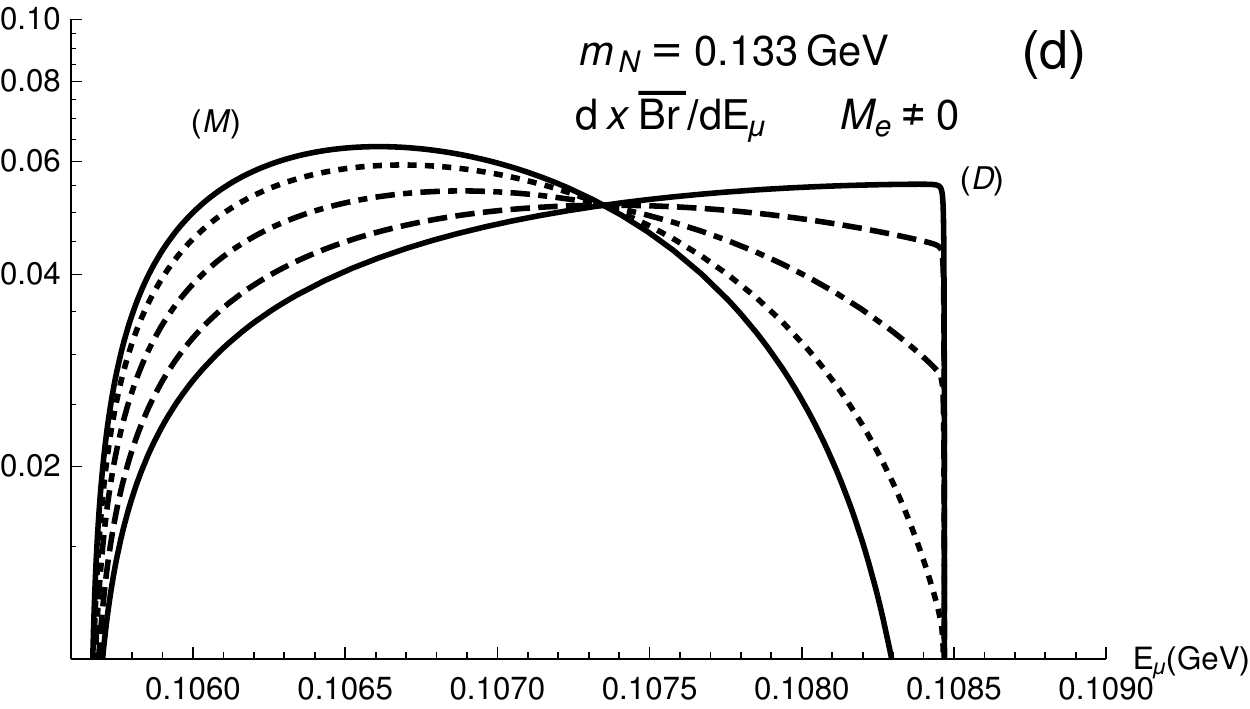}
\end{minipage}
\caption{The canonical differential branching ratio
$d {\overline {\rm Br}}_{\pi}(\alpha_M)/d E_{\mu}$ as a function of
the muon energy in the neutrino $N$ rest frame, $E_{\mu}$, as defined via
Equations~(\ref{dbBral}), for the decays
$\pi^{\pm} \to e^{\pm} e^{\pm} \mu^{\mp} {\nu}$
mediated by a Majorana neutrino $N$, for various neutrino
masses: (\textbf{a}) $M_N=0.112$ GeV; (\textbf{b}) $M_N=0.119$ GeV;
(\textbf{c}) $M_N=0.126$ GeV; (\textbf{d}) $M_N=0.133$ GeV.
In each graph there are five curves, corresponding to different values of the admixture parameter $\alpha_M$ [Equation~(\ref{alM})]:
$\alpha_M=1.0$ is the solid (M) curve; $0.8$ (dotted); $0.5$ (dot-dashed); $0.2$ (dashed). The case mediated by a Dirac neutrino ($\alpha_M=0$)
is also presented as the solid line labelled (D). In comparison with Fig.~5 of Ref.~\cite{CDK} (where $M_e=0$ was taken and linear $y$-scale was used), the true value of $M_e = 0.511$ MeV is taken here and logarithmic $y$-scale is used.}
\label{dGdEelCMN}
\end{figure}

The differential (and full) branching ratios for the process
$\pi^{\pm} \to e^{\pm} e^{\pm} \mu^{\mp} \nu$ differ in the cases when
the on-shell $N$ is Majorana or Dirac. When $N$ is Dirac, only the
X = LNC process contributes. When $N$ is Majorana, both LNC and LNV
processes contribute. We can write the differential and full branching ratios
in these cases in terms of their canonical counterparts, by defining
first the combined canonical differential branching ratios
\be
\frac{d {\overline {\rm Br}}_{\pi}(\alpha)}{d E_{\mu}} \equiv
\alpha \frac{d {\overline {\rm Br}}_{\pi}^{\rm (LNV)}}{d E_{\mu}} +
(1 - \alpha)  \frac{d {\overline {\rm Br}}_{\pi}^{\rm (LNC)}}{d E_{\mu}} \
\label{dbBral}
\ee
where $0 \leq \alpha \leq 1$.
Then it is straightforward to check that in the cases of
Dirac and Majorana $N$ neutrino, the branching ratios can
be expressed in terms of the above quantites (\ref{dbBral})
\bes
\label{dBrDirMaj}
\bea
\frac{d {\rm Br}^{\rm (Dir.)}(\pi^{\pm} \to e^{\pm} e^{\pm} \mu^{\mp} \nu)}{d E_{\mu}}
& = & \frac{|k^{\rm (LNC)}|^2}{\K^{\rm (Dir.)}}
\frac{d {\overline {\rm Br}}_{\pi}(\alpha=0)}{d E_{\mu}} \
\label{dBrDir}
\\
\frac{d {\rm Br}^{\rm (Maj.)}(\pi^{\pm} \to e^{\pm} e^{\pm} \mu^{\mp} \nu)}{d E_{\mu}}
& = & \frac{(|k^{\rm (LNV)}|^2 + |k^{\rm (LNC)}|^2)}{\K^{\rm (Maj.)}}
\frac{d {\overline {\rm Br}}_{\pi}(\alpha_M)}{d E_{\mu}} \
\label{dBrMaj}
\eea
\ees
where we recall the definition (\ref{kpi}) of the coefficients $k^{\rm (X)}$,
and the ``Majorana LNV admixture'' parameter $\alpha_M$
appearing in Equation~(\ref{dBrMaj}) is defined as
\be
\alpha_M = \frac{ |k^{\rm (LNV)}|^2 }{(|k^{\rm (LNV)}|^2 + |k^{\rm (LNC)}|^2)}
= \frac{ |B_{e N}|^2 }{ (|B_{e N}|^2 + |B_{\mu N}|^2) } \
\label{alM}
\ee
Integration of the relations (\ref{dBrDirMaj}) over $E_{\mu}$ leads to
the full branching ratios
\bes
\label{BrDirMaj}
\bea
{\rm Br}^{\rm (Dir.)}(\pi^{\pm} \to e^{\pm} e^{\pm} \mu^{\mp} \nu)
& = & \frac{|k^{\rm (LNC)}|^2}{\K^{\rm (Dir.)}}
{\overline {\rm Br}}_{\pi} =
\frac{|B_{e N}|^2 |B_{\mu N}|^2}
{\sum_{\ell=e,\mu,\tau} {\cal N}^{\rm (Dir.)}_{\ell N} |B_{\ell N}|^2}
{\overline {\rm Br}}_{\pi},
\label{BrDir}
\\
{\rm Br}^{\rm (Maj.)}(\pi^{\pm} \to e^{\pm} e^{\pm} \mu^{\mp} \nu)
& = & \frac{(|k^{\rm (LNV)}|^2\!+\!|k^{\rm (LNC)}|^2)}{\K^{\rm (Maj.)}}
{\overline {\rm Br}}_{\pi} =
\frac{|B_{e N}|^2 (|B_{e N}|^2\!+\!|B_{\mu N}|^2)}
{\sum_{\ell=e,\mu,\tau} {\cal N}^{\rm (Maj.)}_{\ell N} |B_{\ell N}|^2}
{\overline {\rm Br}}_{\pi}
\label{BrMaj}
\eea
\ees
where we used the fact that the $E_{\mu}$-integrated expression
${\overline {\rm Br}}_{\pi}(\alpha)$ is independent
of $\alpha$, \textit{cf}.~Equations~(\ref{bBr}).
In Figures~\ref{dGdEelCMN} we present these canonical differential branching
ratios as a function of the muon energy $E_{\mu}$ in the rest frame of $N$,
for four different values of mass $M_N$ in the on-shell interval
(\ref{MNint}), and for five different values of the
admixture parameter ($\alpha_M = 1,$ 0.8, 0.5, 0.2 and 0),
The value $\alpha_M=0$ corresponds to the case of Dirac $N$.
From the curves of Figure~\ref{dGdEelCMN} we can conclude:
if $N$ is Majorana neutrino with a significant value of the
admixture parameter $\alpha_M$ and with mass in the interval (\ref{MNint}),
then the measurement of such differential
branching ratios may be able to confirm the Majorana nature of $N$.
If the differential branching ratios are studied with respect to the
muon energy $E^{'}_{\mu}$ in the pion rest frame, the distinction between
the Dirac and Majorana case is more difficult, \textit{cf.}~Ref.~\cite{CDK}.


\subsection{Effect of the Long Neutrino Lifetime on the Observability of  $\pi^{\pm} \to e^{\pm} e^{\pm} \mu^{\mp} \nu$}
\label{sec:effBrPisub}

The branching ratios presented in this Section so far, Equations~(\ref{BrDirMaj})
in conjunction with Equations~(\ref{bBr}),
should be multiplied by the probability $P_N$
for the on-shell neutrino $N$ to decay within the detector (of length $L$),
as explained in Section~\ref{sec:effBrMsub}, Equations~(\ref{PN1})--(\ref{bPN})
and Figure~\ref{bLNfig}. As a consequence, the effective (true, measurable)
branching ratios are not ${\rm Br}$ ($\sim |B_{\ell N}|^2$) of
Equations~(\ref{BrDirMaj}), but ${\rm Br}_{\rm eff} = P_N {\rm Br}$
\bes
\label{BreffDir}
\bea
{\rm Br}_{\rm eff}^{\rm (Dir.)}(\pi^{\pm} \to e^{\pm} e^{\pm} \mu^{\mp} \nu)
&=&  P_N^{\rm (Dir.)} {\rm Br}^{\rm (Dir.)}(\pi^{\pm} \to e^{\pm} e^{\pm} \mu^{\mp} \nu)
\nonumber\\
&=& \left[ {\overline P}_N \left( \frac{L}{1 {\rm m}} \right) \K^{\rm (Dir.)} \right]
\left[ \frac{|B_{e N}|^2 |B_{\mu N}|^2}{\K^{\rm (Dir.)}}
{\overline {\rm Br}}_{\pi} \right]
\label{BreffDira}
\\
& = & |B_{e N}|^2 |B_{\mu N}|^2 \left( \frac{L}{1 {\rm m}} \right)
{\overline P}_N {\overline {\rm Br}}_{\pi}
= |B_{e N}|^2 |B_{\mu N}|^2 \left( \frac{L}{1 {\rm m}} \right)
{\overline {\rm Br}}_{\pi,{\rm eff}}
\label{BreffDirb}
\eea
\ees
\bes
\label{BreffMaj}
\bea
\lefteqn{
{\rm Br}_{\rm eff}^{\rm (Maj.)}(\pi^{\pm} \to e^{\pm} e^{\pm} \mu^{\mp} \nu)
=  P_N^{\rm (Maj.)} {\rm Br}^{\rm (Maj.)}(\pi^{\pm} \to e^{\pm} e^{\pm} \mu^{\mp} \nu)
}
\nonumber\\
& = &
\left[ {\overline P}_N \left( \frac{L}{1 {\rm m}} \right) \K^{\rm (Maj.)} \right]
\left[ \frac{|B_{e N}|^2 (|B_{e N}|^2+|B_{\mu N}|^2)}{\K^{\rm (Maj.)}}
{\overline {\rm Br}}_{\pi} \right]
\label{BreffMaja}
\\
& = & |B_{e N}|^2 (|B_{e N}|^2+|B_{\mu N}|^2) \left( \frac{L}{1 {\rm m}} \right)
{\overline P}_N {\overline {\rm Br}}_{\pi}
=|B_{e N}|^2 (|B_{e N}|^2+|B_{\mu N}|^2) \left( \frac{L}{1 {\rm m}} \right)
{\overline {\rm Br}}_{\pi,{\rm eff}}
\label{BreffMajb}
\eea
\ees

\begin{figure}[htb]
\centering\includegraphics[width=110mm]{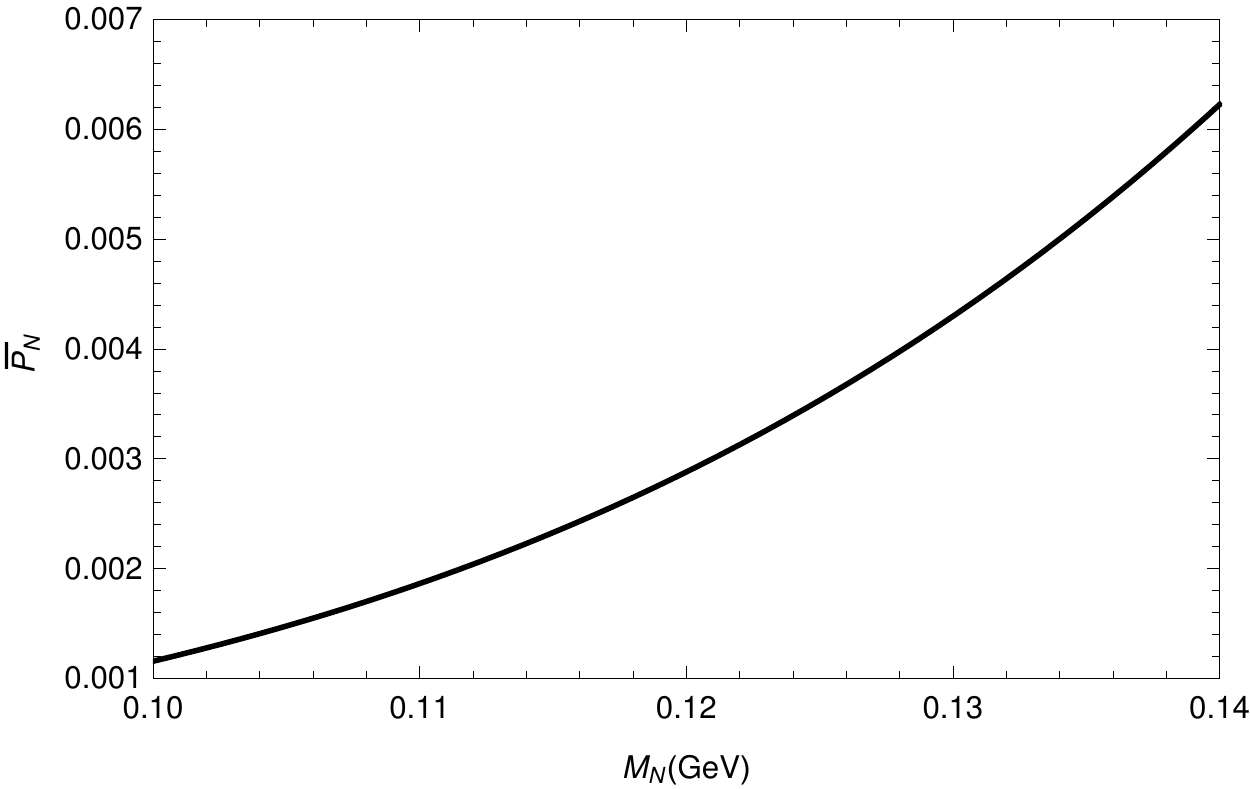}
\caption{The canonical probability
${\overline P}_N \equiv (1 {\rm m})/\bL_N$, Equation~(\ref{bPN}),
as a function of the neutrino mass $M_N$ in the approximate
interval where it is on shell (\ref{MNint})
for decays $\pi^{\pm} \to e^{\pm} e^{\pm} \mu^{\mp} \nu$.
The Lorentz lab time dilation factor is chosen to be
$\gamma_{N}$ [$\equiv (1 - \beta_N^2)^{-1/2}$] $=2$.}
\label{bLNPifig}
\end{figure}

In Equations~(\ref{BreffDira}) and (\ref{BreffMaja}) we used the expressions
(\ref{BrDirMaj}) for ${\rm Br}$ and Equation~(\ref{bPN}) for $P_N$.
In Equations~(\ref{BreffDirb}) and (\ref{BreffMajb}) we introduced canonical
(\emph{i.e}., without any mixing dependence) effective branching ratio
${\overline {\rm Br}}_{\pi,{\rm eff}}$
\be
{\overline {\rm Br}}_{\pi,{\rm eff}} \equiv
{\overline P}_N {\overline {\rm Br}}_{\pi} \
\label{bBrPieff}
\ee
with $ {\overline {\rm Br}}_{\pi}$ given in Equations~(\ref{bBr}),
and the canonical nonsurvival probability ${\overline P}_N$ presented
in Figure~\ref{bLNfig} in Section~\ref{sec:effBrMsub} for a wide range of
$N$ neutrino masses. In Figure~\ref{bLNPifig} we present ${\overline P}_N$
in the here relevant narrower mass interval (\ref{MNint}).

As in the case of semileptonic LNV decays of Section~\ref{sec:effBrMsub},
we notice that also here in the effective branching ratios,
Equations~(\ref{BreffDir}) and (\ref{BreffMaj}), the complicated dependence on the mixing
parameters entailed by the factors
$\K = \sum_{\ell} {\cal N}_{\ell N} |B_{\ell N}|^2$ of $\Gamma_N$
[\textit{cf.}~Equation~(\ref{calK})],
cancels out, and there remains only simple dependence on the mixing
parameters, in the form $|B_{e N}|^2 |B_{\mu N}|^2$ or
$|B_{e N}|^2 (|B_{e N}|^2+|B_{\mu N}|^2)$. Further, in comparison with
the branching ratios ${\rm Br}$ of Equations~(\ref{BrDirMaj}),
which are $\sim |B_{\ell N}|^2$, the effective (true)
branching ratios ${\rm Br}_{\rm eff}$ of Equations~(\ref{BreffDir}) and (\ref{BreffMaj})
are unfortunately significantly more suppressed by the mixing parameters,
namely ${\rm Br}_{\rm eff} \sim |B_{\ell N}|^4$.

The presently known experimental bounds on the mixing parameters
$|B_{\ell N}|^2$ ($\ell = e, \mu, \tau$)
in the here relevant narrow mass range (\ref{MNint}),
are: $|B_{e N}|^2 \lesssim 10^{-8}$ \cite{PIENU:2011aa};
$|B_{\mu N}|^2 \lesssim 10^{-6}$
\cite{BmuN,BmuN2,BmuN3}; $|B_{\tau N}|^2 \lesssim 10^{-4}$ \cite{BtauN};
\textit{cf.}~also
Refs.~\cite{Atre,Ruchayskiy:2011aa,1502.00477,Devlimits}.
%

The future pion factories, such as the Project X at Fermilab,
will be designed to produce charged pions with lab energies
$E_{\pi}$ of a few GeV
and luminosities $\sim 10^{22} \ {\rm cm}^{-2} {\rm s}^{-1}$ \cite{ProjX,Geer},
and $\sim 10^{29}$ charged pions could be expected per year.

The canonical effective branching ratio (\ref{bBrPieff}) can be estimated as
\be
{\overline {\rm Br}}_{\pi,{\rm eff}} \lesssim 10^{-6}
\label{bBrPieffUB}
\ee
as can be inferred from Equation~(\ref{bBrPieff}) and
Figures~\ref{bLNPifig} and \ref{bBrfig}.
Equations (\ref{BreffDir}) and (\ref{BreffMaj}) then imply that the effective
(true) branching ratios for the considered reactions are (we assume
$L=1 {\rm\,m}$ for the detector~length)
\bes
\label{UBPi1}
\bea
{\rm Br}_{\rm eff}^{\rm (Dir.)} &\lesssim & |B_{e N}|^2 |B_{\mu N}|^2 10^{-6},
\label{UBPi1Dir}
\\
{\rm Br}_{\rm eff}^{\rm (Maj.)} &\lesssim &
|B_{e N}|^2 (|B_{e N}|^2 + |B_{\mu N}|^2) 10^{-6}
\label{UBPi1Maj}
\eea
\ees

If the larger among the mixing elements ($|B_{\ell N}|^2$, $\ell=e, \mu$)
is $|B_{\mu N}|^2$ ($\lesssim 10^{-6}$), the LNC processes dominate,
and the effective branching ratios
(\ref{UBPi1}) have the common upper bounds
\be
\label{UBPi2}
{\rm Br}_{\rm eff}^{\rm (Dir.,Maj.)} \lesssim  |B_{e N}|^2 |B_{\mu N}|^2 10^{-6}
\lesssim |B_{e N}|^2 10^{-12}
\ee

If in this case $|B_{e N}|^2$ is close to its present upper bound,
$|B_{e N}|^2 \sim 10^{-8}$, we obtain ${\rm Br}_{\rm eff}^{\rm (Dir.,Maj.)}
\lesssim 10^{-20}$. This implies that up to $10^9$ events
$\pi^{\pm} \to e^{\pm} e^{\pm} \mu^{\mp} \nu$ could be detected per year
in such a scenario.

On the other hand, if the larger among the mixing elements
($B_{\ell N}|^2$, $\ell=e, \mu$) is $|B_{e N}|^2$ ($\lesssim 10^{-8}$),
the LNV processes dominate, and
the effective branching ratios (\ref{UBPi1}) have the following
upper bounds:
\bes
\label{UBPi3}
\bea
{\rm Br}_{\rm eff}^{\rm (Dir.)} &\lesssim & |B_{\mu N}|^2 10^{-14}
\label{UBPi3Dir}
\\
{\rm Br}_{\rm eff}^{\rm (Maj.)} &\lesssim &
10^{-22}
\label{UBPi3Maj}
\eea
\ees
In such a case we have $|B_{\mu N}|^2 < |B_{e N}|^2 \lesssim 10^{-8}$,
and up to $10^7$ events could be detected per year.

The present upper bounds on $|B_{\ell N}|^2$
suggest that the first scenario, Equation~(\ref{UBPi2}),
is more likely.

The measurement of the effective branching ratios alone cannot
distinguish between the Dirac and the Majorana character of intermediate
neutrino $N$. However, as argued in Section~\ref{sec:BrPisub}
and presented in Figure~\ref{dGdEelCMN}, the measurement of the
differential branching ratios for the considered processes
is a promising way to discern the character of the neutrinos.
However, the differential branching ratios (\ref{dBrDirMaj})
must be multiplied by the nonsurvival probability $P_N$, in
order to obtain the effective (true, measurable)
differential branching ratios $d {\rm Br}_{\rm eff}/d E_{\mu}$.
In analogy with Equations~(\ref{BreffDir}) and (\ref{BreffMaj}) we obtain,
using Equation~(\ref{dBrDirMaj})
\bes
\label{dBReffDirMaj}
\bea
\frac{d {\rm Br}_{\rm eff}^{\rm (Dir.)}(\pi^{\pm} \to e^{\pm} e^{\pm} \mu^{\mp} \nu)}{d E_{\mu}} &=& P_N \frac{d {\rm Br}^{\rm (Dir.)}(\pi^{\pm} \to e^{\pm} e^{\pm} \mu^{\mp} \nu)}{d E_{\mu}}
\nonumber\\
& = &
|B_{e N}|^2 |B_{\mu N}|^2 \left( \frac{L}{1 {\rm m}} \right)
\frac{d {\overline {\rm Br}}_{\pi,{\rm eff}}(\alpha=0)}{d E_{\mu}}
\label{dBreffDir}
\\
\frac{d {\rm Br}_{\rm eff}^{\rm (Maj.)}(\pi^{\pm} \to e^{\pm} e^{\pm} \mu^{\mp} \nu)}{d E_{\mu}} &=&  P_N \frac{d {\rm Br}^{\rm (Maj.)}(\pi^{\pm} \to e^{\pm} e^{\pm} \mu^{\mp} \nu)}{d E_{\mu}}
\nonumber\\
& = &
|B_{e N}|^2 (|B_{e N}|^2+|B_{\mu N}|^2)
\left( \frac{L}{1 {\rm m}} \right)
\frac{d {\overline {\rm Br}}_{\pi,{\rm eff}}(\alpha_M)}{d E_{\mu}}
\label{dBreffMaj}
\eea
\ees
where the canonical differential effective branching ratios were introduced,
in analogy with Equation~(\ref{dbBral})
\be
\frac{d {\overline {\rm Br}}_{\pi,{\rm eff}}(\alpha)}{d E_{\mu}} \equiv
{\overline P}_N \frac{d {\overline {\rm Br}}_{\pi}(\alpha)}{d E_{\mu}} \
\label{dbBrPieff}
\ee
with $ d {\overline {\rm Br}}_{\pi}(\alpha)/d E_{\mu}$
given in Equation~(\ref{dbBral}). The parameter $\alpha_M$
appearing in Equation~(\ref{dBreffMaj}) was defined in Equation~(\ref{alM}).
In analogy with Figure~\ref{dGdEelCMN}, and using the values of
${\overline P}_N$ from Figure~\ref{bLNPifig}, we deduce that
the values of the $y$-axes of Figure~\ref{dGdEelCMN}a--d,
must be multiplied by ${\overline P}_N \approx 2.0 \times 10^{-3}$,
$2.8 \times 10^{-3}$, $3.7 \times 10^{-3}$, and $4.8 \times 10^{-3}$,
respectively (assuming $L=1$ m and $\gamma_N=2$),
in order to obtain the representation for the
curves of the canonical differential {\em effective\/} ratios
$d {\overline {\rm Br}}_{\pi,{\rm eff}}(\alpha)/d E_{\mu}$.



\section{CP Violation in Charged Meson Decays Mediated by Massive Sterile Neutrinos}
\label{sec:BrCPV}

CP violation in the lepton sector could be measured
by neutrino oscillations \cite{oscCP}.
Here we consider the possibilities of measuring CP violation
in meson decays mediated by sterile neutrinos $N$, such
as the semileptonic LNV decays of charged heavy pseudoscalar mesons
considered in Section~\ref{sec:BrM}, or the leptonic (LNV and LNC)
decays of charged pions considered in Section~\ref{sec:BrPi}.

It turns out that CP violation in all such decays is possible
in scenarios with at least two massive sterile neutrinos
$N_j$ ($j=1,2$).
CP violation in the neutrino sector is expected whether
neutrinos are Dirac or Majorana particles.
However, in the Pontecorvo--Maki--Nakagawa--Sakata (PMNS)
mixing \mbox{matrix \cite{Pontecorvo,Pontecorvo2,MNS}},
the number of possible CP-violating phases
is larger when the neutrinos are Majorana particles.
If $n = 3 +{\cal N}$ is the total number of neutrinos
(${\cal N}$ is the number of sterile neutrinos),
the number of CP-violating phases
is $n(n-1)/2$ if neutrinos are Majorana,
and $(n-1)(n-2)/2$ if neutrinos are Dirac,
\textit{cf.}~Ref.~\cite{Bilenky}.

CP violation in the decays
$\pi^{\pm} \to e^{\pm} N_j \to e^{\pm} e^{\pm} \mu^{\mp} \nu$
was investigated in Ref.~\cite{CKZ}, and in the decays
$M^{\pm} \to \ell_1 N_j \to \ell_1 \ell_2 M^{' \mp}$ in Refs.~\cite{CKZ2,DCK}.
In both cases,
it turns out that, even though the rates are extremely small, the CP violation asymmetry in these charged decays may become appreciable, even close to order unity,
when the two intermediate neutrinos can go on shell and are almost degenerate in mass
$M_{N_1} \approx M_{N_2}$.
This is to be contrasted with CP violation in charged meson decays due to the standard CKM
mechanism in the quarks sector, where the rates are larger but the asymmetries are much smaller, e.g., of order $10^{-4}$ in $K^\pm\to 3\pi$ decays \cite{Gamiz:2003pi}.


\subsection{CP Violation in Semileptonic LNV Decays
$M^{\pm} \to \ell_1^{\pm} \ell_2^{\pm} M^{' \mp}$}
\label{sec:BrCPVsub}

As mentioned above, we will consider the scenario with at least two
sterile neutrinos, $N_j$ ($j=1,2$), both of which can go on shell in the intermediate state, \emph{i.e}., with masses $M_{N_j}$ satisfying the condition shown in Equation~(\ref{MNjint}). The
processes of interest are again those of Figure~\ref{FigMMp}, except that
now in both the direct ($D$) and crossed ($C$) channels there are
two possible neutrinos exchanged: $N_1$ or $N_2$.
The relative measure of CP violation for these processes will be
the asymmetry:
\be
{\cal A}_{\rm CP}(M)  \equiv
\frac{\Gamma(M^- \to \ell_1^- \ell_2^- M^{' +}) -\Gamma(M^+ \to \ell_1^+ \ell_2^+ M^{' -})}
{\Gamma(M^- \to \ell_1^- \ell_2^- M^{' +}) + \Gamma(M^+ \to \ell_1^+ \ell_2^+ M^{' -})}
\label{ACPdef}
\ee

The corresponding LNV decay widths
$\Gamma(M^{\pm} \to \ell_1^{\pm} \ell_2^{\pm} M^{' \mp})$ will be obtained
now, in the scenario of two sterile neutrinos (${\cal N}=2$), in
close analogy with the calculation in Section~\ref{sec:BrMsub} which was performed
for the case of one neutrino $N$ (${\cal N}=1$).
The relations (\ref{GM1}) and (\ref{d3})
and (\ref{Ksqr}) there hold without change now, but the relations
(\ref{GM2N})--(\ref{PN}) obtain the following slightly more general form when
${\cal N}=2$:
\bea
\lefteqn{
\Gamma(M^{\pm} \to \ell_1 \ell_2 M^{' \mp})
=
}
\nonumber\\
&&
 (2 - \delta_{\ell_1 \ell_2} )
\sum_{i=1}^2 \sum_{j=1}^2 k_i^{(\pm)} k_j^{(\pm)*}
{\big [}
\G(DD^{*})_{ij} + \G(CC^{*})_{ij}
+ \G_{\pm}(DC^{*})_{ij} + \G_{\pm}(CD^{*})_{ij} {\big ]} \
\label{GM2}
\eea
Here, $k_j^{(\pm)}$ are the mixing coefficients
\be
k_j^{(-)} = B_{\ell_1 N_j} B_{\ell_2 N_j} \ ,\qquad  k_j^{(+)}= (k_j^{(-)})^{*} \
\label{kj}
\ee
and $\G_{\pm}(XY^{*})_{ij}$ are $2 \times 2$ matrices,
and represent the normalized
(\emph{i.e}., without the explicit mixing dependence) contributions
of $N_i$ exchange in the $X$ channel and
complex-conjugate of the $N_j$ exchange
in the $Y$ channel ($X,Y=C, D$)
\be
\G_{\pm}(XY^{*})_{ij} \equiv K^2 \;
\frac{1}{2!} \frac{1}{2 M_M} \frac{1}{(2 \pi)^5}
\int d_3 \; P_i(X) P_j(Y)^{*} M_{N_i} M_{N_j} T_{\pm}(X) T_{\pm}(Y)^{*} \
\label{bGXY}
\ee

The expressions for $T_{\pm}(X) T_{\pm}(Y)^{*}$ (where $X,Y=D,C$) are the same
as in Equation~(\ref{bGXYN})
and are given in Appendix 1,
and $P_j(X)$ ($X=D,C$) are the propagators of the
exchanged neutrinos $N_j$ in the direct and crossed channels
\bes
\label{Pj}
\ba
P_j(D) &=& \frac{1}{\left[ (p_{M}-p_1)^2 - M_{N_j}^2 + i \Gamma_{N_j} M_{N_j} \right]} \
\label{PjD}
\\
P_j(C) &=& \frac{1}{\left[ (p_{M}-p_2)^2 - M_{N_j}^2 + i \Gamma_{N_j} M_{N_j} \right]} \
\label{PjC}
\ea
\ees

We will disregard effects due to non-diagonal neutrino widths in their mass basis. For these details we refer to Ref.~\cite{DCK}.
The total decay width of $N_j$, $\Gamma_{N_j}$, is given by
Equations~(\ref{GNwidth})--(\ref{calK}), where each $N_j$ has its own
mixing parameter $\K_j$ Equation~(\ref{calK}), \emph{i.e}.,
$\K_j = \sum_{\ell} {\cal N}_{\ell N_j} |B_{\ell N_j}|^2$, where the
coefficients ${\cal N}_{\ell N_j}$ as a function of $M_{N_j}$ are given by
the left-hand Figure~\ref{FigcNellN} in Section~\ref{sec:BrMsub}.

As we will see below, the CP asymmetry parameter ${\cal A}_{\rm CP}(M)$,
Equation~(\ref{ACPdef}), may acquire a significant
nonzero value if simultaneously:
(a) the phases $\phi_{\ell j}$ of the PMNS heavy-light mixing elements
\linebreak$B_{\ell N_j} = |B_{\ell N_j}| \exp(i \phi_{\ell j})$
fulfill certain conditions: $|\sin \theta_{21}| \equiv
|\sin (\phi_{\ell_1 2}+ \phi_{\ell_2 2} - \phi_{\ell_1 1}-\phi_{\ell_2 1})|
\not\ll 1$;
and (b) the mass difference $\Delta M_N \equiv M_{N_2} - M_{N_1}$
is sufficiently small ($|\Delta M_N| \not \gg \Gamma_{N_j}$) .

First let us calculate the quantities
\be
S_{\mp}(M) \equiv
\Gamma(M^- \to \ell_1^- \ell_2^- M^{' +}) \mp \Gamma(M^+ \to \ell_1^+ \ell_2^+ M^{' -})
\label{Smpdef}
\ee
appearing in the numerator and the denominator of the
CP violation parameter ${\cal A}_{\rm CP}(M)$ Equation~(\ref{ACPdef}).
We introduce the following notations which will be needed below:
\bes
\label{not}
\ba
\kappa_{\ell_1} & = & \frac{|B_{\ell_1 N_2}|}{|B_{\ell_1 N_1}|} \ ,
\quad
\kappa_{\ell_2} =   \frac{|B_{\ell_2 N_2}|}{|B_{\ell_2 N_1}|} \
\label{kap}
\\
B_{\ell_k N_j} & \equiv & |B_{\ell_k N_j}| e^{i \phi_{\ell_k j}}  \qquad (k,j = 1,2) \
\label{phi}
\\
\theta_{ij} & \equiv & (\phi_{\ell_1 i} + \phi_{\ell_2 i} - \phi_{\ell_1 j} - \phi_{\ell_2 j}) \quad
(i,j=1,2) \
\label{theta}
\ea
\ees

For example, in the specific case when $\ell_1=\ell_2=\mu$, we have
$\theta_{21}=2 (\phi_{\mu 2} - \phi_{\mu 1})= 2 ( {\rm arg}(B_{\mu N_2})
- {\rm arg}(B_{\mu N_1}))$.
As in Section~\ref{sec:BrMsub}, when both $N_j$ are on-shell,
it turns out that the interference contributions to the
quantities $S_{\mp}(M)$ from the
direct ($D$) and crossed ($C$) channels ($DC^*$ and $CD^*$)
are suppressed by several orders of magnitude in comparison with the
contributions from the direct ($DD^*$) and crossed ($CC^*$) channels,
and we will neglect them (we refer to \cite{CKZ2} for details on this point).
Then it follows from the expression
(\ref{GM2})
\bea
\lefteqn{
S_{-}(M) \equiv
\Gamma(M^- \to \ell_1^- \ell_2^- M^{' +}) - \Gamma(M^+ \to \ell_1^+ \ell_2^+ M^{' -})
}
\nonumber\\
&= &
4 (2 - \delta_{\ell_1 \ell_2} ) |B_{\ell_1 N_1}| |B_{\ell_2 N_1}| |B_{\ell_1 N_2}| |B_{\ell_2 N_2}|
\left\{ \sin \theta_{21} \left[
{\rm Im} \G(DD^{*})_{12} + {\rm Im} \G(CC^{*})_{12} \right]
\right\} \
\label{Smi}
\eea
and
\bea
S_{+}(M) & \equiv &
\Gamma(M^- \to \ell_1^- \ell_2^- M^{' +}) + \Gamma(M^+ \to \ell_1^+ \ell_2^+ M^{' -})
\nonumber\\
&= &
2 (2 - \delta_{\ell_1 \ell_2} ) |B_{\ell_1 N_1}|^2 |B_{\ell_2 N_1}|^2
{\bigg \{}
\G(DD^{*})_{11}
\left[ 1 + \kappa_{\ell_1}^2 \kappa_{\ell_2}^2 \frac{\G(DD^{*})_{22}}{\G(DD^{*})_{11}}
+ 2 \kappa_{\ell_1} \kappa_{\ell_2} \cos \theta_{21} \delta_1 \right]
\nonumber\\
&&
+ \G(CC^{*})_{11}
\left[ 1 + \kappa_{\ell_1}^2 \kappa_{\ell_2}^2 \frac{\G(CC^{*})_{22}}{\G(CC^{*})_{11}}
+ 2 \kappa_{\ell_1} \kappa_{\ell_2} \cos \theta_{21} \delta_1 \right]
{\bigg \}}
\label{Spl} \
\eea
In the sum (\ref{Spl}), the coefficient $\delta_1$ represents the
effect of $N_1$-$N_2$ overlap contributions
\be
\delta_j \equiv \frac{{\rm Re} \G(XX^{*})_{12}}{ \G(XX^{*})_{jj}} \ , \quad
(X=D; C; \quad j=1; 2) \
\label{delX}
\ee

We expect $\delta_1 \approx 0$ when $\Delta M_N \gg \Gamma_{N_j}$
(where: $\Delta M_N \equiv M_{N_2} - M_{N_1} > 0$); numerical calculations
(see later) confirm this expectation and show that $\delta_j$ is practically
independent of the channel $X=D,C$.
The normalized decay widths $\G(DD^*)_{jj}$ and $\G(CC^*)_{jj}$
are those of Equations~(\ref{GDDN}) and (\ref{GCCN}) of Section~\ref{sec:BrMsub},
with the substitutions $M_N \mapsto M_{N_j}$,
$y_N \mapsto y_{N_j} \equiv M_{N_j}^2/M_M^2$ [\textit{cf.}~Equation~(\ref{yNs})]
and $\Gamma_N \mapsto \Gamma_{N_j}$ ($= \K_j \bG_N$)
\ba
\G(DD^{*})_{jj} & = &
 \frac{K^2 M_M^5}{128 \pi^2} \frac{M_{N_j}}{\Gamma_{N_j} }
\lambda^{1/2}(1, y_{N_j},y_{\ell_1})
\lambda^{1/2} \left( 1, \frac{y'}{y_{N_j}},\frac{y_{\ell_2}}{y_{N_j}} \right)
Q(y_{N_j}; y_{\ell_1}, y_{\ell_2},y')
\label{GDD}
\ea
where $j=1$ or $j=2$; $\G(CC^{*})_{jj}$ is obtained from $\G(DD^{*})_{jj}$
by the simple exchange  $y_{\ell_1} \leftrightarrow y_{\ell_2}$
[\textit{cf.} Equation~(\ref{GCCN})].

For evaluation of the CP-violating difference $S_{-}(M)$, Equation~(\ref{Smi}),
the quantity ${\rm Im} \G(XX^{*})_{12}$
\linebreak($X=D; C$) is of central importance.
In the integrand of ${\rm Im} \G(XX^{*})_{12}$
appears as a factor the following combination
 of the propagators of $N_1$ and $N_2$ [\textit{cf.}~Equation~(\ref{bGXY})]:
\bes
\label{ImP1P2gen}
\ba
{\rm Im} \left( P_1(D) P_2(D)^{*} \right)
&= &
\frac{
\left( p_N^2 - M_{N_1}^2 \right)  \Gamma_{N_2} M_{N_2}
- \Gamma_{N_1} M_{N_1} \left( p_N^2 - M_{N_2}^2 \right)
}
{
\left[ \left( p_N^2 - M_{N_1}^2 \right)^2 + \Gamma_{N_1}^2 M_{N_1}^2
\right]
\left[ \left( p_N^2 - M_{N_2}^2 \right)^2 + \Gamma_{N_2}^2 M_{N_2}^2
\right]
}
\label{ImP1P2ex}
\\
& = &
\eta \times \frac{\pi}{M^{2}_{N_2}-M^{2}_{N_1}} \left [
\delta  ( p^{2}_{N}-M^{2}_{N_2})+ \delta  ( p^{2}_{N}-M^{2}_{N_1})  \right ] \
\label{ImP1P2eta}
\ea
\ees
In Equation~(\ref{ImP1P2eta}) we used the narrow width approximation:
\be
\frac{\Gamma_{N_j} M_{N_j}}
{\left[ \left( p_N^2 - M_{N_j}^2 \right)^2 + \Gamma_{N_j}^2 M_{N_j}^2
\right]} = \pi \delta(p_N^2 - M_{N_j}^2) \quad
({\rm for} \; \Gamma_{N_j} \ll M_{N_j})
\label{deltas}
\ee
and in Equation~(\ref{ImP1P2eta}) the parameter $\eta$ was introduced
which parametrizes
any deviation from the naive expectation $\eta =1$. We expect
$\eta \approx 1$ when $\Delta M_N^2 \gg \Gamma_{N_1}, \Gamma_{N_2}$,
where $\Delta M_N^2 \equiv M_{N_2}^2 - M_{N_1}^2 > 0$.
In Appendix 6 we
argue that this parameter $\eta$, in the case of near
degeneracy $\Delta M_N \ll M_{N_1}$, is a simple
function of only one variable $y \equiv \Delta M/\Gamma_N$,
where $\Delta M_N \equiv M_{N_2}-M_{N_1} > 0$ and
\linebreak$\Gamma_N \equiv (1/2) (\Gamma_{N_1} + \Gamma_{N_2})$
\be
\eta(y) =  \frac{y^2}{(y^2+1)}{\Big |}_{y=\Delta M_N/\Gamma_N} \
\label{etay}
\ee
when $\Delta M_N \ll  M_{N_1}\equiv M_N$ ($\Rightarrow$ $\Delta M_N^2 =
2 M_N \Delta M_N$), and where
\be
\Gamma_N  \equiv  \frac{1}{2} (\Gamma_{N_1} + \Gamma_{N_2}),
\quad y \equiv \frac{\Delta M_N}{\Gamma_N}
\label{GNy}
\ee

This implies that in the case of two  almost degenerate
sterile neutrinos ($\Delta M_N \ll M_{N_1}$) we have for
the factor in Equation~(\ref{ImP1P2eta}) the following identities:
\be
 \eta \times \frac{1}{\Delta M_N^2}  =
\frac{1}{2 M_N \Gamma_N} \frac{\eta(y)}{y}
=\frac{ \Delta M_N^2 }{ (\Delta M_N^2)^2 + 4 M_N^2 \Gamma_N^2}
\label{etadM2}
\ee

We note that the factor $4$ in the denominator
on the right-hand side of Equation~(\ref{etadM2})
is nontrivial, because a somewhat different result would have been obtained
by a more simple and direct consideration of the expression (\ref{ImP1P2ex})
in the limit $\Gamma_{N_j} \ll M_{N_j}$ ($j=1,2$). The result
(\ref{etay}) [or equivalently, Equation~(\ref{etadM2})]
has been confirmed also by numerical evaluation of
${\rm Im} \G(XX^{*})_{12}$, Refs.~\cite{CKZ2,CKZ} (see below).
The mechanism (\ref{ImP1P2gen}) [with the
identity (\ref{etadM2})] is of central importance for the
CP violation in the processes considered here.
The quantity $\eta/\Delta M_N^2$, at fixed $\Gamma_N$ and
fixed $M_N\equiv M_{N_1}$,
achieves its maximum when $\Delta M_N^2 = 2 M_N \Gamma_N$,
\emph{i.e}., $y=1$ [$\Leftrightarrow$ $\Delta M_N = \Gamma_N$ ($\ll M_N$)],
\emph{i.e}., when the two sterile neutrinos are almost degenerate.
If $\Delta M_N \not \sim \Gamma_N$ (\emph{i.e}., $y \not \sim 1$), then the quantity
$\eta(y)/y = y/(y^2+1)$ in Equation~(\ref{etadM2})
is very small and CP violation effects disappear.

The mechanism (\ref{ImP1P2gen}) was used in Ref.~\cite{CKZ} in the
context of CP violation of leptonic decays of charged pions,
and in Refs.~\cite{CKZ2,DCK} in the here presented context of
CP violation of semileptonic LNV decays of heavy pseudoscalars.

We recall that the expression (\ref{ImP1P2gen}) has the
same structure with Dirac delta functions as Equation~(\ref{P1P1})
in Section~\ref{sec:BrMsub}; however,
the factors in front of these Dirac delta
functions are different now. Therefore, integration over the
final particles' phase space can be performed now
in the same way as in Section~\ref{sec:BrMsub}, \emph{i.e}., analytically.
This leads, in analogy with Equation~(\ref{GDDN}), to the result
\bes
\label{ImG12}
\ba
\lefteqn{
{\rm Im} \G(DD^{*})_{12} =
 \eta(y) \times \frac{1}{\Delta M_N^2} \;
\frac{K^2 M_M^5}{128 \pi^2}
M_{N_1} M_{N_2}
}
\nonumber\\
&& \times
\sum_{j=1}^2
\lambda^{1/2}(1, y_{N_j},y_{\ell_1}) \,
\lambda^{1/2} \left( 1, \frac{y'}{y_{N_j}},\frac{y_{\ell_2}}{y_{N_j}} \right)
Q(y_{N_j}; y_{\ell_1}, y_{\ell_2},y') \
\label{ImGDD12}
\\
{\rm Im} \G(CC^{*})_{12} & = &
{\rm Im} \G(DD^{*})_{12}(y_{\ell_1} \leftrightarrow y_{\ell_2}) \
\label{ImGCC12}
\ea
\ees
where we recall the notations Equations~(\ref{notGDDN}) and (\ref{GNy}),
$y_{N_j} \equiv M_{N_j}^2/M_M^2$, the function $Q$
is presented in Appendix 2,
and we denoted $\Delta M_N \equiv M_{N_2} - M_{N_1} > 0$ and
$\Delta M_N^2 \equiv M_{N_2}^2 - M_{N_1}^2 > 0$.
We note that in Equation~(\ref{ImG12}) we have not yet assumed
the near degeneracy of the two sterile neutrinos.

From here on in this Section, we will consider the case of near degeneracy
of the two on-shell sterile neutrinos ($\Delta M_N \ll M_N$,
where $M_N \equiv M_{N_1}$),
in which case Equations~(\ref{etay}) and (\ref{GNy}) hold and the quantities
Equations~(\ref{etadM2}) and (\ref{ImG12}) become appreciable
and CP violation can thus become significant. Therefore,
we have
\be
y_{N_2} \approx y_{N_1} \equiv y_N \equiv \frac{M_N^2}{M_M^2}
\label{notyN}
\ee
where $M_N \equiv M_{N_1} \approx M_{N_2}$. In this case
the identity (\ref{etay}) holds, \textit{cf.}~Appendix 6, and the
expression (\ref{ImGDD12}) becomes simpler
\bes
\label{ImGDD12deg}
\bea
{\rm Im} \G(DD^{*})_{12} &=&
 \frac{\eta(y)}{y} \times
\frac{K^2 M_M^5 M_N}{128 \pi^2 \Gamma_N}
\lambda^{1/2}(1, y_{N},y_{\ell_1}) \,
\lambda^{1/2} \left( 1, \frac{y'}{y_{N}},\frac{y_{\ell_2}}{y_{N}} \right)
Q(y_{N}; y_{\ell_1}, y_{\ell_2},y')
\label{ImGDD12dega}
\\
& = &
\frac{\eta(y)}{y} \G(DD^{*})_{11} \frac{2 \K_1}{(\K_1+\K_2)}
\label{ImGDD12degb}
\eea
\ees
where in the last identity we used the expression (\ref{GDDN}),
the identity (\ref{GNy}),
and the fact that \linebreak$\Gamma_{N_2}/\Gamma_{N_1} =
\K_2/\K_1$ [\textit{cf.}~Equation~(\ref{calK}) for $N_1$ and $N_2$].

The normalized decay matrix elements $\G_{\pm}(XY^{*})_{ij}$, Equation~(\ref{bGXY}),
were evaluated in Ref.~\cite{CKZ2}
also numerically, by Monte Carlo integration and
using finite small widths $\Gamma_{N_j}$ in the propagators.
The numerical calculations confirmed the presented formulas,
among them the expressions (\ref{GDD}), (\ref{ImGDD12deg}),
and (\ref{etay})--(\ref{etadM2}).
The form (\ref{etay}) of $\eta(y)$ was confirmed
numerically with a precision better than a few per mille.
The numerical evaluations also confirmed that the
direct-crossed interference terms ($DC^*$ and $CD^*$)
are really negligible. We refer to \cite{CKZ2} for details.

Further, the mentioned numerical evaluations gave us values of the
$N_1$-$N_2$ overlap parameter $\delta_1$ as defined in Equation~(\ref{delX}),
\emph{i.e}., the parameter which appears in the expression (\ref{Spl}) and represents
the $N_1$-$N_2$ overlap effects.
It turned out that the numerical values of the
parameters $\delta_j$ ($j=1,2$), as well as of $\eta$, are
practically independent of: the channel contribution considered
($DD^{*}$ or $CC^{*}$), of the type of pseudoscalar mesons
($M^{\pm}$, $M^{' \mp}$), and of the light leptons
($\ell_1, \ell_2 = e, \mu$) involved in the considered decays.
The numerical results show that the parameter $\delta \equiv
(1/2) (\delta_1 + \delta_2)$ is a function of only one variable,
namely $y \equiv \Delta M_N/\Gamma_N$ (the same is true for $\eta$)
\bes
\label{del}
\ba
\delta &=&\delta(y) \ , \quad
\delta \equiv  \frac{1}{2} (\delta_1 + \delta_2) \
\label{dela}
\\
\frac{\delta_1}{\delta_2} &=& \frac{\G(DD^*)_{22}}{\G(DD^*)_{11}} =
\frac{\Gamma_{N_1}}{\Gamma_{N_2}} =
\frac{\K_1}{\K_2}
\label{delb}
\ea
\ees

\begin{figure}[htb] 
\begin{minipage}[b]{.49\linewidth}
\centering\includegraphics[width=85mm]{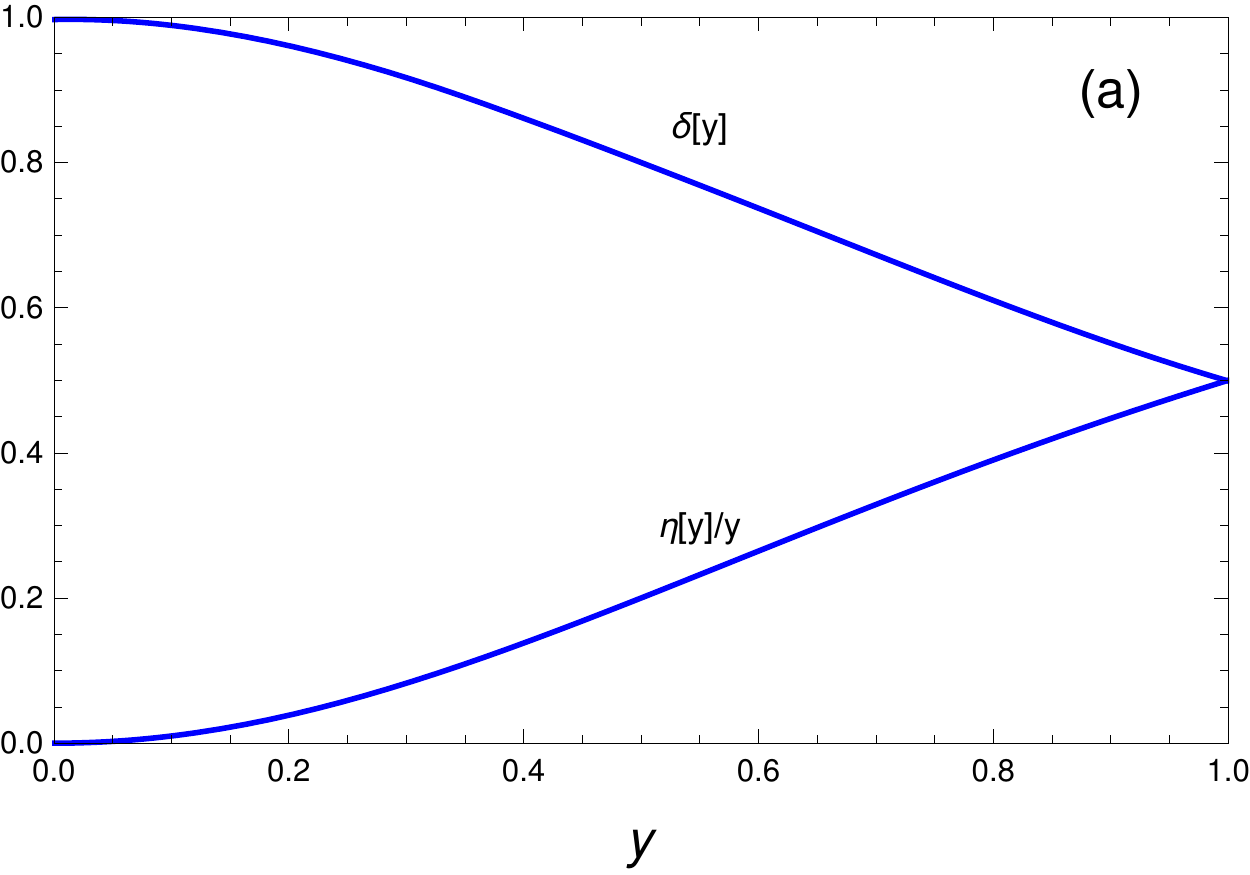}
\end{minipage}
\begin{minipage}[b]{.49\linewidth}
\centering\includegraphics[width=85mm]{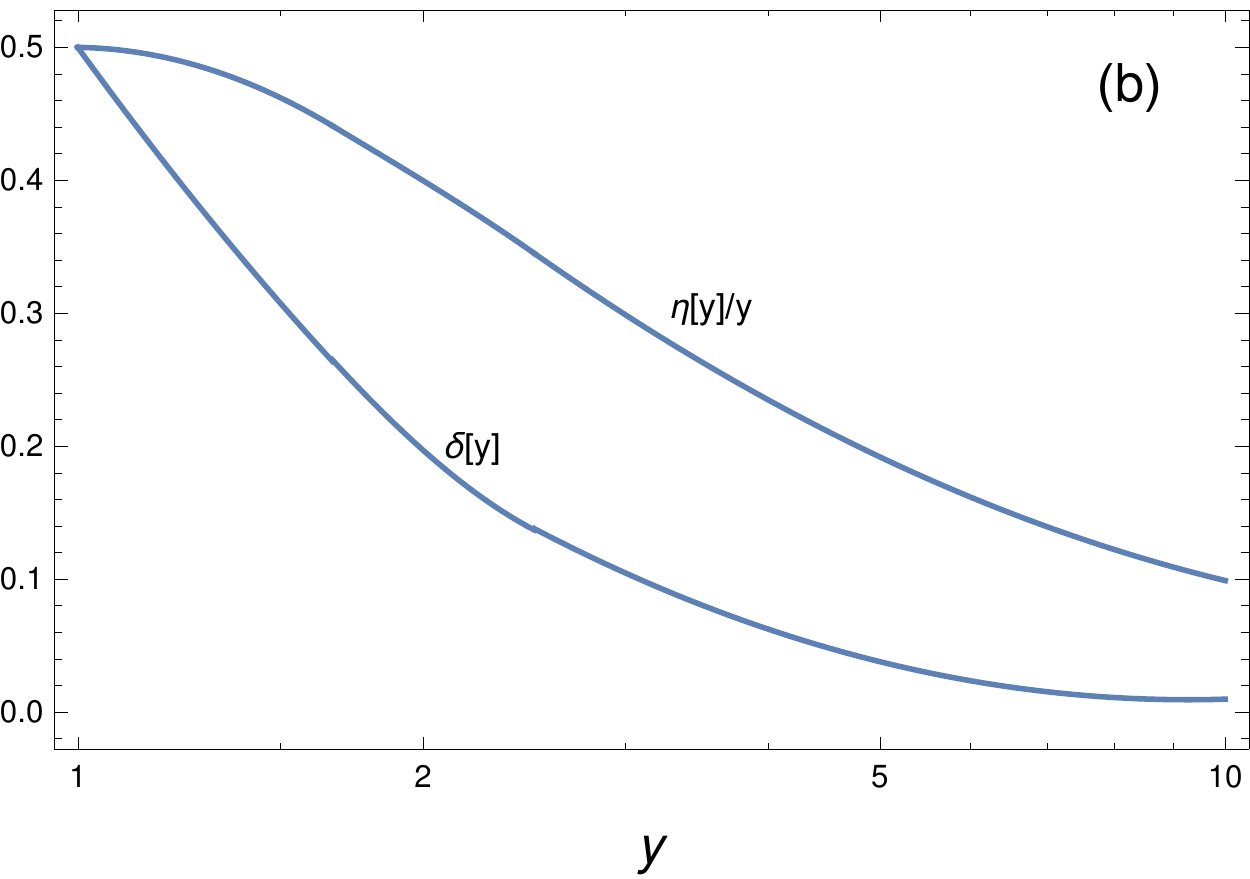}
\end{minipage}
\vspace{6pt}
\caption{The suppression factors $\eta(y)/y$ and $\delta(y)$
as a function of $y \equiv \Delta M_N/\Gamma_N$:
(\textbf{a}) for $y < 1$ (on the linear $y$ scale);
(\textbf{b}) for $1 < y < 10$ (on the logarithmic $y$ scale).}
\label{etadelfig}
\end{figure}

The numerical values of the parameter $\delta$ are given in
Table \ref{T4} as a function of $y$. 
\begin{table} \centering
\caption{Values of the$N_1$-$N_2$ overlap parameter $\delta(y)$
as a function of $y \equiv  \Delta M_N/\Gamma_N$.}
\label{T4}
\begin{tabular}{lr|l}
\hline
$y \equiv \frac{\Delta M_N}{\Gamma_N}$ & $\log_{10} y$ &
$\delta(y)$
\\
\hline
0.10 & $-$1.000 & $0.989 \pm 0.001$
\\
0.30 & $-$0.523 & $0.917 \pm 0.001$
\\
0.50 & $-$0.301 & $0.800 \pm 0.001$
\\
0.70 & $-$0.155 & $0.673 \pm 0.001$
\\
0.80 & $-$0.097 & $0.610 \pm 0.001$
\\
0.90 & $-$0.046 & $0.551 \pm 0.002$
\\
1.00 & 0.000 & $0.499 \pm 0.002$
\\
1.25 & 0.097 & $0.390 \pm 0.003$
\\
1.67 & 0.222 & $0.264 \pm 0.003$
\\
2.50 & 0.398 & $0.138 \pm 0.001$
\\
5.00 & 0.699 & $0.038 \pm 0.001$
\\
10.0 & 1.000 & $0.0098 \pm 0.0010$
\\
\hline
\end{tabular}
\end{table}
It is not clear whether
there exists a simple analytic expression for $\delta$ as
a function of $y$. Further, in Figure~\ref{etadelfig}
we present the quantities $\eta/y$ and $\delta$ as a function
of $y$.
The values for $\delta$ in Table \ref{T4} are practically equal
to the values of the corresponding $\delta$ parameter
in the rare leptonic decays of the
charged pions $\pi^{\pm} \to e^{\pm} N \to e^{\pm} e^{\pm} \mu^{\mp} \nu$,
\textit{cf.}~next Section~\ref{sec:BrCPVsub} and Ref.~\cite{CKZ}.

Branching ratios of experimental significance here can be defined by
dividing the expressions $S_{\mp}(M)$, Equations~(\ref{Smpdef})
and (\ref{Smi}) and (\ref{Spl}), by the corresponding sum of
total decay widths $[ \Gamma(M^+ \to {\rm all}) +
\Gamma(M^- \to {\rm all}) ]$, which is practically equal to
$2 \Gamma(M^{\pm} \to {\rm all})$
\bes
\label{Brplmidef}
\bea
{\rm Br}_{+}(M) & \equiv &
\frac{S_{+}(M)}{2 \Gamma(M^{\pm} \to {\rm all})} \
\label{Brpldef}
\\
{\rm Br}_{-}(M) &\equiv& {\cal A}_{\rm CP}(M) {\rm Br}_{+}(M) =
\frac{S_{-}(M)}{2 \Gamma(M^{\pm} \to {\rm all})}
\label{Brmidef}
\eea
\ees

If employing the canonical (independent of mixing) branching ratio
${\overline {\rm Br}}(y_N; y_{\ell_1}, y_{\ell_2},y') \equiv
{\overline {\rm Br}}(DD^*)$
defined in Equation~(\ref{bBrM}) of Section~\ref{sec:BrMsub},
and its $CC^*$ analog,
{${\overline {\rm Br}}(CC^*) \equiv
{\overline {\rm Br}}(y_N; y_{\ell_2}, y_{\ell_1},y')$,}
%
then it turns out that the branching ratios ${\rm Br}_{\pm}(M)$
of Equation~(\ref{Brplmidef}) can be rewritten in terms
of  ${\overline {\rm Br}}$, of the heavy-light mixing
parameters $|B_{\ell N_j}|$ and $\K_j = \sum_{\ell} {\cal N}_{\ell N_j}
|B_{\ell N_j}|^2$, of function $\eta(y)/y = y/(y^2+1)$
and the overlap function $\delta(y)$ tabulated in Table \ref{T4}
\bes
\label{Brplmi}
\bea
\lefteqn{
{\rm Br}_{+}(M) \equiv
\frac{\Gamma(M^- \to \ell_1^- \ell_2^- M^{' +}) + \Gamma(M^+ \to \ell_1^+ \ell_2^+ M^{' -})}{2 \Gamma(M^{\pm} \to {\rm all})}
=  2 (2 - \delta_{\ell_1 \ell_2})
{\Bigg [} \sum_{j=1}^2
\frac{|B_{\ell_1 N_j}|^2 |B_{\ell_2 N_j}|^2}{\K_j}
}
\nonumber\\
&&
+ 4 \delta(y)
\frac{|B_{\ell_1 N_1}||B_{\ell_2 N_1}||B_{\ell_1 N_2}||B_{\ell_2 N_2}|}{(\K_1+\K_2)}
\cos \theta_{21} {\Bigg ]}
\left( {\overline {\rm Br}}(DD^*) +{\overline {\rm Br}}(CC^*) \right)
\label{Brpl}
\\
\lefteqn{
{\rm Br}_{-}(M) \equiv {\cal A}_{\rm CP}(M) {\rm Br}_{+}(M) \equiv
 \frac{\Gamma(M^- \to \ell_1^- \ell_2^- M^{' +}) - \Gamma(M^+ \to \ell_1^+ \ell_2^+ M^{' -})}{2 \Gamma(M^{\pm} \to {\rm all})}
}
\nonumber\\
&=&  8 (2 - \delta_{\ell_1 \ell_2})
\frac{|B_{\ell_1 N_1}||B_{\ell_2 N_1}||B_{\ell_1 N_2}||B_{\ell_2 N_2}|}{(\K_1+\K_2)}
\; \sin \theta_{21}
\; \frac{\eta(y)}{y}
\left( {\overline {\rm Br}}(DD^*) +{\overline {\rm Br}}(CC^*) \right)
\label{Brmi}
\eea
\ees

This leads to an expression for the CP violation parameter
${\cal A}_{\rm CP}(M)$ defined in Equation~(\ref{ACPdef}),
which now involves only the heavy-light mixing parameters
$|B_{\ell N_j}|$ and $\K_j$ [\textit{cf.}~Equation~(\ref{calK})],
the function $\eta(y)/y = y/(y^2+1)$,
{
where $y \equiv \Delta M_N/\Gamma_N$,
and the overlap function $\delta(y)$ tabulated in Table \ref{T4}:
}
%
\bes
\label{ACP}
\bea
\lefteqn{
{\cal A}_{\rm CP}(M) \equiv  \frac{{\rm Br}_{-}(M)}{{\rm Br}_{+}(M)}
\equiv
\frac{ \Gamma(M^- \to \ell_1^- \ell_2^- M^{' +}) - \Gamma(M^+ \to \ell_1^+ \ell_2^+ M^{' -})}
{ \Gamma(M^- \to \ell_1^- \ell_2^- M^{' +}) + \Gamma(M^+ \to \ell_1^+ \ell_2^+ M^{' -})}
}
\nonumber\\
&=& \frac{ 4 \sin \theta_{21} }
{
\left[
\sum_{j=1}^2
\frac{|B_{\ell_1 N_j}|^2 |B_{\ell_2 N_j}|^2}
{|B_{\ell_1 N_1}||B_{\ell_2 N_1}||B_{\ell_1 N_2}||B_{\ell_2 N_2}|}
\frac{(\K_1+\K_2)}{\K_j} + 4 \delta(y) \cos \theta_{21}
\right]
}
\;  \frac{y}{(y^2+1)}
\label{ACPa}
\\
&=& \frac{ 4\sin \theta_{21} }{
\left\{
 \left[
\kappa_{\ell_1} \kappa_{\ell_2} \left(1 + \frac{\K_1}{\K_2} \right)
+ \frac{1}{\kappa_{\ell_1} \kappa_{\ell_2}} \left(1 + \frac{\K_2}{\K_1} \right)
\right]  +  4\delta(y) \cos \theta_{21}
\right\}
}
\;  \frac{y}{(y^2+1)} \
\label{ACPb}
\eea
\ees

In the usually considered case $\ell_1=\ell_2$ ($\equiv \ell$),
\emph{i.e}., when the considered decays are $M^{\pm} \to \ell^{\pm} \ell^{\pm} M^{' \mp}$,
the formulas (\ref{Brplmi}) and (\ref{ACP}) get simpler,
because in such a case ${\overline {\rm Br}}(CC^*) =
{\overline {\rm Br}}(DD^*) \equiv
{\overline {\rm Br}}$, and \linebreak$B_{\ell_2 N_j} = B_{\ell_1 N_j} \equiv B_{\ell N_j}$,
and $\kappa_{\ell_2} = \kappa_{\ell_1} = \kappa_{\ell}$
\bes
\label{Brplmiell}
\bea
{\rm Br}_{+}(M) &=&  4
\left[ \sum_{j=1}^2
\frac{|B_{\ell N_j}|^4}{\K_j}
+ 4 \delta(y)
\frac{|B_{\ell N_1}|^2 |B_{\ell N_2}|^2}{(\K_1+\K_2)}
\cos \theta_{21} \right]
{\overline {\rm Br}} \
\label{Brplell}
\\
{\rm Br}_{-}(M) &\equiv& {\cal A}_{\rm CP}(M) {\rm Br}_{+}(M)
=  16 \frac{ |B_{\ell N_1}|^2 |B_{\ell N_2}|^2}{(\K_1+\K_2)}
\; \sin \theta_{21}
\; \frac{\eta(y)}{y}
{\overline {\rm Br}} \
\label{Brmiell}
\\
{\cal A}_{\rm CP}(M)
&=& \frac{ 4 \sin \theta_{21} }{
\left\{
 \left[
\kappa_{\ell}^2 \left(1 + \frac{\K_1}{\K_2} \right)
+ \frac{1}{\kappa_{\ell}^2} \left(1 + \frac{\K_2}{\K_1} \right)
\right]  + 4 \delta(y) \cos \theta_{21}
\right\}
}
\;  \frac{y}{(y^2+ 1)} \
\label{ACPell}
\eea
\ees

These formulas become even simpler if the absolute values
of the heavy-light mixings of $N_1$ and $N_2$ are equal (but not their phases),
\emph{i.e}., when
\be
|B_{\ell^{'} N_2}|^2 \approx |B_{\ell^{'} N_1}|^2 \equiv |B_{\ell^{'} N}|^2
\qquad (\ell^{'}=e, \mu, \tau)
\label{BN1N2}
\ee
In such a case, we have all $\kappa_{\ell^{'}} \approx 1$,
and $\K_2 \approx \K_1
\equiv \K$,
and therefore the expressions (\ref{Brplmiell}) reduce to
\bes
\label{BrplmiellN1N2}
\bea
{\rm Br}_{+}(M) & \approx &  8 \frac{|B_{\ell N}|^4}{\K}
{\overline {\rm Br}} \left( 1 + {\cal O}(\delta) \right) \
\label{BrplellN1N2}
\\
{\rm Br}_{-}(M) &\equiv& {\cal A}_{\rm CP}(M) {\rm Br}_{+}(M)
 \approx   8 \frac{ |B_{\ell N}|^4}{\K}
\; \sin \theta_{21}
\; \frac{\eta(y)}{y}
{\overline {\rm Br}} \
\label{BrmiellN1N2}
\\
{\cal A}_{\rm CP}(M)
& \approx & \sin \theta_{21} \frac{y}{y^2+1}
 \left( 1 + {\cal O}(\delta) \right)
\leq \frac{1}{2} \sin \theta_{21}  \left( 1 + {\cal O}(\delta) \right)
\label{ACPellN1N2}
\eea
\ees
In these expressions, we assumed, in addition, that the
$N_1$-$N_2$ overlap terms are small (${\cal O}(\delta)$).

In order to obtain the corresponding effective (true) branching ratios,
we have to multiply the above branching ratios ${\rm Br}_{\pm}$
by the decay-within-the-detector probability
$P_{N_j} \equiv {\overline P}_N \K_j (L/1 {\rm m})$,
as in Section~\ref{sec:effBrMsub} ($L$ is the length of the detector).
Again, the complicated mixing dependence entailed in the
parameters $\K_j$ gets cancelled in this multiplication.
When adopting the simplifying assumption Equation~(\ref{BN1N2}),
\emph{i.e}., the validity of Equations~(\ref{BrplmiellN1N2}), we obtain
the following effective branching ratio ${\rm Br}_{\rm eff}$
and the CP violation effective branching ratio
${\cal A}_{\rm CP} {\rm Br}_{\rm eff}$:
\bes
\label{PNBr}
\bea
{\rm Br}_{\rm eff}(M^{\pm} \to \ell^{\pm} \ell^{\pm} M^{' \mp})
&=&
P_N {\rm Br}_{+}(M) \approx
\left[ {\overline P}_N \left( \frac{L}{1 {\rm m}} \right) \K \right]
\left[ 8 \frac{|B_{\ell N}|^4}{\K} {\overline {\rm Br}} \right]
\nonumber\\
& = &  \left( \frac{L}{1 {\rm m}} \right)
8   |B_{\ell N}|^4  {\overline P}_N {\overline {\rm Br}} \equiv
 2 |B_{\ell N}|^4
\left( \frac{L}{1 {\rm m}} \right) {\overline {\rm Br}}_{\rm eff}
\label{PNBrpl}
\\
{\cal A}_{\rm CP} {\rm Br}_{\rm eff}(M^{\pm} \to \ell^{\pm} \ell^{\pm} M^{' \mp})
& = &  \sin \theta_{21} \frac{y}{y^2+1}
\times
 2 |B_{\ell N}|^4
\left( \frac{L}{1 {\rm m}} \right) {\overline {\rm Br}}_{\rm eff}
\nonumber\\
&\lesssim&
|B_{\ell N}|^4 \left( \frac{L}{1 {\rm m}} \right) \sin \theta_{21}
{\overline {\rm Br}}_{\rm eff}
\label{PNBrmi}
\eea
\ees

In these formulas, we used the canonical effective branching ratio
$ {\overline {\rm Br}}_{\rm eff}$ as defined via Equations~(\ref{bBreff})
and (\ref{bBrM}) and depicted in Figures~\ref{brKfig}--\ref{brBcfig}
%
%
as a function of $M_N$ ($ \equiv M_{N_1} \approx M_{N_2}$).
{
The values of the Lorentz factors in the lab system are taken to be $\gamma_N = 2$ for both $N_1$ and $N_2$, keeping in mind that
${\overline {\rm Br}}_{\rm eff}$
scales as $1/\gamma_N$.
}
We recall that
$\theta_{21}= 2 (\phi_{\ell 2} - \phi_{\ell 1}) = 2 ({\rm arg}(B_{\ell N_2})
- {\rm arg}(B_{\ell N_1}))$. We notice that on the right-hand side of
Equation~(\ref{PNBrpl}) there is an additional factor two in comparison with
Equation~(\ref{Breffb}) of Section~\ref{sec:effBrMsub}; this factor two
comes from the fact that we now have contributions of two intermediate
neutrinos $N_1$ and $N_2$, and we neglected the contributions from
the $N_1$-$N_2$ overlap (${\cal O}(\delta)$).

As at the end of Section~\ref{sec:effBrMsub},
let us consider now as an illustrative example
the decays $D_s^{\pm} \to \mu^{\pm} \mu^{\pm} \pi^{\mp}$.
In addition, let us assume that
$|B_{\mu N}|^2$ is the dominant mixing.
In such a case, the estimate Equation~(\ref{BreffDs})
is still valid, and the CP-violating difference of the
effective branching ratios, ${\cal A}_{\rm CP} {\rm Br}_{\rm eff}$,
is obtained by comparison of Equations~(\ref{PNBrpl}) and (\ref{PNBrmi})
\be
{\cal A}_{\rm CP} {\rm Br}_{\rm eff}(D^{\pm}_s \to  \mu^{\pm} \mu^{\pm} \pi^{\mp})
\sim 10^2 |B_{\mu N}|^4 \sin \theta_{21}
\frac{y}{y^2+1}{\Big |}_{y \equiv \Delta M_N/\Gamma_N}
\lesssim  10^2 |B_{\mu N}|^4 \sin \theta_{21}
\label{BrCPeffDs}
\ee

Since for $M_N \approx 1$ GeV we have at present
$|B_{\mu N}|^2 \lesssim 10^{-7}$,
\textit{cf.}~Table \ref{T2},
Equation~(\ref{BrCPeffDs}) means that
${\cal A}_{\rm CP} {\rm Br}^{\rm (eff)} \lesssim 10^{-12}$
for such decays.
As already mentioned at the end of Section~\ref{sec:effBrMsub},
the proposed CERN-SPS experiment \cite{CERN-SPS,CERN-SPS2} could
produce $D$ and $D_s$ mesons in numbers
by several orders of magnitude higher than $10^{12}$,
and production of the sterile Majorana neutrinos $N_j$
could be explored. Further, if there exist two
almost degenerate sterile neutrinos of mass $M_N \sim 1$ GeV
(this is so in the  $\nu$MSM model \cite{nuMSM,nuMSM2,Shapo,Shapo2,Shapo3,Shapo4,Shapo5,Shapo6}),
such that $y \equiv \Delta M_N/\Gamma_N
\sim 1$, then we would have $\eta(y)/y \equiv y/(y^2+1) \sim 1$.
In such a case the estimate (\ref{BrCPeffDs}) would imply that the CP-violating
difference of effective branching ratios,
$ {\cal A}_{\rm CP} {\rm Br}_{\rm (eff)}(D_s)$,
is of the same order as the effective branching ratio
 ${\rm Br}_{\rm (eff)}(D_s)$
(if the phase difference $|\theta_{21}| \not\ll 1$).
Therefore, if experiments can discover the mentioned
$\nu$MSM-type Majorana neutrinos, they will possibly detect
also CP violation effects coming from the Majorana neutrinos.

The case of $B_c^{\pm} \to \mu^{\pm} \mu^{\pm} \pi^{\mp}$ is similar to
the case of $D_s^{\pm} \to \mu^{\pm} \mu^{\pm} \pi^{\mp}$ described above,
\textit{cf.}~Equations~(\ref{BreffDs}) and (\ref{BreffBc}) at the end of
Section~\ref{sec:effBrMsub}. Therefore, Equation~(\ref{BrCPeffDs}) is valid also for
CP violation in such decays of $B_c^{\pm}$. For the relative advantages and
disadvantages of $D_s^{\pm}$ and $B_c^{\pm}$ decays, we refer to
the comments at the end of Section~\ref{sec:effBrMsub}.
%


\subsection{CP Violation in Pion Decays
$\pi^{\pm} \to e^{\pm} e^{\pm} \mu^{\mp} \nu$}
\label{sec:BrCPVPisub}

In this Section, we will only briefly outline the calculation of the
CP violation asymmetry in the (LNC and LNV) semileptonic decays
$\pi^{\pm} \to e^{\pm} e^{\pm}  \mu^{\mp} \nu$ as described in Section~\ref{sec:BrPi}.
We will assume the presence of at least two nearly degenerate
sterile neutrinos $N_j$ ($j=1,2$) that can go on shell in the intermediate state, as
in Section~\ref{sec:BrCPVsub}.
The present Section is a similar extension of the analysis of
the decays $\pi^{\pm} \to e^{\pm} e^{\pm} \nu$ of Section~\ref{sec:BrPi}
to two sterile neutrinos. The results of the present Section are
largely based on Ref.~\cite{CKZ}. Only few details will
be presented here, for further details we refer to Ref.~\cite{CKZ}.

Similarly to the previous Section~\ref{sec:BrCPVsub}, the quantities relevant for
the CP violation will be
\bes
\label{defs}
\ba
{\rm Br}_{\pi,\pm}^{\rm (X)} &=&
\frac{S^{\rm (X)}_{\pm}(\pi)}{2 \Gamma(\pi^{\pm} \to {\rm all})} \equiv
\frac{ \Gamma^{(\rm X)}(\pi^- \to e^- e^- \mu^+ \nu) \pm
\Gamma^{\rm (X)}(\pi^+ \to e^+ e^+ \mu^- \nu)}{2 \Gamma(\pi^{\pm} \to {\rm all})} \
\label{Brdef}
\\
{\cal A}^{\rm (X)}_{\pi,\rm CP} &=& \frac{{\rm Br}_{\pi,-}^{\rm (X)}}{{\rm Br}_{\pi,+}^{\rm (X)}}=
 \equiv \frac{ \Gamma^{\rm (X)}(\pi^- \to e^- e^- \mu^+ \nu) - \Gamma^{\rm (X)}(\pi^+ \to e^+ e^+ \mu^- \nu)}{ \Gamma^{\rm (X)}(\pi^- \to e^- e^- \mu^+ \nu) + \Gamma^{\rm (X)}(\pi^+ \to e^+ e^+ \mu^- \nu)}
\label{Adef}
\ea
\ees
where X = LNC, LNV. The total branching ratios are
${\rm Br}_{\pm} ={\rm Br}_{\pm}^{\rm (LNV)}+{\rm Br}_{\pm}^{\rm (LNC)}$
when $N_j$ are Majorana neutrinos, and ${\rm Br}_{\pm} ={\rm Br}_{\pm}^{\rm (LNC)}$
when $N_j$ are Dirac neutrinos.
We adopt the same conventions and the same notations as
in the previous Section~\ref{sec:BrCPVsub}.
In addition, since we have
now LNV and LNC processes, we introduce the additional notations
\bes
\label{thPi}
\bea
\theta^{\rm (LNV)} & = & 2 (\phi_{e 2} - \phi_{e 1})
\label{thPiLNV}
\\
\theta^{\rm (LNC)} &=&  (\phi_{e 2} - \phi_{e 1}) -
 (\phi_{\mu 2} - \phi_{\mu 1})
\label{thPiLNC}
\ea
\ees

As in Section~\ref{sec:BrCPVsub}, the requirement that the
quantities $\sin \theta^{\rm (X)}$ (here: X = LNV, LNC) be nonzero,
and the requirement of the near degeneracy
of the two neutrinos ($\Delta M_N \ll M_{N_1} \equiv M_N$)
in conjunction with
the expressions Equations~(\ref{ImP1P2gen})--(\ref{etadM2})
for ${\rm Im}(P_1(D) P_2(D)^*)$,
are needed in order that the
CP violation parameters
${\cal A}^{\rm (X)}_{\pi,\rm CP} \not= 0$
acquire nonnegligible values.
Analysis similar to that of the previous Section~\ref{sec:BrCPVsub}
(but algebraically more complicated) leads then to the
results for the quantities defined in Equations~(\ref{defs}).
More specifically, the results for the Dirac case,
${\rm Br}_{\pi,+}^{\rm (Dir.)} \equiv {\rm Br}_{\pi,+}^{\rm (LNC)}$ and
${\cal A}^{\rm (Dir.)}_{\pi,\rm CP} \equiv {\cal A}^{\rm (LNC)}_{\pi,\rm CP}$,
are the~following:
\bes
\label{BrPiDir}
\bea
{\rm Br}_{\pi,+}^{\rm (Dir.)} & \equiv &
\frac{ \Gamma^{\rm (LNC)}(\pi^- \to e^- e^- \mu^+ \nu) +
\Gamma^{\rm (LNC)}(\pi^+ \to e^+ e^+ \mu^- \nu)}{2 \Gamma(\pi^{\pm} \to {\rm all})}
\nonumber\\
&=&  {\bigg [} \sum_{j=1}^2 \frac{|B_{e N_j}|^2 |B_{\mu N_j}|^2}{\K_j}
+ 4 \delta(y) \frac{|B_{e N_1}||B_{e N_2}||B_{\mu N_1}||B_{\mu N_2}|}{(\K_1+\K_2)}
\cos \theta^{\rm (LNC)} {\bigg ]} {\overline {\rm Br}}_{\pi} \
\label{BrplPiDir}
\\
{\cal A}_{\pi,\rm CP}^{\rm (Dir.)}  &\equiv&
\frac{ \Gamma^{\rm (LNC)}(\pi^- \to e^- e^- \mu^+ \nu) -
\Gamma^{\rm (LNC)}(\pi^+ \to e^+ e^+ \mu^- \nu)}
{ \Gamma^{\rm (LNC)}(\pi^- \to e^- e^- \mu^+ \nu) +
\Gamma^{\rm (LNC)}(\pi^+ \to e^+ e^+ \mu^- \nu)}
\nonumber\\
& = &
\frac{ 4 \sin \theta^{\rm (LNC)}}{\left[
\frac{|B_{e N_1}|}{|B_{e N_2}|}\frac{|B_{\mu N_1}|}{|B_{\mu N_2}|}
\left(1 + \frac{\K_2}{\K_1} \right) +
\frac{|B_{e N_2}|}{|B_{e N_1}|}\frac{|B_{\mu N_2}|}{|B_{\mu N_1}|}
\left(1 + \frac{\K_1}{\K_2} \right) + 4 \delta(y) \cos \theta^{\rm (LNC)}
\right]
} \; \frac{\eta(y)}{y}
\label{ACPPiDir}
\eea
\ees

The expression for the canonical quantity ${\overline {\rm Br}}_{\pi}$,
appearing in Equation~(\ref{BrplPiDir}), is given in Equation~(\ref{bBr}) in
conjunction with the notation (\ref{notGXDD}) in
Section~\ref{sec:BrPisub}
The results for the Majorana case,
\linebreak${\rm Br}_{\pi,+}^{\rm (Maj.)}$ $\equiv {\rm Br}_{\pi,+}^{\rm (LNV)}+{\rm Br}_{\pi,+}^{\rm (LNC)}$ and
${\cal A}^{\rm (Maj.)}_{\pi,\rm CP}$
are the following:
\bes
\label{BrPiMaj}
\bea
{\rm Br}_{\pi,+}^{\rm (Maj.)} & \equiv &
\frac{ \sum_{{\rm X}={\rm LNV,LNC}}
\left( \Gamma^{\rm (X)}(\pi^-\to e^- e^- \mu^+ \nu) + \Gamma^{\rm (X)}(\pi^+ \to e^+ e^+ \mu^- \nu) \right)}{2 \Gamma(\pi^{\pm} \to {\rm all})}
\nonumber\\
&=&
 {\bigg [} \sum_{j=1}^2
\frac{|B_{e N_j}|^2 ( |B_{e N_j}|^2 + |B_{\mu N_j}|^2)}{\K_j}
+ 4 \delta(y) \frac{|B_{e N_1}||B_{e N_2}|}{(\K_1+\K_2)}
\nonumber\\
&& \times
\left( |B_{e N_1}||B_{e N_2}| \cos \theta^{\rm (LNV)} +
 |B_{\mu N_1}||B_{\mu N_2}| \cos \theta^{\rm (LNC)} \right) {\bigg ]}
 {\overline {\rm Br}}_{\pi} \
\label{BrplPiMaj}
\eea
\bea
{\cal A}_{\pi,\rm CP}^{\rm (Maj.)}  &\equiv&
\frac{ \sum_{{\rm X}={\rm LNV,LNC}} \left(\Gamma^{\rm (X)}(\pi^- \to e^- e^- \mu^+ \nu) - \Gamma^{\rm (X)}(\pi^+ \to e^+ e^+ \mu^- \nu) \right)}
{ \sum_{{\rm X}={\rm LNV,LNC}} \left(\Gamma^{\rm (X)}(\pi^- \to e^- e^- \mu^+ \nu) + \Gamma^{\rm (X)}(\pi^+ \to e^+ e^+ \mu^- \nu) \right)}
\nonumber\\
& = &
4 \left( \sin \theta^{\rm (LNV)} +  \frac{|B_{\mu N_1}||B_{\mu N_2}|}{|B_{e N_1}||B_{e N_2}|}\sin \theta^{\rm (LNC)} \right)
\nonumber\\
&& \times
{\Bigg [}
\frac{(|B_{e N_1}|^2 +|B_{\mu N_1}|^2) }{|B_{e N_2}|^2}
\left(1 + \frac{\K_2}{\K_1} \right) +
\frac{(|B_{e N_2}|^2 +|B_{\mu N_2}|^2) }{|B_{e N_1}|^2}
\left(1 + \frac{\K_1}{\K_2} \right)
\nonumber\\
&&
+ 4 \delta(y) \left(
\cos \theta^{\rm (LNV)} +
\frac{|B_{\mu N_1}||B_{\mu N_2}|}{|B_{e N_1}||B_{e N_2}|} \cos \theta^{\rm (LNC)}
\right)
{\Bigg ]}^{-1}
\times \frac{\eta(y)}{y} \
\label{ACPPiMaj}
\ea
\ees

The function $\eta(y)/y=y/(y^2+1)$ is the same as in Section~\ref{sec:BrCPVsub}
(with: $y \equiv \Delta M_N/\Gamma_N$).
Even more so, numerical evaluations give for the $N_1$-$N_2$ overlap
parameter $\delta(y)$ the same values as in the semileptonic decays of
Section~\ref{sec:BrCPVsub}, \textit{cf.}~Table \ref{T4} and Figure~\ref{etadelfig}
there.

If we assume that
$|B_{\ell N_2}| \approx |B_{\ell N_1}|$ (for $\ell = e, \mu, \tau$),
\emph{i.e}., Equation~(\ref{BN1N2}),
then we have $\K_1 \approx \K_2 \equiv \K$,
and the expressions for ${\cal A}_{\pi,\rm CP}$ simplify significantly
\bes
\label{ACPPisimp}
\bea
{\cal A}_{\pi,\rm CP}^{\rm (Dir.)} & = & \frac{\sin \theta^{\rm (LNC)}}
{\left(1 + \delta(y) \cos \theta^{\rm (LNC)} \right)} \; \frac{\eta(y)}{y}
= \sin \theta^{\rm (LNC)} \; \frac{\eta(y)}{y} \left(1 + {\cal O}(\delta) \right) \
\label{ACPPisimpDir}
\\
{\cal A}_{\pi,\rm CP}^{\rm (Maj.)} &=&
\left( \frac{|B_{e N_1}|^2 \sin \theta^{\rm (LNV)} +
|B_{\mu N_1}|^2 \sin \theta^{\rm (LNC)}}{|B_{e N_1}|^2+|B_{\mu N_1}|^2}
\right) \frac{\eta(y)}{y}
\left(1 + {\cal O}(\delta) \right) \
\label{ACPPisimpMaj}
\eea
\ees

As in the case of semileptonic LNV decays of the previous
Section~\ref{sec:BrCPVsub}, we see that the CP asymmetry
parameter ${\cal A}_{\pi,\rm CP}$ can become appreciable
and even of order one if the following two conditions are
fulfilled simultaneously:
(a) at least one of the angles $\theta^{\rm (X)}$ (X=LNC,LNV),
defined in Equation~(\ref{thPi}), is appreciable;
(b) the quantity $y \equiv \Delta M_N/\Gamma_N$ is $y \sim 1$
(near degeneracy). In such cases, the estimates for the
effective (true) branching ratios ${\rm Br}_{\rm eff}^{\rm (X)}$
of Equations~(\ref{UBPi1}) and (\ref{UBPi2})
would apply also to the CP-violating difference of effective branching ratios,
${\cal A}^{\rm (X)}_{\pi, \rm CP} {\rm Br}_{\rm eff}^{\rm (X)}$,
where (X) = (Dir.),(Maj.).


\section{Conclusions}
\label{sec:summ}

We have studied
lepton number violating (LNV)
semileptonic decays of charged pseudoscalar
mesons, specifically  $\pi^\pm$, $K^{\pm}$, $D^\pm$, $D_s^\pm$, $B^\pm$ and $B_c^{\pm}$, mediated by heavy neutrinos that can go on their mass shell.

We first presented the LNV semileptonic decays of charged Kaons and of the heavier mesons
$D^\pm$, $D_s^\pm$, $B^\pm$ and $B_c^\pm$, in processes of the form
$M^{\pm} \to \ell_1^{\pm} \ell_2^{\pm} M^{\prime \mp}$,
mediated by on-shell massive neutrinos,
where $M$ is the decaying meson and  $M^{'}$ a correspondingly lighter meson.
We estimated the branching ratios as functions of the neutrino masses and mixing parameters, and found the scenarios where upper limits on the mixing parameters can be obtained. We also studied the effect on the observability of these decays due to the long neutrino lifetime, as the secondary decay vertex is likely to fall outside the detector for the range of neutrino masses that are relevant to these processes.

We then presented our corresponding study of charged pion decays, which in this case are purely leptonic since pions are the lightest mesons. Here we can have modes that conserve lepton number (LNC) as well as modes that violate lepton number (LNV), if the intermediate neutrinos are of Majorana type, while only the former modes occur if the intermediate neutrino is of Dirac type.
However, these modes are not distinguished by the final state because the latter involves a standard neutrino, which is not experimentally observable.
We find that it could be possible to discern the Majorana or Dirac nature of neutrinos if one is able to observe features in the final state distribution.

We finally explored the possibility of observing CP
violation in the lepton sector using these meson decays mediated by massive neutrinos on shell. The CP signal in charged meson decays is the usual asymmetry between the decays of opposite charge mesons. We found that leptonic CP violation
may show in semileptonic LNV decays of  charged Kaons and charged B mesons, as well
as in LNC and LNV decays of charged pions, depending on the mass of the intermediate neutrinos.
It turns out that such CP violation becomes appreciable and possibly detectable if there are
at least two heavy neutrinos almost degenerate in mass that can go on their mass shell.
The neutrino mass splitting that gives maximal CP asymmetries is close to the neutrino decay width.  This type scenario fits well into the so called \emph{neutrino
minimal standard model} ($\nu$MSM), which
contains two almost degenerate Majorana neutrinos of mass
near 1 GeV and another lighter neutrino of mass of order $10^1$ keV, a model that
can explain simultaneously neutrino
oscillations, the dark matter and the baryon asymmetry of the Universe.



\acknowledgments{
This work was supported in part by FONDECYT, Chile Grants No. 1130599 (G.C. and C.S.K.) and No. 1130617 (C.D.), and projects PIIC 2014 and Mecesup FSM1204 (J.Z.S.).
The work of C.S.K. was supported by the NRF
grant funded by the Korean government of the MEST
(No. 2011-0017430) and (No. 2011-0020333).}


\appendix
\section*{\noindent Appendix}
\vspace{12pt}
\noindent{\textit{A.1. Explicit Formulas for Amplitudes of
the $N_j$-mediated Decay $M^{\pm} \to \ell_1^{\pm} \ell_2^{\pm} M^{' \mp}$}}
\label{appBrM}

In this Appendix we provide, for completeness, formulas which are used
in Sections~\ref{sec:BrM} and \ref{sec:BrCPVsub}. The formulas were presented
in Ref.~\cite{CKZ2} for the case of exchange of two different neutrinos
$N_1$ and $N_2$. Here we present them in a slightly more general
form, when the number of exchanged neutrinos is ${\cal N}$ ($N_1, \ldots,
N_{\cal N}$). In Section~\ref{sec:BrM} the simpler case of ${\cal N}=1$ is taken,
because such a case is representative enough for the consideration of the
branching ratios there. On the other hand, the case ${\cal N}=2$
(or: \linebreak${\cal N} \geq 2$) is taken in Section~\ref{sec:BrCPV}, with
two of the neutrinos ($N_1$, $N_2$) considered on-shell and almost degenerate,
because in such a case significant CP violation effects can arise in the
Majorana neutrino~sector.

The amplitude squared $|{\cal T}(M^{\pm})|^2$ for the decay of Figure~\ref{FigMMp}
appears in the expression Equation~(\ref{GM1}) for the decay width
$\Gamma(M^{\pm} \to \ell_1^{\pm} \ell_2^{\pm} M^{' \mp})$,
and can be written in the form
\ba
|{\cal T}(M^{\pm})|^2 &=&
K^2 \sum_{i=1}^{{\cal N}}  \sum_{j=1}^{{\cal N}} k_i^{(\pm)} k_j^{(\pm) *} M_{N_i} M_{N_j}
\nonumber\\
&& \times
{\big [} P_i(D) P_j(D)^* T_{\pm}(D) T_{\pm}(D)^* +
P_i(C) P_j(C)^* T_{\pm}(C) T_{\pm}(C)^*
\nonumber\\
&& +
\left( P_i(D) P_j(C)^* T_{\pm}(D) T_{\pm}(C)^* +  P_i(C) P_j(D)^* T_{\pm}(C) T_{\pm}(D)^* \right) {\big ]} \
\label{TMsqr}
\ea
Here, $i, j=1,\ldots,{\cal N}$ are indices of
contributions of the exchanges of intermediate neutrinos
$N_i, N_j$, and $X=D, C$ denote contributions of amplitudes of the
direct and crossed channels, respectively, cf.~Figure~\ref{FigMMp}.
Further, $k_j^{(\pm)}$ are the heavy-light mixing factors for $N_j$ defined
in Equation~(\ref{kj}); $P_j(X)$ \linebreak($j=1,2; X=D, C$) are the propagator
functions of $N_j$ neutrino for the $D$ and $C$ channel, Equation~(\ref{Pj}).
$K^2$ is the constant originating from the vertices and is given in
Equation~(\ref{Ksqr}).
These expressions appear in the normalized decay
widths $\G_{\pm}(XY^{*})$ in Equation~(\ref{bGXYN}) when ${\cal N}=1$,
and in  $\G_{\pm}(XY^{*})_{ij}$ in Equation~(\ref{bGXY}) when ${\cal N} \geq 2$.
The quadratic expressions of $T_{\pm}(X) T_{\pm}(Y)^*$ in Equation~(\ref{TMsqr})
get simplified after summation over the final helicities of the leptons
$\ell_1$ and $\ell_2$, and acquire the following form:
\bes
\label{TT}
\bea
 T_{\pm}(D) T_{\pm}(D)^{*} & = & 8 {\big [} M_M^2 M_{M'}^2 (p_1 \cdot p_2)
- 2 M_M^2 (p_1 \cdot p_{M'}) (p_2 \cdot p_{M'}) -2  M_{M'}^2 (p_1 \cdot p_{M}) (p_2 \cdot p_{M})
\nonumber\\
&&
+ 4 (p_1 \cdot p_{M}) (p_2 \cdot p_{M'}) (p_M \cdot  p_{M'}) {\big ]}
\equiv T(D) T(D)^{*} \
\label{TDTD}
\\
 T_{\pm}(C) T_{\pm}(C)^{*} & = & 8 {\big [} M_M^2 M_{M'}^2 (p_1 \cdot p_2)
- 2 M_M^2 (p_1 \cdot p_{M'}) (p_2 \cdot p_{M'}) -2  M_{M'}^2 (p_1 \cdot p_{M}) (p_2 \cdot p_{M})
\nonumber\\
&&
+ 4 (p_2 \cdot p_{M}) (p_1 \cdot p_{M'}) (p_M \cdot  p_{M'})  {\big ]}
\equiv T(C) T(C)^{*} \
\label{TCTC}
\\
T_{\pm}(D) T_{\pm}(C)^{*} & = & 16 {\Big \{}
M_M^2 (p_1 \cdot p_{M'}) (p_2 \cdot p_{M'})  +  M_{M'}^2 (p_1 \cdot p_{M}) (p_2 \cdot p_{M})
- \frac{1}{2} M_M^2 M_{M'}^2 (p_1 \cdot p_2)
\nonumber\\
&&
+ (p_M \cdot  p_{M'})
\left[
-(p_1 \cdot p_{M}) (p_2 \cdot p_{M'})
-(p_2 \cdot p_{M}) (p_1 \cdot p_{M'})
+  (p_M \cdot  p_{M'}) (p_1 \cdot p_2) \right]
\nonumber\\
&&
\mp i (p_M \cdot  p_{M'}) \epsilon(p_M,p_1,p_2,p_{M'})
{\Big \}}
\label{TDTC}
\\
T_{\pm}(C) T_{\pm}(D)^{*} & = & \left( T_{\pm}(D) T_{\pm}(C)^{*} \right)^{*}
= T_{\mp}(D) T_{\mp}(C)^{*} = \left(  T_{\mp}(C) T_{\mp}(D)^{*}  \right)^{*} \
\label{TCTD}
\eea
\ees
where we used the notation
\be
\epsilon(q_1, q_2, q_3, q_4) \equiv \epsilon^{\eta_1 \eta_2 \eta_3 \eta_4}
(q_1)_{\eta_1} (q_2)_{\eta_2} (q_3)_{\eta_3} (q_4)_{\eta_4} \
\label{eps}
\ee
Here, $\epsilon^{\eta_1 \eta_2 \eta_3 \eta_4}$ is the totally antisymmetric
Levi-Civita tensor with the sign convention $\epsilon^{0123}=+1$.

The expression (\ref{TT}),
together with the definition (\ref{bGXY}), imply for
the normalized decay widths $\G_{\pm}(XY^{*})_{ij}$ of Equation~(\ref{bGXY})
various symmetry relations, namely that $\G_{\pm}(DD^{*})$ and
$\G_{\pm}(CC^{*})$ are self-adjoint (${\cal N} \times {\cal N}$) matrices,
and that elements of the $D$-$C$ interference matrices
$\G_{\pm}(CD^{*})$ and $\G_{\pm}(DC^{*})$ are simply related
\bes
\label{symm}
\bea
\G(DD^{*})_{ij} & = & \left( \G(DD^{*})_{ji} \right)^{*} \ ,
\qquad
\G(CC^{*})_{ij} =  \left( \G(CC^{*})_{ji} \right)^{*} \ ,
\label{symmDD}
\\
\G_{\pm}(CD^{*})_{ij} & = &  \left( \G_{\pm}(DC^{*})_{ji} \right)^{*} \ 
\label{symmCD}
\eea
\ees
If the two final leptons are
of the same flavor ($\ell_1=\ell_2$), one can use the
property that the integration $d_3$ over the final particles is symmetric under
exchange of $p_1$ and $p_2$ (because $M_{\ell_1} = M_{\ell_2}$), and we have
the following additional symmetries:
\bes
\label{symmadd}
\bea
\G(DD^{*})_{ij} & = & \G(CC^{*})_{ij} \
\label{symmaddDD}
\\
\G_{\pm}(CD^{*})_{ij} & = & \G_{\pm}(DC^{*})_{ij} \
\label{symmaddCD}
\eea
\ees
and the ${\cal N} \times {\cal N}$ $D$-$C$ interference matrices
$\G_{\pm}(CD^{*})$ become self-adjoint, too.

When ${\cal N}=1$, as in Section~\ref{sec:BrM}, then we have
in the case of $\ell_1=\ell_2$
\be
\G(DD^{*}) =  \G(CC^{*}) \quad ({\cal N}=1 \; {\rm and} \; \ell_1=\ell_2)
\label{symmN}
\ee

\vspace{12pt}
\noindent{\textit{A.2. Explicit Expression for the Function $Q$}}
\label{appBrMexpl}
\vspace{12pt}

The expression in Equations~(\ref{GDDN}) and (\ref{GDD})
is arrived at by using in the integration
over the phase space of three final particles [Equations~(\ref{GM1}) and (\ref{d3})],
for the contribution of the $N$ neutrino, the identity
\bes
\label{d3d2}
\bea
\lefteqn{
d_3 \left( M(p_M) \to  \ell_1(p_1) \ell_2(p_2) M^{'}(p_{M'}) \right) =
}
\nonumber\\
&& d_2 \left( M(p_M) \to  \ell_1(p_1) N_j(p_N) \right) d p_N^2
d_2 \left( N_j(p_N) \to  \ell_2(p_2) M^{'}(p_{M'}) \right)
\label{d3d2D}
\\
&& d_2 \left( M(p_M) \to  \ell_2(p_2) N_j(p_N) \right) d p_N^2
d_2 \left( N_j(p_N) \to  \ell_1(p_1) M^{'}(p_{M'}) \right) \
\label{d3d2C}
\eea
\ees

The first identity can be used for the $DD^{*}$ contribution
(where $p_N=p_M-p_1$) and the second for the $CC^{*}$ contribution
(where $p_N=p_M-p_2$). When one uses the identity (\ref{P1P1}) in the
$DD^{*}$ contribution, and the analogous identity for the $CC^{*}$
contribution, the integration over $d p_N^2$ becomes trivial,
and the $d_2$-type of integrations can be performed.
{Notice that this is equivalent to the factorization approach
$\Gamma(M \to \ell_1 N)
{\rm Br}(N \to \ell_2 M^{'})$, which holds when $N$ is on-shell.
}
The obtained expression for $\G(DD^{*})$ is then the
expression Equation~(\ref{GDDN}) when ${\cal N}=1$
[Equation~(\ref{GDD}) when ${\cal N} \geq 2$]
with the notations (\ref{notGDDN}),
where the obtained function $Q$ has the following form:
\bea
Q(y_N; y_{\ell_1}, y_{\ell_2}, y') & = &
{\bigg \{} \frac{1}{2}(y_N - y_{\ell_1})(y_N - y_{\ell_2})(1 - y_N - y_{\ell_1})
\left( 1 - \frac{y'}{y_N} +\frac{y_{\ell_2}}{y_N} \right)
\nonumber\\
&& +  {\big [}
- y_{\ell_1} y_{\ell_2} (1 + y' + 2 y_N - y_{\ell_1} - y_{\ell_2} )
- y_{\ell_1}^2 (y_N - y') + y_{\ell_2}^2 (1 - y_N)
\nonumber\\
&&
+ y_{\ell_1} (1+y_N) (y_N - y') - y_{\ell_2} (1-y_N)(y_N+y') {\big ]}
{\bigg \}} \
\label{Q}
\eea
In the limit of massless charged leptons ($y_{\ell_1}=y_{\ell_2}=0$),
this reduces to
\be
Q(y_N; 0,0, y') = \frac{1}{2} y_N^2 (1 - y_N) \left( 1 - \frac{y'}{y_N} \right) \
\label{Q0}
\ee

\vspace{12pt}
\noindent{\textit{A.3. Calculation of the Total Decay Width of Neutrino $N$}}
\label{appGN}
\vspace{12pt}

In this Appendix, for completeness, we summarize the formulas
needed for evaluation of the total decay width of a massive
sterile neutrino $N$, \textit{cf.}~Equations~(\ref{GNwidth})--(\ref{calK}) and
Figure~\ref{FigcNellN}.

In Ref.~\cite{Atre} (Appendix C there), the formulas for the
leptonic decay and semimesonic decay widths of a sterile neutrino
$N$ have been obtained, for the masses $M_N \lesssim 1$ GeV.
For higher values of the masses $M_N$, the
calculation of the semileptonic decay widths becomes difficult
because not all the resonances are known. Hence, for such masses
the authors of Refs.~\cite{HKS,GKS} proposed
an inclusive approach, based on duality, for the calculation of the total
contribution of the semileptonic decay width of $N$.
It consists of representing the various
(pseudoscalar and vector) meson channels
by quark-antiquark channels.
This approach was applied for
$M_N \geq M_{\eta^{'}} \approx 0.958$ GeV. Here we summarize
the formulas given in Ref.~\cite{HKS} for the decay width channels
(\textit{cf.}~also: \cite{Atre}).
In some of these formulas, twice the decay width is given
[$2 \Gamma(N \to \ldots)$], signalling the fact that
for each possible decay of Majorana neutrino in charged particles,
there is an equally possible decay into charge conjugate channel
(something not possible if $N$ is Dirac~particle).
\bes
\label{GNlept}
\bea
2 \Gamma(N \to \ell^- \ell^{'+} \nu_{\ell^{'}}) & = &
|B_{\ell N}|^2 \frac{G_F^2}{96 \pi^3} M_N^5 I_1(x_{\ell},0, x_{\ell^{'}})
(1 - \delta_{\ell \ell^{'}} ) \ ,
\label{GNlepta}
\\
\Gamma(N \to \nu_{\ell} \ell^{'-} \ell^{'+}) & = &
|B_{\ell N}|^2 \frac{G_F^2}{96 \pi^3} M_N^5
{\big [}
(g_L^{(\rm lept)} g_R^{(\rm lept)}
+ \delta_{\ell \ell^{'}} g_R^{(\rm lept)}) I_2(0,x_{\ell^{'}},x_{\ell^{'}})
\nonumber\\
&&
+ \left( (g_L^{(\rm lept)})^2 + (g_R^{(\rm lept)})^2
+  \delta_{\ell \ell^{'}}
(1 + 2 g_L^{(\rm lept)}) \right) I_1(0,x_{\ell^{'}},x_{\ell^{'}})
{\big ]}
\label{GNleptb}
\\
\sum_{\nu_{\ell}} \sum_{\nu^{'}} \Gamma(N \to \nu_{\ell} \nu^{'} {\bar \nu}^{'})
& = &
\sum_{{\ell}} |B_{\ell N}|^2 \frac{G_F^2}{96 \pi^3} M_N^5 \
\label{GNleptc}
\eea
\ees

In Equation~(\ref{GNlepta}) factor $2$  was included because for
Majorana neutrino $N$ both decays
$N \to \ell^- \ell^{'+} \nu_{\ell^{'}}$ and $N \to \ell^+ \ell^{'-} \nu_{\ell^{'}}$
contribute ($\ell \not= \ell^{'}$).

When $M_N < M_{\eta^{'}} \approx 0.968$ GeV, the following semimesonic
decays contribute, which involve presudoscalar ($P$) and vector ($V$)
mesons:
\bes
\label{GNmes}
\bea
2 \Gamma(N \to  \ell^- P^+) & = &
|B_{\ell N}|^2 \frac{G_F^2}{8 \pi} M_N^3 f_P^2 |V_P|^2 F_P(x_{\ell}, x_P) \
\label{GNMesPch}
\\
\Gamma(N \to  \nu_{\ell} P^0) & = &
|B_{\ell N}|^2 \frac{G_F^2}{64 \pi} M_N^3 f_P^2 (1 - x_P^2)^2 \
\label{GNMesP0}
\\
2 \Gamma(N \to  \ell^- V^+) & = &
|B_{\ell N}|^2 \frac{G_F^2}{8 \pi} M_N^3 f_V^2 |V_V|^2 F_V(x_{\ell}, x_V) \
\label{GNMesVch}
\\
\Gamma(N \to  \nu_{\ell} V^0) & = &
|B_{\ell N}|^2 \frac{G_F^2}{2 \pi} M_N^3 f_V^2 \kappa_V^2 (1 - x_V^2)^2
(1 + 2 x_V^2) \
\label{GNMesV0}
\eea
\ees
where factor $2$ in the charged meson channels appears because
both decays $N \to  \ell^- M^{'+}$ and $N \to  \ell^+ M^{'-}$ contribute
($M^{'}=P, V$) if $N$ is Majorana.
The factors $V_P$ and $V_V$ are the CKM
matrix elements involving the valence quarks of the mesons;
and $f_P$ and $f_V$ are the corresponding decay constants,
{
whose values are given, e.g., in
Table 1 in Ref.~\cite{HKS}.
}
The pseudoscalar mesons which contribute here are:
$P^{\pm} = \pi^{\pm}, K^{\pm}$; $P^0 = \pi^0, K^0, {\bar K}^0, \eta$.
The vector mesons which contribute are:
$V^{\pm} = \rho^{\pm}, K^{* \pm}$; $V^0= \rho^0, \omega, K^{*0}, {\bar K}^{*0}$.
%
If $M_N \geq M_{\eta^{'}}$ ($=$0.9578 GeV), due to duality
the (many) semimesonic decay modes are represented
by the following quark-antiquark decay modes \cite{HKS}:
{\small
\bes
\label{GNquark}
\bea
2 \Gamma(N \to  \ell^- U {\bar D}) & = &
|B_{\ell N}|^2 \frac{G_F^2}{32 \pi^3} M_N^5 |V_{UD}|^2 I_1(x_{\ell},x_U,x_D) \
\label{GNquarka}
\\
\Gamma(N \to  \nu_{\ell} q {\bar q}) & = &
|B_{\ell N}|^2 \frac{G_F^2}{32 \pi^3} M_N^5
\left[
g_L^{(q)} g_R^{(q)} I_2(0,x_q,x_q)\!+\!
\left( (g_L^{(q)})^2\!+\!(g_R^{(q)})^2 \right) I_1(0,x_q,x_q)
\right] \
\label{GNquarkb}
\eea
\ees}
\vspace{-6pt}

In the formulas (\ref{GNlept})--(\ref{GNquark}) the notations
$x_{Y} \equiv M_Y/M_N$ ($Y = \ell, \nu_{\ell},  P, V, q$) are used; and
in Equation~(\ref{GNquark}) we denoted: $U=u,c$; $D=d,s,b$; $q=u,d,c,s,b$.
The values of quark masses which we used in our evaluations are:
$M_u=M_d = 3.5$ MeV; $M_s=105$ MeV; $M_c=1.27$ GeV; $M_b=4.2$ GeV.

We note that in the evaluation of the total decay width $\Gamma_N$,
the expressions (\ref{GNquarka}) and (\ref{GNquarkb}) should be added
when $N$ is Majorana; if $N$ is Dirac, the same summation should be taken,
but the expressions (\ref{GNquarka})
should be multiplied by $1/2$.
The same approach is valid also in the case of summation of
expressions (\ref{GNlept}) and (\ref{GNmes}).

In Equations~(\ref{GNleptb}) and (\ref{GNquarkb}) there appear the following
SM neutral current couplings:
\bes
\label{NCc}
\bea
g_L^{(\rm lept)} &=& - \frac{1}{2} + \sin^2 \theta_W \ ,
\quad
g_R^{(\rm lept)} =\sin^2 \theta_W \
\label{NCcl}
\\
g_L^{(U)} & = & \frac{1}{2} - \frac{2}{3} \sin^2 \theta_W \ ,
\quad
g_R^{(U)} = - \frac{2}{3} \sin^2 \theta_W \
\label{NCcU}
\\
g_L^{(D)} & = & - \frac{1}{2} + \frac{1}{3} \sin^2 \theta_W \ ,
\quad
g_R^{(U)} =  \frac{1}{3} \sin^2 \theta_W \
\label{NCcD}
\eea
\ees

Further, the neutral current couplings $\kappa_V$ of the neutral vector mesons
in Equation~(\ref{GNMesV0}) are
\bes
\label{kV}
\bea
\kappa_V & = & \frac{1}{3} \sin^2 \theta_W  \quad (V=\rho^0, \omega) \
\label{kVa}
\\
\kappa_V & = & - \frac{1}{4} +\frac{1}{3} \sin^2 \theta_W
\quad (V=K^{*0}, {\bar K}^{*0}) \
\label{kVb}
\eea
\ees

The following kinematical expressions $I_1$, $I_2$, $F_P$ and $F_V$ were used:
\bes
\label{kinex}
\bea
I_1(x,y,z) & = & 12 \int_{(x+y)^2}^{(1-z)^2} \; \frac{ds}{s}
(s - x^2 - y^2) (1 + z^2 -s)
\lambda^{1/2}(s,x^2,y^2) \lambda^{1/2}(1,s,z^2) \
\label{I1}
\\
I_2(x,y,z) & = & 24 y z \int_{(y+z)^2}^{(1-x)^2} \; \frac{ds}{s}
(1 + x^2 - s)
\lambda^{1/2}(s,y^2,z^2) \lambda^{1/2}(1,s,x^2) \
\label{I2}
\\
F_P(x,y) & = & \lambda^{1/2}(1,x^2,y^2) \left[(1 + x^2)(1 + x^2-y^2) - 4 x^2
\right] \
\label{FP}
\\
F_V(x,y) & = & \lambda^{1/2}(1,x^2,y^2) \left[(1 - x^2)^2 + (1 + x^2) y^2 - 2 y^4 \right] \
\label{FV}
\eea
\ees
with $\lambda$ function defined in Equation~(\ref{lambda}).
These formulas allow us to obtain
the total decay width $\Gamma(N \to {\rm all})$ as a function of $M_N$.
Using these formulas, we evaluated the coefficients ${\cal N}_{\ell N}$,
appearing in Equation~(\ref{calK}) at the mixing terms $|B_{\ell N}|^2$,
and presented them in Figure~\ref{FigcNellN} as a function of $M_N$
for the cases of Majorana and Dirac neutrino $N$.
One may notice a small kink in the curves
of Figure~\ref{FigcNellN} at $M_N=M_{\eta^{'}}$ ($=$0.9578 GeV). This kink
appears because at  $M_N \geq M_{\eta^{'}}$ the use of duality is made
(the replacement of the semileptonic decay channel contributions
by the quark-antiquark channel contributions).
We can see that the
duality works quite well at $M_N \geq M_{\eta^{'}}$,
with the possible exception for the case $\ell = \tau$
because $\tau$ lepton has a large mass.

\vspace{12pt}
\noindent{\textit{A.4. Explicit Amplitudes for $N_j$-Mediated Decays
$\pi^{\pm} \to e^{\pm} e^{\pm} \mu^{\mp} \nu$}}
\label {appBrPi}
\vspace{12pt}

In this Appendix we summarize, for completeness,
formulas needed in Sections~\ref{sec:BrPi} and \ref{sec:BrCPVPisub}.
These formulas were derived
and presented in Ref.~\cite{CKZ}, for the case of
exchange of two different neutrinos $N_1$ and $N_2$.
Here we summarize them in a slightly more general form, for the
case of ${\cal N}$ different
neutrinos $N_j$ ($j=1,\ldots, {\cal N}$).
In Section~\ref{sec:BrPi} the simpler case ${\cal N}=1$ is taken,
as it is sufficiently representative for the branching ratios
considered there.
In Section~\ref{sec:BrCPV} the case ${\cal N}=2$ (or: ${\cal N} \geq 2$)
is considered, with two (on-shell) neutrinos $N_1$ and $N_2$ almost degenerate,
as in such a case significant CP violation effects can occur in the
neutrino sector.

The squared amplitude $| {\cal T}_{\pi,\pm}^{\rm (X)} | ^2$
for the $N_j$-mediated leptonic decays of neutrinos,
appearing, for example, in Equation~(\ref{GpiX1}) (where X = LNC, LNV),
is a combination of contributions
from the two channels $D$ (direct) and $C$ (crossed)
(\textit{cf.}~Figures~\ref{FigLNCpi} and \ref{FigLNVpi}), and, in general,
of the contributions of ${\cal N}$ neutrinos~$N_j$
\ba
\lefteqn{
| {\cal T}_{\pi,\pm}^{(X)} | ^2 =
K_{\pi}^2 \sum_{i=1}^{{\cal N}} \sum_{j=1}^{{\cal N}} k_{i,\pm}^{(X) *} k_{j,\pm}^{(X)} }
\nonumber\\
&&
\times {\bigg [} P_i^{(X)}(D) P_j^{(X)}(D)^{*} T_{{\pi},\pm}^{(X)}(DD^{*})
+  P_i^{(X)}(C) P_j^{(X)}(C)^{*} T_{{\pi},\pm}^{(X)}(CC^{*})
\nonumber\\
&&+
\left( P_i^{(X)}(D) P_j^{(X)}(C)^{*} T_{{\pi},\pm}^{(X)}(DC^{*}) +
P_i^{(X)}(C) P_j^{(X)}(D)^{*} T_{{\pi},\pm}^{(X)}(CD^{*}) \right) {\bigg ]} \
\label{calTXsqr}
\ea
The constant $K_{\pi}^2$ is given in Equation~(\ref{Kpisqr}),
and the mixing  factors $k_{j,\pm}^{(X)}$ are
\be
\label{kjpi}
k_{j,+}^{{\rm (LNV)}}  =  B_{e N_j}^2 \ , \qquad k_{j,+}^{{\rm (LNC)}} = B_{e N_j} B^{*}_{\mu N_j} \ , \qquad k_{j,-}^{(X)} = \left( k_{j,+}^{(X)} \right)^{*} \
\ee

In the case of ${\cal N}=1$, these coefficients are in Equation~(\ref{kpi}).
$P_j^{(X)}(Y)$ are the $N_j$-propagator functions
[when ${\cal N}=1$: $N$-propagator functions of Equation~(\ref{Ps})]
of the direct and crossed channels ($Y=D, C$).

Explicit expressions for the direct ($DD^{*}$),
crossed ($CC^{*}$)
and direct-crossed interference
($DC^{*}$ and $CD^{*}$) terms
[$T_{{\pi},\pm}^{\rm (X)}( DD^{*})$, $T_{{\pi},\pm}^{\rm (X)}( CC^{*})$,
$T_{{\pi},\pm}^{\rm (X)}( DC^{*})$, $T_{{\pi},\pm}^{(X)}( CD^{*})$],
appearing in Equations~(\ref{calTXsqr}), get simplified when summed over the
helicities of all the final leptons. In the case of the X = LNV processes
(\textit{cf.}~Figure~\ref{FigLNVpi}) they acquire the following form:
\bes
\label{TLV}
\ba
T_{{\pi},\pm}^{\rm (LNV)}(DD^{*})&=& 256 (p_2 \cdot p_{\nu}) \left[
- M_{\pi}^2 (p_1 \cdot p_{\mu}) + 2 (p_1 \cdot p_{\pi}) (p_{\mu} \cdot p_{\pi}) \right] \equiv T_{\pi}^{\rm (LNV)}(DD^{*})  \ ,
\label{TLVDD}
\\
T_{{\pi},\pm}^{\rm (LNV)}(CC^{*})&=& 256 (p_1 \cdot p_{\nu}) \left[
- M_{\pi}^2 (p_2 \cdot p_{\mu}) + 2 (p_2 \cdot p_{\pi}) (p_{\mu} \cdot p_{\pi}) \right] \equiv T_{\pi}^{\rm (LNV)}(CC^{*}) \ ,
\label{TLVCC}
\\
T_{{\pi},\pm}^{\rm (LNV)}(DC^{*})&=& 128 {\bigg \{}
(p_1 \cdot p_{\nu})
\left[ M_{\pi}^2 (p_2 \cdot p_{\mu}) - 2 (p_2 \cdot p_{\pi})(p_{\mu} \cdot p_{\pi}) \right]
\nonumber\\
&&+
(p_2 \cdot p_{\nu})
\left[ M_{\pi}^2 (p_1 \cdot p_{\mu}) - 2 (p_1 \cdot p_{\pi})(p_{\mu} \cdot p_{\pi}) \right]
\nonumber\\
&&- (p_1 \cdot p_2)
\left[ M_{\pi}^2 (p_{\nu} \cdot p_{\mu}) - 2 (p_{\nu} \cdot p_{\pi})(p_{\mu} \cdot p_{\pi}) \right] {\bigg \}}
\nonumber\\
&&
\mp i {\bigg \{}
- (p_1 \cdot p_{\pi}) \epsilon(p_2,p_{\nu},p_{\mu},p_{\pi})
+ (p_2 \cdot p_{\pi}) \epsilon(p_1,p_{\nu},p_{\mu},p_{\pi})
\nonumber\\
&&- (p_{\nu} \cdot p_{\pi}) \epsilon(p_1,p_2,p_{\mu},p_{\pi})
- (p_{\mu} \cdot p_{\pi}) \epsilon(p_1,p_2,p_{\nu},p_{\pi}) {\bigg \}} \
\label{TLVDC}
\\
T_{{\pi},\pm}^{\rm (LNV)}(CD^{*})&=& \left( T_{{\pi},\pm}^{\rm (LNV)}(DC^{*}) \right)^{*}
= T_{{\pi},\mp}^{\rm (LNV)}(DC^{*}) \
\label{TLVCD}
\ea
\ees
where we used the notation Equation~(\ref{eps}) for $\epsilon$.

In the case of X = LNC processes (\textit{cf.}~Figure~\ref{FigLNCpi}),
the expressions are
\bes
\label{TLC}
\ba
T_{{\pi},\pm}^{\rm (LNC)}(DD^{*})&=& 256 (p_{\mu} \cdot p_{\nu}) {\bigg [}
(p_1 \cdot p_2) \left( M_{\pi}^4 - M_{\pi}^2 M_e^2
- 4 M_{\pi}^2 (p_1 \cdot p_{\pi}) + 4 (p_1 \cdot p_{\pi})^2 \right)
\nonumber\\ &&
+ 2 M_e^2 (p_2 \cdot p_{\pi}) (M_{\pi}^2 - p_1 \cdot p_{\pi}) {\bigg ]}
\equiv T_{\pi}^{\rm (LNC)}(DD^{*}) \
\label{TLCDD}
\\
T_{{\pi},\pm}^{\rm (LNC)}(CC^{*})&=& 256 (p_{\mu} \cdot p_{\nu}) {\bigg [}
(p_1 \cdot p_2) \left( M_{\pi}^4 - M_{\pi}^2 M_e^2
- 4 M_{\pi}^2 (p_2 \cdot p_{\pi}) + 4 (p_2 \cdot p_{\pi})^2 \right)
\nonumber\\ &&
+ 2 M_e^2 (p_1 \cdot p_{\pi}) (M_{\pi}^2 - p_2 \cdot p_{\pi}) {\bigg ]}
\equiv T_{\pi}^{\rm (LNC)}(CC^{*}) \
\label{TLCCC}
\\
T_{{\pi},\pm}^{\rm (LNC)}(DC^{*})&=& 256 (p_{\mu} \cdot p_{\nu})
{\bigg [} (p_1 \cdot p_2) (M_{\pi}^2 - 2 p_1 \cdot p_{\pi}) (M_{\pi}^2 - 2 p_2 \cdot p_{\pi})
\nonumber\\
&& + M_e^2 \left(-2 (p_1 \cdot p_{\pi})^2 -2 (p_2 \cdot p_{\pi})^2
+ M_{\pi}^2 (p_1 +p_2) \cdot p_{\pi} + M_{\pi}^2 M_e^2 \right) {\bigg ]}
\label{TLCDC}
\\
&\equiv& T_{\pi}^{\rm (LNC)}(DC^{*}) \ ,
\nonumber\\
T_{\pi}^{\rm (LNC)}(CD^{*})&=& \left( T_{\pi}^{\rm (LNC)}(DC^{*}) \right)^{*} \
\label{TLCCD}
\ea
\ees
These expressions appear in
the definition of the normalized
(\emph{i.e}., without explicit mixing dependence) decay width matrices
$\G_{\pi,\pm}^{\rm (X)}(YZ^{*})_{ij}$ (X = LNV, LNC; $Y, Z = D, C$;
$i,j=1,\ldots,{\cal N}$)
\be
\label{GXij}
\G_{\pi,\pm}^{\rm (X)}(YZ^{*})_{ij} =
K^2 \; \frac{1}{2!} \frac{1}{2 M_{\pi}} \frac{1}{(2 \pi)^8} \int d_4 \;
P_i^{\rm (X)}(Y) P_j^{\rm (X)}(Z)^{*} \; T_{\pm}^{(X)}(YZ^{*}) \
\ee
where $P_j^{\rm (X)}$ are the propagator functions of neutrino $N_j$
(with mass $M_{N_j}$), \textit{cf.}~the definitions (\ref{Ps}) written when ${\cal N}=1$.
When ${\cal N}=1$, the definition (\ref{GXij}) reduces
to the definition (\ref{GpiX}) in Section~\ref{sec:BrPisub}.

When we use the symmetry of the $d_4$ integration under the
exchange $p_1 \leftrightarrow p_2$ (we note: \linebreak $M_1=M_2=M_e$ in our
considered case), this leads to the
following identities:
\bes
\label{symmPi}
\ba
\G_{\pi}^{(X)}(DD^{*})_{ij} & = & \G_{\pi}^{(X)}(CC^{*})_{ij} \ , \quad
\G_{\pi}^{(X)}(DD^{*})_{ji} = \left( \G_{\pi}^{(X)}(DD^{*})_{ij} \right)^{*} \
\label{symmDDpi}
\\
\G_{{\pi},\pm}^{(X)}(CD^{*})_{ij} & = & \G_{{\pi},\pm}^{(X)}(DC^{*})_{ij} =
 \left( \G_{\pm}^{(X)}(CD^{*})_{ji} \right)^{*} \
\label{symmCDpi}
\ea
\ees
In the case of ${\cal N}=1$ this reduces simply to
\be
\G_{\pi}^{\rm (X)}(DD^{*})  =  \G_{\pi}^{\rm (X)}(CC^{*}) \
\label{symmpicalN1}
\ee

\vspace{12pt}
\noindent{\textit{A.5. Explicit Expression for $\G_{\pi}^{\rm (X)}$ and $d \G_{\pi}^{\rm (X)}/d E_{\mu}$
for $\pi^{\pm} \to e^{\pm} e^{\pm} \mu^{\mp} \nu$ with On-Shell $N$}}
\label{appBrPiexpl}
\vspace{12pt}

Equation (\ref{GXDDN}) refers to the expression obtained by performing
the integration (\ref{GpiX1})  over the phase space
of the four final particles [\textit{cf.}~Equation~(\ref{d4})], of the integrand
written explicitly in Appendix 4.
In the integration, the on-shellness (\ref{PP}) is assumed, which
makes the integration over $p_N^2$ trivial.
At the final stage of integration,
the differential decay width $d \G^{\rm (X)}/d E_{\mu}$ over the muon
energy $E_{\mu}$, in the rest frame of the $N$ neutrino, is performed.
The expressions for $d \G^{\rm (X)}/d E_{\mu}$ were
written in Refs.~\cite{CDK,CKZ}, and we write them down here
for completeness. In the case of X = LNV it is
\bea
\lefteqn{
\frac{ d \G^{\rm (LNV)}(\pi^{\pm} \to e^{\pm} e^{\pm} \mu^{\mp} \nu)}{d E_{\mu}}
}
\nonumber\\
&&
 =  \frac{K_{\pi}^2}{2(2 \pi)^4} \frac{1}{\Gamma_{N} M_{\pi}^3}
\lambda^{1/2}(M_{\pi}^2,M_{N}^2,M_e^2) \times
\left[ M_{\pi}^2 M_{N}^2 - M_N^4 + M_e^2 (M_{\pi}^2 + 2 M_N^2 - M_e^2) \right]
\nonumber\\
&& \times
E_{\mu} \sqrt{E_{\mu}^2 - M_{\mu}^2}
\frac{ ( M_{N}^2 - 2 M_{N} E_{\mu} + M_{\mu}^2 - M_e^2)^2}
{(M_{N}^2 - 2 M_{N} E_{\mu} + M_{\mu}^2)} \quad
\left( M_{\mu} \leq E_{\mu} \leq \frac{(M_N^2+M_{\mu}^2-M_e^2)}{2 M_e} \right).
\label{dbGLVEmu}
\eea

The integration of this expression over $E_{\mu}$ can be performed explicitly
(in Ref.~\cite{CDK} it was performed only in the limit $M_e=0$).
The result is Equation~(\ref{GXDDN}) with notations (\ref{notGXDD}), where
the function ${\cal F}(x{\mu},x_{e})$ was obtained in Ref.~\cite{CKZ}.
We write it down here again, for completeness.
\bea
\lefteqn{
{\cal F}(x_{\mu},x_{e}) =
{\Bigg \{}
\lambda^{1/2} (1, x_{\mu}, x_{e}) {\big [} (1 + x_{\mu}) (1 -8 x_{\mu} + x_{\mu}^2)  -
x_{e} (7 - 12 x_{\mu} + 7 x_{\mu}^2)
}
\nonumber\\
&&
- 7 x_{e}^2 (1 + x_{\mu})  + x_{e}^3 {\big ]}
- 24 (1 - x_{e}^2) x_{\mu}^2 \ln 2
\nonumber\\
&&
+  12 {\bigg [} - x_{\mu}^2 (1 - x_{e}^2) \ln x_{\mu}
+ (2 x_{\mu}^2 -x_{e}^2 (1 + x_{\mu}^2)) \ln (1 + x_{\mu}
+ \lambda^{1/2} (1, x_{\mu}, x_{e})  - x_{e})
\nonumber\\
&&
+ x_{e}^2 (1 - x_{\mu}^2)
\ln \left( \frac{(1 - x_{\mu})^2 + (1-x_{\mu}) \lambda^{1/2} (1, x_{\mu}, x_{e}) - x_{e} (1+x_{\mu})}{x_{e}}
\right) {\bigg ]}
{\Bigg \}} \
\label{calF}
\eea

We can obtain the LNV canonical differential decay width, according to
Equation~(\ref{dbBrX}), from the normalized differential decay width
Equation~(\ref{dbGLVEmu}). For this, it turns out to be convenient to
use the following identity:
\be
\frac{2 \K}{\Gamma(\pi^{\pm} \to {\rm all})}
\frac{K_{\pi}^2}{(2 \pi)^4 \Gamma_N M_{\pi}^3} =
96 \frac{1}{M_N^5 M_{\mu}^2 (M_{\pi}^2 - M_{\mu}^2)^2 (1 + \delta g_{\pi})} \
\label{const1}
\ee
which is obtained by using Equations~(\ref{Kpisqr}), (\ref{GNwidth}) and (\ref{barGN})
and (\ref{GPiall}). This then gives us
\bea
\lefteqn{
\frac{d {\overline {\rm Br}_{\pi}^{\rm (LNV)}}}{d E_{\mu}} \equiv
2 \frac{\K}{\Gamma_{\pi}}
\frac{ d \G^{\rm (LNV)}(\pi^{\pm} \to e^{\pm} e^{\pm} \mu^{\mp} \nu)}{d E_{\mu}}
}
\nonumber\\
&&=
48 \frac{1}{M_N^5 M_{\mu}^2 (M_{\pi}^2 - M_{\mu}^2)^2 (1 + \delta g_{\pi})}
\lambda^{1/2}(M_{\pi}^2,M_{N}^2,M_e^2)
\times
\left[ M_{\pi}^2 M_{N}^2 - M_N^4 + M_e^2 (M_{\pi}^2 + 2 M_N^2 - M_e^2) \right]
\nonumber\\
&& \times
E_{\mu} \sqrt{E_{\mu}^2 - M_{\mu}^2}
\frac{ ( M_{N}^2 - 2 M_{N} E_{\mu} + M_{\mu}^2 - M_e^2)^2}
{(M_{N}^2 - 2 M_{N} E_{\mu} + M_{\mu}^2)} \qquad
\left( M_{\mu} \leq E_{\mu} \leq (E_{\mu})_{\rm max} \right)
\label{dbBrLVEmu}
\eea
where
\be
(E_{\mu})_{\rm max} = \frac{(M_N^2+M_{\mu}^2-M_e^2)}{2 M_N}
\label{Emumax}
\ee
\newpage
The LNC canonical differential decay width turns out to be
\ba
\lefteqn{
\frac{d {\overline {\rm Br}_{\pi}^{\rm (LNC)}}}{d E_{\mu}} \equiv
2 \frac{\K}{\Gamma_{\pi}}
\frac{ d \G^{\rm (LNC)}(\pi^{\pm} \to e^{\pm} e^{\pm} \mu^{\mp} \nu)}{d E_{\mu}}
}
\nonumber\\
&&
= \frac{1}{M_N^6 M_{\mu}^2 (M_{\pi}^2 - M_{\mu}^2)^2 (1 + \delta g_{\pi})}
\lambda^{1/2}(M_{\pi}^2,M_{N}^2,M_e^2)
\frac{1}{ \left[ M_{\mu}^2 + M_{N} (-2 E_{\mu} + M_{N}) \right]^3}
\nonumber\\
&& \times {\bigg \{} 8 \sqrt{(E_{\mu}^2 - M_{\mu}^2)} M_{N}
\left[ (2 E_{\mu} - M_{N}) M_{N} - M_{\mu}^2 + M_e^2 \right]^2
\nonumber\\
&& \times
\left[M_{\pi}^2 M_{N}^2  -  M_{N}^4 + M_e^2 (M_{\pi}^2+2 M_{N}^2)  -  M_e^4
\right]
\nonumber\\
&&
\times
{\Big [}  8 E_{\mu}^3 M_{N}^2 - 2 M_{\mu}^2 M_{N} (M_{\mu}^2 +M_{N}^2+ 2 M_e^2)
 - 2 E_{\mu}^2 M_{N} \left( 5 (M_{\mu}^2 + M_{N}^2)+M_e^2 \right)
\nonumber\\
&&
+ E_{\mu}
\left( 3 M_{\mu}^4 + 10 M_{\mu}^2 M_{N}^2 + 3 M_{N}^4 + 3 M_e^2 (M_{\mu}^2 + M_{N}^2) \right) {\Big ]}
{\bigg \}}
 \qquad
\left( M_{\mu} \leq E_{\mu} \leq (E_{\mu})_{\rm max} \right)
\label{dbBrLCEmu}
\ea

It turns out that, upon integration of this expression over $E_{\mu}$,
we obtain the same result as in the X = LNV case, \emph{i.e}.,
Equations~(\ref{bBrc}) with (\ref{notGXDD}) and (\ref{calF}), or equivalently,
Equations~(\ref{GXDDN}) with (\ref{notGXDD}) and
(\ref{calF}).
{We must add that we found a typographical error in  Equation~(A.16) of Ref.~\cite{CDK}, where
$E_{\ell}^2$ must be replaced by $2 E_{\ell}^2$, and in Equations~(B.1c)
and (B3) of Ref.~\cite{CKZ}, where
$(E_{\mu})_{\rm max}$ should read  $(M_{N_j}^2 + M_{\mu}^2 - M_e^2)/(2 M_N)$.}

In the limit $M_e=0$ (which is a good approximation), the canonical differential
decay widths (\ref{dbBrLVEmu}) and (\ref{dbBrLCEmu}) get simplified
\bes
\label{dbBrLVLCMe0}
\bea
\frac{d {\overline {\rm Br}_{\pi}^{\rm (LNV)}}}{d E_{\mu}}{\Bigg |}_{M_e=0} & = &
\frac{48 (M_{\pi}^2 - M_N^2)^2}{M_N^3 M_{\mu}^2 (M_{\pi}^2 - M_{\mu}^2)^2}
\sqrt{E_{\mu}^2 - M_{\mu}^2} E_{\mu} (M_N^2 - 2 M_N E_{\mu} +M_{\mu}^2),
\label{dbBrLVMe0}
\\
\frac{d {\overline {\rm Br}_{\pi}^{\rm (LNC)}}}{d E_{\mu}}{\Bigg |}_{M_e=0} & = &
\frac{48 (M_{\pi}^2 - M_N^2)^2}{M_N^3 M_{\mu}^2 (M_{\pi}^2 - M_{\mu}^2)^2}
\sqrt{E_{\mu}^2 - M_{\mu}^2}
\left[ \frac{1}{2} E_{\mu} (M_N^2+M_{\mu}^2)
- \frac{1}{3} M_N (2 E_{\mu}^2+M_{\mu}^2) \right]
\label{dbBrLCMe0}
\eea
\ees
The full (integrated) canonical branching ratio in the $M_e=0$ limit is
obtained by taking the $x_e =0$ limit of Equation~(\ref{bBrc})
\be
{\overline {\rm Br}_{\pi}} {\big |}_{M_e=0} =
\frac{1}{2} \frac{M_N^2}{M_{\mu}^2} \frac{ (M_{\pi}^2 - M_N^2)^2}{(M_{\pi}^2 - M_{\mu}^2)^2} f \left(\frac{M_{\mu}^2}{M_N^2} \right) \
\label{bBrMe0}
\ee
where the function $f$ is written in Equation~(\ref{fxmu}).

\vspace{12pt}
\noindent{\textit{A.6. Delta Function Approximation for the Imaginary Part of the Propagator Product}}
\label{appdelta}
\vspace{12pt}

In this Appendix we investigate the expression for the imaginary part of the
propagator product, ${\rm Im}(P_1(D) P_2(D)^*)$,
Equation~(\ref{ImP1P2ex}) of Section~\ref{sec:BrCPVsub}.
For convenience we introduce in this Appendix the following
simplified notations $x$, $M^2$, $\Delta$ and $\xi$:
\bes
\label{notsimp}
\bea
x &\equiv& p_N^2, \quad M^2  \equiv M_{N_1}^2
\label{xM2}
\\
\Delta &\equiv& \Delta M_N^2 \equiv M_{N_2}^2 - M_{N_1}^2
\label{Del}
\\
\Gamma_{N_1} & = & \xi \Gamma_N, \quad \Gamma_{N_2} = (2 - \xi) \Gamma_N
\eea
\ees

We note that $\Delta>0$ by convention; and $0 < \xi < 2$.
Further, $\Gamma_{N_1}+\Gamma_{N_2} = 2 \Gamma_N$, in
accordance with the definition of $\Gamma_N$ Equation~(\ref{GNy}).
Since we always have $\Gamma_{N_j} \ll M_{N_j}$ (the neutrinos $N_j$ are
sterile), the relation (\ref{deltas}) holds, \emph{i.e}.,
\be
\frac{ \Gamma_{N_j} M_{N_j}}{(x - M_{N_j}^2)^2 + \Gamma_{N_j}^2 M_{N_j}^2}
= \pi \delta(x - M_{N_j}^2)
\label{deltaA}
\ee

We can write the right-hand side
of Equation~(\ref{ImP1P2ex}) for ${\rm Im}(P_1(D) P_2(D)^*)$ as
\be
{\rm Im}\left( P_1(D) P_2(D)^* \right) = {\cal R}_1 + {\cal R}_2
\label{ImP1P21}
\ee
where ${\cal R}_1$ and ${\cal R}_2$ can be written, in our notation, as
\bes
\label{R1R2}
\bea
 {\cal R}_1 &=& \frac{(x - M^2) (2 - \xi) \Gamma_N \sqrt{M^2+\Delta}}
{\left[ (x-M^2)^2 + \xi^2 \Gamma_N^2 M^2 \right]
 \left[ (x-M^2-\Delta)^2 + (2-\xi)^2 \Gamma_N^2 (M^2+\Delta) \right]}
\label{R1a}
\\
&=&
\eta_1 \times \frac{\pi}{\Delta} \delta(x - M^2 - \Delta)
\label{R1b}
\\
 {\cal R}_2 &=& - \frac{\xi \Gamma_N M (x - M^2-\Delta)}
{\left[ (x-M^2)^2 + \xi^2 \Gamma_N^2 M^2 \right]
 \left[ (x-M^2-\Delta)^2 + (2-\xi)^2 \Gamma_N^2 (M^2+\Delta) \right]}
\label{R2a}
\\
&=&
\eta_2 \times \frac{\pi}{\Delta} \delta(x - M^2)
\label{R2b}
\eea
\ees

In Equations~(\ref{R1b}) and (\ref{R2b}), the identity (\ref{deltaA}) was used,
and we introduced two (dimensionless) parameters $\eta_j$ ($j=1,2$).
We want to obtain these two parameters $\eta_j$.
They can be obtained by integrating analytically the explicit expressions
(\ref{R1a}) and (\ref{R2a}) for ${\cal R}_j(x)$ over $x$.
For example, integration of ${\cal R}_1(x)$ gives
\be
\label{intg}
\int_{-\infty}^{+\infty} dx \frac{(x - M^2) (2 - \xi) \Gamma_N \sqrt{M^2+\Delta}}
{\left[ (x-M^2)^2 + \xi^2 \Gamma_N^2 M^2 \right]
 \left[ (x-M^2-\Delta)^2 + (2-\xi)^2 \Gamma_N^2 (M^2+\Delta) \right]}
= \frac{\pi \Delta}{( \Delta^2 + 4 \Gamma_N^2 M_{*}^2)}
\ee
where
\bes
\label{M2ast}
\bea
M_{*}^2 &=& \frac{1}{2} M^2 \left[ \left( 2 - \xi(2 - \xi) \right)
+ \xi(2 - \xi)\sqrt{1 + \Delta/M^2} \right]
+ \frac{1}{4} (2 - \xi)^2 \Delta
\label{M2asta}
\\
& = & M^2 \left[ 1 + (1 - \xi/2) \frac{\Delta}{M^2}
+ {\cal O} \left( \frac{\Delta^2}{M^4} \right) \right]
\label{M2astb}
\eea
\ees

Therefore, in the case of near degeneracy ($\Delta \ll M^2$) we have
$M_{*}^2 = M^2$.
If we now use in the integration over $d x$ the expression (\ref{R1b}) instead,
take into account $M_{*}^2 = M^2$ in the case of near degeneracy,
and compare with (\ref{intg}), we obtain the following expression for the
parameter $\eta_1$ by comparison with (\ref{intg}):
\bes
\label{eta1}
\bea
\eta_1 \frac{1}{\Delta} & = & \frac{\Delta}{( \Delta^2 + 4 \Gamma_N^2 M^2)}
\quad (\Delta \ll M^2)
\label{eta1a}
\\
\eta_1 &=& \frac{y^2}{y^2+1} \quad
\left( y \equiv \frac{\Delta}{(2 M \Gamma_N)}, \; \Delta \ll M^2 \right)
\label{eta1b}
\eea
\ees
where in Equation~(\ref{eta1b}) we use the usual notation in this paper
$y \equiv (M_{N_2} - M_{N_1})/\Gamma_N = \Delta/(2 M \Gamma_N)$.
Here we note that $\Delta \equiv (M_{N_2}^2-M_{N_1}^2) = (M_{N_2}-M_{N_1}) 2 M_{N_1}$
in the case of near degeneracy \mbox{$\Delta \ll M^2 \equiv M_{N_1}^2$}.

Doing the same procedure with the quantity ${\cal R}_2$, we obtain
for $\eta_2$ the very same result as for $\eta_1$
\be
\eta_1 = \eta_2 = \frac{y^2}{y^2+1} \quad   (\Delta \ll M^2)
\label{eta12}
\ee



\begin{thebibliography}{------}


\bibitem{0nubb}
  Racah, G.
  On the symmetry of particle and antiparticle.
  {\it Nuovo Cim.\/}  {\bf 1937}, \emph{14}, 322--328.
\bibitem{0nubb2}
  Furry, W.H.
  On transition probabilities in double beta-disintegration.
  {\it Phys.\ Rev.\/}  {\bf 1939}, \emph{56}, 1184--1193.
\bibitem{0nubb3}
Primakoff, H.; Rosen, S.P. Double beta decay. {\it Rep. Prog. Phys.\/} {\bf 1959}, \emph{22}, 121--166.
\bibitem{0nubb31}
Primakoff, H.; Rosen, S.P.
  Nuclear double-beta decay and a new limit on lepton nonconservation.
  {\it Phys.\ Rev.\/}  {\bf 1969}, \emph{184}, 1925--1933.
\bibitem{0nubb4}
Primakoff, H.; Rosen, S.P.
  Baryon number and lepton number conservation laws.
  {\it Ann.\ Rev.\ Nucl.\ Part.\ Sci.\/}  {\bf 1981}, \emph{31}, 145--192.
\bibitem{0nubb5}
  Schechter, J.;  Valle, J.W.F.
  Neutrinoless double beta decay in $SU(2) \times U(1)$ theories.
  {\it Phys.\ Rev.\ D\/} {\bf 1982}, \emph{25}, 2951--2954.
\bibitem{0nubb6}
 Doi, M.; Kotani, T.; Takasugi, E.
  Double beta decay and Majorana neutrino.
  {\it Prog.\ Theor.\ Phys.\ Suppl.\/}  {\bf 1985}, \emph{83}, 1--368.
\bibitem{0nubb7}
  Elliott, S.R.; Engel, J.
  Double beta decay.
  {\it J.\ Phys.\ G\/}  {\bf 2004}, \emph{30}, R183--R215.
\bibitem{0nubb8}
  Rodin, V.A.; Faessler, A.; \v{S}imkovi\'c, F.; Vogel, P.
  Assessment of uncertainties in QRPA $0\nu\beta\beta$-decay nuclear matrix elements.
  {\it Nucl.\ Phys.\ A\/} {\bf 2006}, \emph{766}, 107--131.

\bibitem{0nubb9}
 Rodin, V.A.; Faessler, A.; \v{S}imkovi\'c, F.; Vogel, P.
Erratum to: ``Assessment of uncertainties in QRPA $0\nu\beta\beta$-decay nuclear matrix elements'' [Nucl. Phys. A 766 (2006) 107]. {\it Nucl.\ Phys.\ A\/} {\bf 2007}, \emph{793}, 213--215.

\bibitem{scatt1}
  Keung, W.Y.; Senjanovi\'c, G.
  Majorana neutrinos and the production of the right-handed charged gauge boson.
  {\it Phys.\ Rev.\ Lett.\/}  {\bf 1983}, \emph{50}, 1427--1430.
\bibitem{scatt12}
  Tello, V.; Nemev\v{s}ek, M.; Nesti, F.; Senjanovi\'c, G.; Vissani, F.
  Left-right symmetry: From LHC to neutrinoless double beta decay.
  {\it Phys.\ Rev.\ Lett.\/}  {\bf 2011}, \emph{106}, doi:10.1103/PhysRevLett.106.151801.
\bibitem{scatt13}
  Nemev\v{s}ek, M.; Nesti, F.; Senjanovi\'c, G.; Tello, V.
  Neutrinoless double beta decay: Low left-right symmetry scale?
Available online: http://arxiv.org/abs/1112.3061 (accessed on 27 April 2015).
\bibitem{scatt14}
  Senjanovi\'c, G.
  Neutrino mass: From LHC to grand unification.
  {\it Riv.\ Nuovo Cim.\/}  {\bf 2011}, \emph{34}, 1--68.

\bibitem{scatt2}
  Buchm\"uller, W.; Greub, C.
  Heavy Majorana neutrinos in electron-positron and electron-proton collisions.
  {\it Nucl.\ Phys.\ B\/} {\bf 1991}, \emph{363}, 345--368.

\bibitem{scatt3}
  Helo, J.; Hirsch, M.; Kovalenko, S.
  Heavy neutrino searches at the LHC with displaced vertices.
  {\it Phys.\ Rev.\ D\/} {\bf 2014}, \emph{89}, doi:10.1103/PhysRevD.89.073005.

\bibitem{scatt4}
  Kohda, M.; Sugiyama, M.; Tsumura, K.
  Lepton number violation at the LHC with leptoquark and diquark.
  {\it Phys.\ Lett.\ B\/} {\bf 2013}, \emph{718}, 1436--1440.

%
\bibitem{DevscattLR}
Chen, C.Y.; Bhupal Dev, P.S. Multi-lepton collider signatures of heavy Dirac and Majorana neutrinos. {\it  Phys.\ Rev.\ D\/} {\bf 2012}, \emph{85}, doi:10.1103/PhysRevD.85.093018.
\bibitem{DevscattLR2}
Chen, C.Y.; Bhupal Dev, P.S.; Mohapatra, R.N. Probing heavy-light neutrino mixing in Left-Right Seesaw models at the LHC. {\it Phys.\ Rev.\ D\/} {\bf 2013}, \emph{88}, doi:10.1103/PhysRevD.88.033014.
%

%
\bibitem{Devscattssaw}
Bhupal Dev, P.S.; Pilaftsis, A.; Yang, U.-K. New production mechanism for heavy neutrinos at the LHC. {\it Phys.\ Rev.\ Lett.\/}  {\bf 2014}, \emph{112}, 081801:1--081801:5.
\bibitem{Devscattssaw1}
Das, A.; Bhupal Dev, P.S.; Okada, N. Direct bounds on electroweak scale pseudo-Dirac neutrinos from $\sqrt{s}=8$ TeV LHC data. {\it Phys.\ Lett.\ B\/} {\bf 2014}, \emph{735}, 364--370.
%

\bibitem{LittSh}
 Littenberg, L.S.; Shrock, R.E.
  Upper bounds on lepton number violating meson decays.
  {\it Phys.\ Rev.\ Lett.\/}  {\bf 1992}, \emph{68}, 443--446.
\bibitem{LittSh2}
 Littenberg, L.S.; Shrock, R.E.
  Implications of improved upper bounds on $|\Delta L| = 2$ processes.
  {\it Phys.\ Lett.\ B\/} {\bf 2000}, \emph{491}, 285--290.

\bibitem{DGKS}
 Dib, C.; Gribanov, V.; Kovalenko, S.; Schmidt, I.
  K meson neutrinoless double muon decay as a probe of neutrino masses and mixings.
  {\it Phys.\ Lett.\ B\/} {\bf 2000}, \emph{493}, 82--87.


\bibitem{Ali}
  Ali, A.; Borisov, A.V.; Zamorin, N.B.
  Majorana neutrinos and same sign dilepton production at LHC and in rare meson decays.
  {\it Eur.\ Phys.\ J.\ C\/} {\bf 2001}, \emph{21}, 123--132.

 \bibitem{CDKK}
  Cveti\v{c}, G.; Dib, C.; Kang, S.K.; Kim, C.S.
  Probing Majorana neutrinos in rare $K$ and $D, D_s, B, B_c$ meson decays.
  {\it Phys.\ Rev.\ D\/} {\bf 2010}, \emph{82}, doi:10.1103/PhysRevD.82.053010.


\bibitem{IvKo}
  Ivanov, M.A.; Kovalenko, S.G.
  Hadronic structure aspects of $K^+ \to \pi^- + \ell_1^{+} + \ell_2^{+}$ decays.
  {\it Phys.\ Rev.\ D\/} {\bf 2005}, \emph{71}, doi:10.1103/PhysRevD.71.053004.

\bibitem{GoJe}
  De Gouvea, A.; Jenkins, J.
  A Survey of lepton number violation via effective operators.
  {\it Phys.\ Rev.\ D\/} {\bf 2008}, \emph{77}, doi:10.1103/PhysRevD.77.013008.

 \bibitem{Atre}
  Atre, A.; Han, T.; Pascoli, S.; Zhang, B.
  The Search for heavy Majorana neutrinos.
  {\it JHEP\/} {\bf 2009}, \emph{0905}, doi:10.1088/1126-6708/2009/05/030.
and references therein.

\bibitem{HKS}
  Helo, J.C.; Kovalenko, S.; Schmidt, I.
  Sterile neutrinos in lepton number and lepton flavor violating decays.
  {\it Nucl.\ Phys.\ B\/} {\bf 2011}, \emph{853}, 80--104.

%
\bibitem{QLD}
Quintero, N.; L\'opez Castro, G.; Delepine, D. Lepton number violation in top quark and neutral B meson decays. {\it Phys.\ Rev.\ D\/} {\bf 2011}, \emph{84}, doi:10.1103/PhysRevD.84.096011.

\bibitem{QLD2}
Quintero, N.; L\'opez Castro, G.; Delepine, D.
Erratum: Lepton number violation in top quark and neutral B meson decays [Phys. Rev. D 84, 096011 (2011)].
{\it Phys.\ Rev.\ D\/} {\bf 2012}, \emph{86}, doi:10.1103/PhysRevD.86.079905.

\bibitem{QLD1}
L\'opez Castro, G.; Quintero, N. Bounding resonant Majorana neutrinos from four-body B and D decays. {\it Phys.\ Rev.\ D\/} {\bf 2013}, \emph{87}, doi:10.1103/PhysRevD.87.077901. 
%

%
\bibitem{Abada} Abada, A.; Teixeira, A.M.; Vicente, A.; Weiland, C.
Sterile neutrinos in leptonic and semileptonic decays. {\it JHEP\/} {\bf 2014}, \emph{1402}, doi:10.1007/JHEP02(2014)091.
%

%
\bibitem{Wang} Wang, Y.; Bao, S.S.; Li, Z.H.; Zhu, N.; Si, Z.G. Study Majorana neutrino contribution to B-meson semi-leptonic rare decays. {\it Phys.\ Lett.\ B\/} {\bf 2014}, \emph{736}, 428--432.
%

%
\bibitem{Boya} Boyanovsky, D. Nearly degenerate heavy sterile neutrinos in cascade decay: Mixing and oscillations. {\it Phys.\ Rev.\ D\/} {\bf 2014}, \emph{90}, doi:10.1103/PhysRevD.90.105024.
%


\bibitem{CDK}
  Cveti\v{c}, G.; Dib, C.; Kim, C.S.
  Probing Majorana neutrinos in rare $\pi^+ \to e^+ e^+ \mu^- \nu$ decays.
  {\it JHEP\/} {\bf 2012}, \emph{1206}, doi:10.1007/JHEP06(2012)149.

\bibitem{CKZ}
  Cveti\v{c}, G.; Kim, C.S.; Zamora-Sa\'a, J.
  CP violations in $\pi^{\pm}$ meson decay.
  {\it J.\ Phys.\ G\/} {\bf 2014}, \emph{41}, doi:10.1088/0954-3899/41/7/075004.

\bibitem{CKZ2}
  Cveti\v{c}, G.; Kim, C.S.; Zamora-Sa\'a, J.
  CP violation in lepton number violating semileptonic decays of $K, D, D_s, B, B_c$.
  {\it Phys.\ Rev.\ D\/} {\bf 2014}, \emph{89}, doi:10.1103/PhysRevD.89.093012.

\bibitem{DCK}
  Dib, C.O.; Campos, M.; Kim, C.S.
  CP Violation with Majorana neutrinos in K meson decays.
  {\it JHEP\/} {\bf 2015}, \emph{1502},  doi:10.1007/JHEP02(2015)108.

\bibitem{oscatm}
  Fukuda, Y.; {\it et al.}  [Super-Kamiokande Collaboration].
  Evidence for oscillation of atmospheric neutrinos.
  {\it Phys.\ Rev.\ Lett.\/}  {\bf 1998}, \emph{81}, 1562--1567.

\bibitem{oscsol}
  Ahmad, Q.R.; {\it et al.}  [SNO Collaboration].
  Direct evidence for neutrino flavor transformation from neutral current interactions in the Sudbury Neutrino Observatory.
{\it Phys.\ Rev.\ Lett.\/}  {\bf 2002}, \emph{89}, doi:10.1103/PhysRevLett.89.011301.
\bibitem{oscsol2}
  Lipari, P.
  CP violation effects and high-energy neutrinos.
  {\it Phys.\ Rev.\ D\/} {\bf 2001}, \emph{64}, doi:10.1103/PhysRevD.64.033002.
\bibitem{oscsol3}
  Rahman, Z.; Dasgupta, A.; Adhikari, R.
  Non-standard interaction effect on CP violation in neutrino oscillation with super-beam.
Available online: http://arxiv.org/abs/1210.2603 (accessed on 28 April 2015).
\bibitem{oscsol4}
Rahman, Z.; Dasgupta, A.; Adhikari, R.
  Discovery reach of CP violation and non-standard interactions in low energy neutrino factory.
  Available online: http://arxiv.org/abs/1210.4801 (accessed on 28 April 2015).

\bibitem{oscnuc}
  Eguchi, K.; {\it et al.}  [KamLAND Collaboration].
  First results from KamLAND: Evidence for reactor anti-neutrino disappearance.
  {\it Phys.\ Rev.\ Lett.\/}  {\bf 2003}, \emph{90}, doi:10.1103/PhysRevLett.90.021802.

\bibitem{PlanckColl}
Planck Collaboration.
Planck 2013 results. XVI. Cosmological parameters.
{\it Astron. Astrophys.\/} {\bf 2014}, \emph{571}, A16 (66pp.).

\bibitem{seesaw}
  Minkowski, P.
 $\mu \to e \gamma$ at a rate of one out of 1-billion muon decays?
  {\it Phys.\ Lett.\ B\/} {\bf 1977}, \emph{67}, 421--428.
\bibitem{seesaw2}
Gell-Mann, M.; Ramond, P.; Slansky, R.
The Family Group in Grand Unified Theories. In {\it Supergravity\/};
van Nieuwenhuizen, P., Freedman, D.Z., Eds.;
North-Holland: Amsterdam, The Netherlands, 1979.
\bibitem{seesaw3}
  Yanagida, T.
  Horizontal symmetry and masses of neutrinos.
  \textit{Prog. Theor. Phys.} \textbf{1980}, \textit{64}, 1103--1105.
\bibitem{seesaw4}
Glashow, S.L.  {\it Quarks and Leptons\/}; L\'evy, M., Basdevant, J.-L., Speiser, D., Weyers, J., Gastmans, R., Jacob, M.; Eds.;
Cargese, Plenum: New York, NY, USA, 1980.
\bibitem{seesaw5}
  Mohapatra, R.N.; Senjanovi\'c, G.
  Neutrino mass and spontaneous parity violation.
  {\it Phys.\ Rev.\ Lett.\/}  {\bf 1980}, \emph{44}, 912--915.

\bibitem{Wyler1}
Wyler, D.; Wolfenstein, L.
Massless neutrinos in left-right symmetric models.
{\it Nucl.\ Phys.\ B\/} {\bf 1983}, \emph{218}, 205--214.

\bibitem{Witten1}
Witten, E.
Symmetry breaking patterns in superstring models.
 {\it Nucl.\ Phys.\ B\/} {\bf 1985}, \emph{258}, 75--100.

\bibitem{Mohapatra1}
Mohapatra, R.N.; Valle, J.W.F.
Neutrino mass and baryon number nonconservation in superstring models.
  {\it Phys.\ Rev.\ D\/} {\bf 1986}, \emph{34}, 1642--1645.

\bibitem{Malinsky1}
Malinsky, M.; Romao, J.C.; Valle, J.W.F.
Novel supersymmetric $SO(10)$ seesaw mechanism.
{\it Phys.\ Rev.\ Lett.\/}  {\bf 2005}, \emph{95}, 161801:1--161801:5.

%
\bibitem{Devseesaw}
Dev, P.S.B.; Mohapatra, R.N. TeV Scale Inverse Seesaw in SO(10) and Leptonic Non-Unitarity Effects. {\it  Phys.\ Rev.\ D\/} {\bf 2010}, \emph{81}, doi:10.1103/PhysRevD.81.013001.
\bibitem{Devseesaw2}
Dev, P.S.B.; Pilaftsis, A. Minimal Radiative Neutrino Mass Mechanism for Inverse Seesaw Models. {\it Phys.\ Rev.\ D\/} {\bf 2012}, \emph{86}, doi:10.1103/PhysRevD.86.113001.
\bibitem{Devseesaw3}
Lee, C.H.; Dev, P.S.B.; Mohapatra, R.N. Natural TeV-scale left-right seesaw mechanism for neutrinos and experimental tests. {\it Phys.\ Rev.\ D\/} {\bf 2013}, \emph{88}, doi:10.1103/PhysRevD.88.093010.
%

\bibitem{nuMSM}
  Asaka, T.; Blanchet, S.; Shaposhnikov, M.
  The $\nu$MSM, dark matter and neutrino masses.
  {\it Phys.\ Lett.\ B\/} {\bf 2005}, \emph{631}, 151--156.
\bibitem{nuMSM2}
  Asaka, T.; Shaposhnikov, M.
  The $\nu$MSM, dark matter and baryon asymmetry of the universe.
  {\it Phys.\ Lett.\ B\/} {\bf 2005}, \emph{620}, 17--26.

\bibitem{HeAAS}
  He, X.G.; Oh, S.; Tandean, S.J.; Wen, C.C.
  Large mixing of light and heavy neutrinos in seesaw models and the LHC.
  {\it Phys.\ Rev.\ D\/} {\bf 2009}, \emph{80}, doi:10.1103/PhysRevD.80.073012.

\bibitem{HeAAS2}
  Del Aguila, F.; Aguilar-Saavedra, J.A.; de Blas, J.; Zralek, M.
  Looking for signals beyond the neutrino Standard Model.
  {\it Acta Phys.\ Polon.\ B\/} {\bf 2007}, \emph{38}, 3339--3348.


\bibitem{KS}
  Kersten, J.; Smirnov, A.Y.
  Right-handed neutrinos at CERN LHC and the mechanism of neutrino mass generation.
  {\it Phys.\ Rev.\ D\/} {\bf 2007}, \emph{76}, doi:10.1103/PhysRevD.76.073005.


\bibitem{AMP}
  Ibarra, A.; Molinaro, E.; Petcov, S.T.
  TeV scale see-saw mechanisms of neutrino mass generation, the Majorana nature of the heavy singlet neutrinos and $(\beta\beta)_{0\nu}$-Decay.
  {\it JHEP\/} {\bf 2010}, \emph{1009}, 108--129.

\bibitem{oscCP}
  Cabibbo, N.
  Time reversal violation in neutrino oscillation.
  {\it Phys.\ Lett.\ B\/} {\bf 1978}, \emph{72}, 333--335.

\bibitem{Shapo}
  Gorbunov, D.; Shaposhnikov, M.
  How to find neutral leptons of the $\nu$MSM?
  {\it JHEP\/} {\bf 2007}, \emph{0710}, doi:10.1088/1126-6708/2007/10/015.
\bibitem{Shapo2}
  Shaposhnikov, M.
  Is there a new physics between electroweak and Planck scales?
Available online: http://arxiv.org/abs/0708.3550 (accessed on 28 April 2015).
\bibitem{Shapo3}
  Boyarsky, A.; Ruchayskiy, O.; Shaposhnikov, M.
  The role of sterile neutrinos in cosmology and astrophysics.
  {\it Ann.\ Rev.\ Nucl.\ Part.\ Sci.\/}  {\bf 2009}, \emph{59}, 191--214.

\bibitem{Shapo4}
Shaposhnikov, M.
  Neutrino physics within and beyond the three flavour oscillation.
  {\it J.\ Phys.\ Conf.\ Ser.\/}  {\bf 2013}, \emph{408}, doi:10.1088/1742-6596/408/1/012015.
\bibitem{Shapo5}
  Canetti, L.; Drewes, M.; Shaposhnikov, M.
  Sterile neutrinos as the origin of dark and baryonic matter.
  {\it Phys.\ Rev.\ Lett.\/}  {\bf 2013}, \emph{110}, doi:10.1103/PhysRevLett.110.061801.
\bibitem{Shapo6}
  Canetti, L.; Drewes, M.; Frossard, T.; Shaposhnikov, M.
  Dark matter, baryogenesis and neutrino oscillations from right handed neutrinos.
  {\it Phys.\ Rev.\ D\/} {\bf 2013}, \emph{87}, doi:10.1103/\linebreak PhysRevD.87.093006.

%
\bibitem{1404.7114} Canetti, L.; Drewes, M.; Garbrecht, B. Probing leptogenesis with GeV-scale sterile neutrinos at LHCb and Belle II. {\it Phys.\ Rev.\ D\/} {\bf 2014},\emph{ 90}, doi:10.1103/PhysRevD.90.125005.
%

%
\bibitem{1502.00477} Drewes, M.; Garbrecht, B. Experimental and cosmological constraints on heavy neutrinos. Available online: http://arxiv.org/abs/1502.00477 (accessed on 28 April 2015).
%

%
\bibitem{Pilaftsis} Pilaftsis, A. CP violation and baryogenesis due to heavy Majorana neutrinos. {\it Phys.\ Rev.\ D\/} {\bf 1997}, \emph{56}, 5431--5451.

\bibitem{Pilaftsis2}
Bray, S.; Lee, J.S.; Pilaftsis, A. Resonant CP violation due to heavy neutrinos at the LHC. {\it Nucl.\ Phys.\ B\/} {\bf 2007}, \emph{786}, 95--118.
%

%
\bibitem{Akhmedov} Akhmedov, E.; Kartavtsev, A.; Lindner, M.; Michaels, L.; Smirnov, J. Improving Electro-Weak Fits with TeV-scale Sterile Neutrinos. {\it JHEP\/} {\bf 2013}, \emph{1305}, doi:10.1007/JHEP05(2013)081.
%

%
\bibitem{Qian} Qian, X.; Zhang, C.; Diwan, M.; Vogel, P. Unitarity Tests of the Neutrino Mixing Matrix. Available online: http://arxiv.org/abs/1308.5700 (accessed on 28 April 2015).
%

%
\bibitem{Basso} Basso, L.; Fischer, O.; van der Bij, J.J. Precision tests of unitarity in leptonic mixing. {\it Europhys.\ Lett.\/}  {\bf 2014}, \emph{105}, doi:10.1209/0295-5075/105/11001.
%

\bibitem{CERN-SPS}
Bonivento,  W.; Boyarsky, A.; Dijkstra, H.; Egede, U.; Ferro-Luzzi, M.; Goddard, B.;  Golutvin,~A.; Gorbunov, D.; Jacobsson, R.; Panman, J.; \textit{et al.} Proposal to Search for Heavy Neutral Leptons at the SPS. Available online: http://arxiv.org/abs/1310.1762 (accessed on 28 April 2015).


\bibitem{CERN-SPS2}
Jacobson, R. Search for heavy neutral neutrinos at the SPS.
Presented at High Energy Physics in the LHC Era, UTFSM,
Valpara\'{\i}so, Chile, 16--20 December 2014. Available online:
https://indico.cern.ch/event/252857/contribution/215
(accessed on April 29, 2015)

\bibitem{commKim}
Dib, C.O.; Kim, C.S.
  Remarks on the lifetime of sterile neutrinos and the effect on detection of rare meson decays $M^+ \to M^{\prime}-\ell^+\ell^+$.
  {\it Phys.\ Rev.\ D\/} {\bf 2014}, \emph{89}, doi:10.1103/PhysRevD.89.077301.

\bibitem{Benes}
  Bene\v{s}, P.; Faessler, A.; \v{S}imkovi\'c, F.; Kovalenko, S.
  Sterile neutrinos in neutrinoless double beta decay.
 {\it  Phys.\ Rev.\ D\/} {\bf 2005}, \emph{71}, doi:10.1103/PhysRevD.71.077901.

\bibitem{Belanger}
  B\'elanger, G.; Boudjema, F.; London, D.; Nadeau, H.
  Inverse neutrinoless double beta decay revisited.
  {\it Phys.\ Rev.\ D\/} {\bf 1996}, \emph{53}, 6292--6301.
\bibitem{Belanger2}
  London, D.
  Inverse neutrinoless double beta decay (and other $\Delta L = 2$ processes). Available online: http://arxiv.org/abs/hep-ph/9907419 (accessed on 28 April 2015).


\bibitem{Kusenko}
  Kusenko, A.; Pascoli, S.; Semikoz, D.
  New bounds on MeV sterile neutrinos based on the accelerator and Super-Kamiokande results.
  {\it JHEP\/} {\bf 2005}, \emph{0511}, doi:10.1088/1126-6708/\linebreak 2005/11/028.

\bibitem{beamdump}
WA66 Collaboration.
 Search for heavy neutrino decays in the BEBC beam dump experiment.
  {\it Phys.\ Lett.\ B\/} {\bf 1985}, \emph{160}, 207--211.
\bibitem{beamdump2}
NA3 Collaboration.
  Mass and lifetime limits on new longlived particles in $300 GeV/c$ $\pi^-$ interactions.
  {\it Z.\ Phys.\ C\/} {\bf 1986}, \emph{31}, 21--32.
\bibitem{beamdump3}
  Bernardi, G.; Carugno, G.; Chauveau, J.; Dicarlo, F.; Dris, M.; Dumarchez, J.; Ferro-Luzzi, M.; Levy, J.-M.; Lukas, D.; Perreau, J.-M.; {\it et al.}
  Further limits on heavy neutrino couplings.
  {\it Phys.\ Lett.\ B\/} {\bf 1988}, \emph{203}, 332--334.
\bibitem{beamdump4}
FMMF Collaboration.
  Search for neutral weakly interacting massive particles in the Fermilab Tevatron wide band neutrino beam.
  {\it Phys.\ Rev.\ D\/} {\bf 1995}, \emph{52}, doi:10.1103/PhysRevD.52.6.
\bibitem{beamdump5}
NuTeV and E815 Collaborations.
  Search for neutral heavy leptons in a high-energy neutrino beam.
  {\it Phys.\ Rev.\ Lett.\/}  {\bf 1999}, \emph{83}, 4943--4946.

\bibitem{l3etal}
 L3 Collaboration.
  Search for isosinglet neutral heavy leptons in Z0 decays.
  {\it Phys.\ Lett.\ B\/} {\bf 1992}, \emph{295}, 371--382.
\bibitem{l3etal2}
  CHARM II Collaboration.
  Search for heavy isosinglet neutrinos.
  {\it Phys.\ Lett.\ B\/} {\bf 1995}, \emph{343}, 453--458
  [{\it Phys.\ Lett.\ B\/} {\bf 1995}, \emph{351}, 387--392].

\bibitem{delphi}
DELPHI Collaboration.
  Search for neutral heavy leptons produced in Z decays.
  {\it Z.\ Phys.\ C\/} {\bf 1997}, \emph{74}, 57--71.

\bibitem{noch}
NOMAD Collaboration.
  Search for heavy neutrinos mixing with tau neutrinos.
  {\it Phys.\ Lett.\ B\/} {\bf 2001}, \emph{506}, 27--38.
\bibitem{noch2}
  Orloff, J.; Rozanov, A.N.; Santoni, C.
  Limits on the mixing of tau neutrino to heavy neutrinos.
  {\it Phys.\ Lett.\ B\/} {\bf 2002}, \emph{550}, 8--15.


\bibitem{PDG2014}
 Particle Data Group Collaboration.
  Review of Particle Physics.
  {\it Chin.\ Phys.\ C\/} {\bf 2014}, \emph{38}, doi:10.1088/1674-1137/38/9/090001.

\bibitem{CKWN}
  Cveti\v{c}, G.; Kim, C.S.; Wang, G.L.; Namgung, W.
  Decay constants of heavy meson of 0-state in relativistic Salpeter method.
  {\it Phys.\ Lett.\ B\/} {\bf 2004}, \emph{596}, 84--89.

\bibitem{PIENU:2011aa}
PIENU Collaboration.
  Search for massive neutrinos in the decay $\pi \to e \nu$.
  {\it Phys.\ Rev.\ D\/} {\bf 2011}, \emph{84}, doi:10.1103/PhysRevD.84.052002.

\bibitem{BmuN}
  Yamazaki, T.; Ishikawa, T.; Akiba, Y.; Iwasaki, M.; Tanaka, K.H.; Ohtake, S.; Tamura, H.; Nakajima, M.; Yamanaka, T.; Arai, I.;  {\it et al.}
  Search for Heavy Neutrinos in Kaon Decay.
  {\it Conf.\ Proc.\ C\/} {\bf 1984}, \emph{840719}, 262--262.
\bibitem{BmuN2}
  Hayano, R.S.; Taniguchi, T.; Yamanaka, T.; Tanimori, T.; Enomoto, R.; Ishibashi, A.; Ishikawa, T.; Sato, S.; Fujii, T.; Yamazaki, T.; {\it et al.}
  Heavy neutrino search using $K_{\mu 2}$ decay.
  {\it Phys.\ Rev.\ Lett.\/}  {\bf 1982}, \emph{49}, doi:10.1103/PhysRevLett.49.1305.
\bibitem{BmuN3}
  Kusenko, A.; Pascoli, S.; Semikoz, D.
  New bounds on MeV sterile neutrinos based on the accelerator and Super-Kamiokande results.
  {\it JHEP\/} {\bf 2005}, \emph{0511}, doi:10.1088/1126-6708/\linebreak 2005/11/028.

\bibitem{BtauN}
  Orloff, J.; Rozanov, A.N.; Santoni, C.
  Limits on the mixing of tau neutrino to heavy neutrinos.
  {\it Phys.\ Lett.\ B\/} {\bf 2002}, \emph{550}, 8--15.

\bibitem{Ruchayskiy:2011aa}
  Ruchayskiy, O.; Ivashko, A.
  Experimental bounds on sterile neutrino mixing angles.
  {\it JHEP\/} {\bf 2012}, \emph{1206}, doi:10.1007/JHEP06(2012)100.


%
\bibitem{Devlimits} Deppisch, F.F.; Dev, P.S.B.; Pilaftsis, A. Neutrinos and Collider Physics. Available online: http://arxiv.org/abs/1502.06541 (accessed on 28 April 2015).
%

\bibitem{ProjX}
Project X and the Science of
the Intensity Frontier. In Proceedings of the
Project X Physics Workshop, Fermilab, USA, 9--10 November 2009.
Available online: http://www.fnal.gov/projectx/pdfs/ProjectXwhitepaperJan.v2.pdf (accessed on April 29, 2015).


\bibitem{Geer}
Geer, S. Private communication,, Fermilab, 2012 (sgeer@fnal.gov).

\bibitem{Pontecorvo}
  Pontecorvo, B.
  Inverse beta processes and nonconservation of lepton charge.
  {\it Sov.\ Phys.\ JETP\/} {\bf 1958}, \emph{7}, 172--173
  [{\it Zh.\ Eksp.\ Teor.\ Fiz.\/}  {\bf 1957}, \emph{34}, 247].
\bibitem{Pontecorvo2}
  Neutrino experiments and the problem of conservation of leptonic charge.
  {\it Sov.\ Phys.\ JETP\/} {\bf 1968}, \emph{26}, 984--988
  [{\it Zh.\ Eksp.\ Teor.\ Fiz.\/}  {\bf 1967}, \emph{53}, 1717].

\bibitem{MNS}
  Maki, Z.; Nakagawa, M.; Sakata, S.
  Remarks on the unified model of elementary particles.
  {\it Prog.\ Theor.\ Phys.\/}  {\bf 1962}, \emph{28}, 870--880.

\bibitem{Bilenky}
Bilenky, S. {\it Introduction to the Physics of Massive and Mixed Neutrinos};
Lecture Notes in Physics 817; Springer Verlag: Berlin, Heidelberg, Germany, 2010.


\bibitem{Gamiz:2003pi}
  Gamiz, E.; Prades, J.; Scimemi, I.
  Charged kaon K $\to$ 3$\pi$ CP violating asymmetries at NLO in CHPT.
  \emph{JHEP} \textbf{2003}, \emph{0310}, doi:10.1088/1126-6708/2003/10/042.

\bibitem{GKS}
  Gribanov, V.; Kovalenko, S.; Schmidt, I.
  Sterile neutrinos in tau lepton decays.
  {\it Nucl.\ Phys.\ B\/} {\bf 2001}, \emph{607}, 355--368.


\end{thebibliography}
\end{document}